\documentclass[11pt]{iopart}

\usepackage{iopams}  
\usepackage{graphicx}
\usepackage{cases}
\usepackage{url}
\usepackage{hyperref}
\usepackage{amssymb}
\usepackage{natbib}
\usepackage{color}
\usepackage[normalem]{ulem}

\begin{document}

\title[Comparing timelike geodesics around a Kerr black hole and a boson star]{Comparing timelike geodesics around a Kerr black hole and a boson star}

\author{M. Grould$^1$, Z. Meliani$^1$, F. H. Vincent$^2$, P. Grandcl\'ement$^1$ \& E. Gourgoulhon$^1$}

\address{$^1$ LUTh, Observatoire de Paris, PSL Research University, CNRS UMR 8109, Universit\'e Pierre et Marie Curie, Universit\'e Paris Diderot, 5 place Jules Janssen, 92190 Meudon, France\\
 $^2$ LESIA, Observatoire de Paris, PSL Research University, CNRS UMR 8109, Universit\'e Pierre et Marie Curie, Universit\'e Paris Diderot, 5 place Jules Janssen, 92190 Meudon, France}
\ead{marion.grould@obspm.fr}
\vspace{10pt}

\begin{abstract}
The second-generation beam combiner at the Very Large Telescope (VLT), GRAVITY, observes the stars orbiting the compact object located at the center of our galaxy, with an unprecedented astrometric accuracy of 10 $\mu$as. The nature of this compact source is still unknown since black holes are not the only candidates explaining the four million solar masses at the Galactic center. Boson stars are such an alternative model to black holes. This paper focuses on the study of trajectories of stars orbiting a boson star and a Kerr black hole. We put in light strong differences between orbits obtained in both metrics when considering stars with sufficiently close pericenters to the compact object, typically $\lesssim 30~M$. Discovery of closer stars to the Galactic center than the S2 star by the GRAVITY instrument would thus be a powerful tool to possibly constrain the nature of the central source.
\end{abstract}

%
\vspace{2pc}

%
%
%

\section{Introduction}

Boson star models have been developed by \cite{1966PhRv..148.1269B}, \cite{1968PhRv..168.1445F}, \cite{1968PhRv..172.1331K} and \cite{1969PhRv..187.1767R}. Various subtypes of boson stars have been introduced, depending on the choice of the interaction potential \citep{1986PhRvL..57.2485C,1987PhRvD..35.3658F,2003CQGra..20R.301S,2013PhRvD..88f4046M}. Boson stars are described by general relativity and they are systems of self-gravitating massive complex scalar field. These bosons thus have zero intrinsic angular momentum.  More details on these objects can be found in the recent review of \cite{2012LRR....15....6L}. Nowadays, only one elementary boson has been discovered and corresponds to the Higgs boson observed in 2012 at CERN, and whose mass reaches approximatively 125 GeV \citep{2012PhLB..716...62A}. Such discovery thus makes boson stars less exotic. Different authors have studied boson stars such as \cite{1992PhR...220..163J}, \cite{1992PhR...221..251L}, or \cite{2003CQGra..20R.301S}. In particular, the two last authors discuss the possibilities of detecting the various subtypes of boson stars through astrophysical observations. The main motivation of studying such objects is their ability to mimic black holes. Indeed, boson stars do not have emissive surface and they can have a gravitational field as intense as black holes. A distinctive feature though is the absence of event horizon.

Mielke and Schunck were the first to obtain numerical solutions for rotating boson stars in weak field regime \citep{1996gpst.conf..391M, 1998PhLA..249..389S,Mielke2016}. Then, an extension to stronger field regime has been done by \cite{1997PhRvD..55.6081R} and \cite{1997PhRvD..56..762Y}. Amount of numerical investigations have been performed over the last fifteen years and in particular by \cite{2005PhRvD..72f4002K,2008PhRvD..77f4025K,2012PhRvD..85b4045K}. The first numerical computation of null and timelike geodesics in a non-rotating boson star metric has been done recently by \cite{2013PhRvD..88d4025D}. The same year \cite{2013PhRvD..88f4046M} studied the geodesics in various subtypes of non-rotating boson stars. The first computation of geodesics around a rotating boson star was obtained by  Grandcl\'ement et al., highlighting the discovery of particular geodesics not encountered in the Kerr metric \citep{2014PhRvD..90b4068G, 2017PhRvD..95h4011G}.

A significant number of studies have shown the presence of a compact source of several million solar masses at the Galactic center, called Sagittarius A* (Sgr A*) \citep{1977ApJ...218L.103W,1996ApJ...472..153G, 1997MNRAS.284..576E, 1998ApJ...509..678G, 2008ApJ...689.1044G, 2009ApJ...692.1075G,2017ApJ...837...30G}. In particular, monitorings of young stars close to Sgr~A* allowed to highly constrain its mass up to $(4.31 \pm 0.42) \times 10^{6} M_{\odot}$ (\cite{2009ApJ...692.1075G}, see \cite{2016ApJ...830...17B} or \cite{2017ApJ...837...30G} for a recent improvement of this mass estimation). Such an important mass suggests that a supermassive black hole lives at the center of our galaxy. This assumption is in particular discussed in \cite{2017FoPh...47..553E} where the authors review all the observations supporting the fact that Sgr~A* could be a black hole. However, others compact objects such as boson stars can also explain the mass at the Galactic center. The two key instruments which are expected to bring answers on the nature of Sgr A* are the Event Horizon Telescope (EHT, \cite{2009astro2010S..68D}) and the second-generation Very Large Telescope Interferometer (VLTI), GRAVITY \citep{2003SPIE.4841.1548E}. The first instrument will obtain in a few years sub-millimetric images of Sgr A* with an unprecedented angular resolution of about $15~\mu$as, which corresponds approximatively to a third of the angular apparent size of a supermassive black hole located at 8 kpc from the Earth. Such a resolution will thus allow to probe the vicinity of Sgr A*. The second instrument has been installed in 2015 at the VLT and observes in the near-infrared the motion of stars and gas orbiting Sgr A* with an astrometric accuracy of about $10~\mu$as. Low- and high-order relativistic effects are expected to be measured in order to better constrain the nature of the central source \citep{Grould2017}.

In the framework of obtaining for the first time highly accurate observations close to a compact object, several studies have been performed to determine whether both instruments will be capable of distinguishing a Kerr black hole from alternative objects also described by general relativity \citep{2000gr.qc....12031T, 2013arXiv1301.1396B, 2014PhRvD..90j4013S, 2014PhRvD..90b4068G, 2015CQGra..32w5022M, 2016CQGra..33j5015V} or from alternative theories of gravitation \citep{2008ApJ...674L..25W,2010PhRvD..81f2002M,2011CQGra..28v5029S,2014ApJ...784....7B,2016CQGra..33k3001J}. Indeed, in addition to Kerr solutions other black holes also described by general relativity can exist \citep{2012LRR....15....7C} such as the black holes with scalar hair or Proca hair \citep{2015PhRvL.115u1102C,2016PhRvD..94j4023C,2017PhRvD..95j4035Z}. In particular, the studies recently performed by \cite{2015PhRvL.115u1102C,2016PhRvD..94j4023C} focused on null geodesics around boson stars and Kerr black holes with scalar hair where the aim was to distinguish both compact objects. The authors showed that the shadow of such objects can be very different in shape and size. However, \cite{2016CQGra..33j5015V} showed that synthetic images of a Kerr black hole and a boson star are very similar, only small structures of size $3~\mu$as appear in the boson star image and are not present in the Kerr black hole one. \cite{2015CQGra..32w5022M} have, nevertheless, demonstrated that the accretion tori around a boson star has different characteristics than in the surroundings of a black hole. We also remind that \cite{2014PhRvD..90b4068G} have shown the existence of pointy petal orbits not encountered in the Kerr metric. Distinction between black holes and boson stars via null geodesics have been also treated in \cite{2006MNRAS.369..485S} and \cite{2017PhRvD..95h4011G}. Other studies on boson stars are made in order to determine whether they could be discriminated from Kerr black holes, for instance, by using the gravitational-wave signal generated by a close boson stars binary \citep{2017arXiv170408651S}.

The aim of this paper is to go further ahead in the study on timelike geodesics obtained in the boson-star metric in the perspective of better grasp whether it could be possible to discriminate a boson star from a Kerr black hole with the GRAVITY instrument. This paper will thus be organized as follows. In Sect.~\ref{KerrBlackHoleBosonStar}, we introduce the Kerr black holes and the boson stars. In Sect.~\ref{TLBosonStar}, we focus on sustainable orbits of stars encountered in the boson-star metric that cannot exist with a Kerr black hole because the star would fall into it. In Sect.~\ref{CompBlackHoleBosonStar}, we compare the energy and the trajectory of a star orbiting a Kerr black hole and a boson star, considering identical initial position for this star in both metrics. A conclusion and a discussion are given in Sect.~\ref{ConclusionsDiscussions}.

\section{Two candidates for the central compact source Sgr A*}
\label{KerrBlackHoleBosonStar}

In this section, we shall review the basics of rotating black holes as described by general relativity and define the boson stars. We also introduce notations to be used throughout this paper. All quantities are expressed in geometrized units $M$ corresponding to the mass of the compact object (the black hole or the boson star), and the Newtonian gravitational constant $G$ and the speed of light $c$ are set to unity ($G~=~c~=~1$). The spacetime metrics signature is $(-,+,+,+)$. Finally, we place in the quasi-isotropic system $(t,r,\theta,\varphi)$ for both the black hole and the boson-star metrics.

\subsection{Kerr black holes}
\label{sec:KBH}

According to the no-hair theorem all stationary and axisymmetric rotating black holes are Kerr black holes, corresponding to the solution of the vacuum Einstein equation. This solution is given, in quasi-isotropic coordinates, by
\begin{eqnarray}
g_{\mu\nu} \mathrm{d}x^\mu \mathrm{d}x^\nu = & - \left(\frac{\rho^2\Delta}{\Sigma} - \beta^{\varphi} \beta_{\varphi}\right) \mathrm{d}t^2 - 2 \beta_{\varphi} \mathrm{d}t \mathrm{d}\varphi \nonumber \\
& + \chi^4 \left[ \left( \frac{\rho^4}{\Sigma} \right)^{1/3} (\mathrm{d}r^2 + r^2 \mathrm{d}\theta^2) +  r^2 \sin^2{\theta} \left(\frac{\Sigma}{\rho^4}\right)^{2/3} \mathrm{d}\varphi^2 \right]
\end{eqnarray}
where
\begin{eqnarray}
& \rho^2 = \psi^4 r^2 + a^2 \cos^2{\theta}, \nonumber \\
& \Delta = \psi^4 r^2 - 2M\psi^2r + a^2, \nonumber \\
& \Sigma = (\psi^4 r^2 + a^2)^2 - a^2 \Delta \sin^2{\theta}, \nonumber \\
& \beta^{\varphi} = -\frac{2 M \psi^2 r a}{\Sigma}, \nonumber \\
& \chi^4 = \left( \frac{\rho^2 \Sigma}{r^6} \right)^{1/3}, \nonumber \\
& \psi^2 = \left(1 + \frac{M + a}{2 r}\right) \left(1 + \frac{M - a}{2 r}\right).
\end{eqnarray} 
The parameter $a$ corresponds to the spin of the black hole varying between $-M$ and $M$. If $|a|$ is superior to $M$, there is no event horizon and the central singularity becomes naked. In what follows, we will note $a$ as being the dimensionless parameter of the spin. In this case, $a$ will vary between $-1$ and 1. This parameter will also be used to denominate the spin of the boson star.

\subsection{Boson stars}

In this paper, we consider boson stars with minimal coupling of the scalar field to gravity such that their action is expressed as
\begin{equation}
\label{eq:actionBS}
S =  \int{ \left( \mathcal{L}_g + \mathcal{L}_\Phi \right) \sqrt{-g} \mathrm{d}^4x}
\end{equation}
where $\mathcal{L}_g$ is the Hilbert-Einstein Lagrangian of gravitational field expressed as
\begin{equation}
\mathcal{L}_g = \frac{1}{16 \pi}R,
\end{equation}
with $R$ the scalar curvature; and $\mathcal{L}_\Phi$ is the Lagrangian of the complex scalar field $\Phi$ given by
\begin{equation}
\label{eq:Lphi}
\mathcal{L}_\Phi = - \frac{1}{2} \left[ \nabla_\mu \Phi \nabla^\mu \bar{\Phi}  + V\left( |\Phi|^2 \right)\right],
\end{equation}
with $V$ the interaction potential depending on $|\Phi|^2$. The complex scalar field $\Phi$ required to describe stationary and axisymmetric rotating boson stars takes the following form
\begin{equation}
\label{eq:Phi}
\Phi = \phi(r,\theta) \exp\left[ i(\omega t - k \varphi) \right].
\end{equation}
Contrary to Kerr black holes, boson stars do not have any event horizon and their spin can be superior to 1 \citep{2014PhRvD..90b4068G}. The parameters $\phi$, $\omega$ and $k$ in the complex scalar field correspond to the modulus of $\Phi$, the frequency and the azimuthal number, respectively. Boson stars are thus defined by only the two parameters $\omega$ and $k$ such that \citep{2014PhRvD..90b4068G}
\begin{eqnarray}
&0 < \omega \leqslant \frac{m}{\hbar}, \nonumber \\
&k \in \mathbb{N} &\mbox{ with } \cases{k = 0&for non-rotating boson stars, \\
k > 0 &for rotating boson stars.\\}
\end{eqnarray}
The parameter $m$ corresponds to the mass of one individual boson composing the boson star, and acts as a scaling parameter. As illustrated by \cite{2014PhRvD..90b4068G}, when $\omega$ tends to $m/\hbar$ the boson star is less compact (less relativistic), and at the limit $\omega = m/\hbar$ the complex scalar field vanishes, no more boson star exists. The angular momentum of the boson star $J_\mathrm{BS}$ depends on the azimuthal number and the total boson number $\mathcal{N}$ as
\begin{equation}
J_\mathrm{BS} = k \hbar \mathcal{N}.
\end{equation}
The spin of the boson star is thus directly proportional to the integer $k$. 

The choice of the potential $V$ in equation~\eref{eq:Lphi} allows to recover various subtypes of boson stars \citep{1986PhRvL..57.2485C,1987PhRvD..35.3658F,2003CQGra..20R.301S,2013PhRvD..88f4046M}. In our study, we focus on simplest boson star models called \textit{mini-boson stars} in which there is no self-interaction potential between bosons \citep{2003CQGra..20R.301S}, and the potential $V$ involves only the mass term
\begin{equation}
V\left( |\Phi|^2 \right) =  \frac{m^2}{\hbar^2} |\Phi|^2.
\end{equation}
The ADM (for Arnowitt--Deser--Misner) mass of the boson star depends on the choice of the potential and thus depends on the parameters $k$ and $\omega$ (see the upper plot of Fig.~6 from \cite{2014PhRvD..90b4068G}). In the particular case of mini-boson stars and considering small azimuthal numbers, the ADM mass of such objects satisfies \citep{2014PhRvD..90b4068G}
\begin{equation}
\label{eq:ADMmass}
M < M_\mathrm{max} = \alpha \frac{m_p^2}{m} = \alpha \frac{\hbar}{m}
\end{equation}
where $\alpha$ is a dimensionless constant depending on the choice of $(k,\omega)$ and varying in the interval $]0,10]$, and $m_p$ is the Planck mass. Massive boson stars that can explain the mass of Sgr~A* are obtained by taking into account the self-interaction between bosons \citep{1986PhRvL..57.2485C,2000NuPhB.564..185M,2014PhRvD..90b4068G}. However, as mentioned by \cite{2016CQGra..33j5015V} mini-boson stars can reach such mass by considering extremely light bosons with $m~\approx~10^{-16}$~eV. By comparison, we know that the mass of the Higgs boson is of about 125 GeV \citep{2012PhLB..716...62A}, which leads to a very weak total ADM mass for mini-boson star of about $10^{-21} M_\odot$. Recovering the mass of Sgr~A* with free-field boson stars thus impose the existence of very light bosons. A study performed by \cite{2010JCAP...11..002A} allows to fix a lower limit for $m$ by using dark matter models. The limit found by these authors is not compatible with the extremely light mass found considering a free-field boson star. However, in this paper we only focus on the mass of the compact source Sgr A*, without considering the surrounded dark matter, the limit imposed by \cite{2010JCAP...11..002A} can be omitted, we will thus assume for simplicity that such light bosons could exist. 

Contrary to the Kerr metric, the stationary and axisymmetric rotating boson-star metric can only be obtained numerically, by solving the coupled Einstein--Klein--Gordon system given by
\begin{equation}
\label{eq:systeme}
\cases{R_{\mu\nu}-\frac{1}{2} R g_{\mu\nu} = 8 \pi T_{\mu\nu}, \\
\nabla_\mu \nabla^\mu \Phi = \frac{\mathrm{d}V}{\mathrm{d}|\Phi|^2} \Phi\\}
\end{equation}
where $T_{\mu\nu}$ is the energy-momentum tensor of the complex scalar field expressed as
\begin{equation}
T_{\mu\nu} =  \nabla_{(\mu} \Phi \nabla_{\nu)}\bar{\Phi} - \frac{1}{2}\left[ \nabla_\alpha \Phi \nabla^\alpha \bar{\Phi} + V \left( |\Phi|^2 \right)  \right] g_{\mu\nu}.
\end{equation}
The second equation of the system~\eref{eq:systeme} is obtained by varying the action $S$ given by equation~\eref{eq:actionBS} with respect to the complex scalar field $\Phi$. The solution of the system~\eref{eq:systeme} is given in the $3+1$ formalism \citep{2012LNP...846.....G} and in the quasi-isotropic coordinates
\begin{eqnarray}
\label{eq:BSmetric}
g_{\mu\nu} \mathrm{d}x^\mu \mathrm{d}x^\nu = &-N^2\mathrm{d}t^2 + A^2 \left( \mathrm{d}r^2 + r^2 \mathrm{d}\theta^2 \right) + B^2 r^2 \sin^2{\theta} \left( \mathrm{d}\varphi + \beta^{\varphi} \mathrm{d}t \right)^2
\end{eqnarray}
where $N$, $A$, $B$ and $\beta^{\varphi}$ are the four functions to determine through the resolution of the system~\eref{eq:systeme}. We specify that the function $N$ is called the lapse, and $\beta^{\varphi} $ is the shift factor. For non-rotating boson stars, we get $\beta^{\varphi} =0$ and $A=B$. The shift factor is equal to $-\omega$ when the boson star is rotating. 

In this paper, we use the highly accurate code \textsc{Kadath} developed by \cite{2010JCoPh.229.3334G} to obtain the spacetime metric $g_{\mu\nu}$ of the boson stars. We consider various pairs $(k,\omega)$: for non-rotating boson stars ($k=0$) the frequencies allowed are superior to $\approx 0.76~m/\hbar$ \citep{2010JCoPh.229.3334G}, we will thus focus on two boson stars at $\omega=[0.86,0.9]~m/\hbar$; for rotating boson stars we focus on those which do not have ergoregion: $k=[1,2,3]$ with $\omega~=~[0.7,0.8,0.9]~m/\hbar$. Indeed, contrary to non-rotating boson stars, rotating ones can have ergoregion (those with low $\omega$, see \cite{2010JCoPh.229.3334G}) and as mentioned by \cite{2014PhRvD..90b4068G} boson stars with ergoregion are probably unstable. This is the reason why we choose boson stars without ergoregion.

\section{Timelike geodesics in the boson-star metric}
\label{TLBosonStar}

Massive particules (e.g. the stars) follow timelike geodesics along which two quantities are conserved due to the spacetime symmetry represented by the two killing vectors $\boldsymbol{\partial}_t$ and $\boldsymbol{\partial}_\varphi$. These constants are given by
\begin{eqnarray}
\label{eq:el}
&\varepsilon = - \boldsymbol{\partial}_{t} \cdot \boldsymbol{u}, \nonumber \\
&l = \boldsymbol{\partial}_{\varphi} \cdot \boldsymbol{u}
\end{eqnarray}
where $\boldsymbol{u}$, $\varepsilon$ and $l$ correspond to the four-velocity of the star, and its energy and angular momentum as observed from infinity, respectively. For this study, we focus on timelike geodesics in the equatorial plane $(\theta=\pi/2)$ implying $u^\theta = 0$. The radial motion of the massive particule in the metric is obtained by solving the equation $\boldsymbol{u}\cdot\boldsymbol{u}=-1$ whose developed expression leads to
\begin{equation}
\label{eq:LBH}
(u^r)^2 = \frac{1}{A^2}\left[ \frac{1}{N^2}(\varepsilon - \omega l)^2 - \frac{l}{B^2r^2} - 1\right].
\end{equation}
If we note $\mathcal{V}$ the right term of the above equation, the effective potential $V_\mathrm{eff}$ can be written as
\begin{equation}
V_\mathrm{eff} = \frac{\varepsilon^2 - 1 - \mathcal{V}}{2}.
\end{equation}
Note that the effective potential expressed here is also valid in the Kerr metric, the only difference will come from the metric coefficients in $\mathcal{V}$.

The timelike geodesic is obtained by fixing the initial coordinates of the star. Its position and velocity are then inputted in the ray-tracing code \textsc{Gyoto} developed by \cite{2011CQGra..28v5011V}. This code allows to compute the trajectory of the star given that its initial coordinates with respect to the studied compact object. We mention that in order to validate the integration of geodesics performed in numerical metrics by the \textsc{Gyoto} code, we compared trajectories of star orbiting a Kerr black hole considering analytical and numerical metrics. The latter is obtained by using the \textsc{Lorene} code developed by \cite{2016ascl.soft08018G}. The highest difference between both trajectories is of about $10^{-5}~\%$ which corresponds to a negligible numerical error on the computation of the \textsc{Gyoto} geodesic in the numerical metric. Computation of orbits of stars in the boson-star metric can thus be obtained confidently. However, we should notice that the study of boson stars performed with the \textsc{Kadath} code in \cite{2014PhRvD..90b4068G} has only been done in strong field regime. This is the reason why we focus on orbits evolving close to boson star.

We choose to fix the initial coordinates of the star at pericenter or apocenter which means that $u^r = 0$. The initial three-position of the star is given by $(r,\pi/2,0)$ and the components $u^t$ and $u^\varphi$ are obtained by using equation~\eref{eq:el}. Finally, the initial velocity of the star in the boson-star metric is defined as
\begin{eqnarray}
\label{eq:CIBS}
\cases{u^t =\frac{\varepsilon g_{\varphi \varphi} + l g_{t \varphi}}{g_{t \varphi}^2 - g_{\varphi \varphi} g_{tt}}, \\
u^r = 0, \\
u^\theta = 0, \\
u^\varphi = \frac{\varepsilon g_{t \varphi} + l g_{t t}}{g_{\varphi \varphi} g_{tt} - g_{t \varphi}^2}.}
\end{eqnarray}
Thus, the quantities $u^t$ and $u^\varphi$ are computed by fixing the radial position $r$ and the angular momentum $l$ of the star. The energy $\varepsilon$ of the star in those quantities is obtained by solving the following equation as in \cite{2014PhRvD..90b4068G}
\begin{equation}
\label{eq:LBH}
\mathcal{V} = 0
\end{equation}
where we considered $(u^r)^2 = 0$ since the initial coordinates is taken at pericenter or apocenter.

Now, it is possible to compute the trajectory of a star orbiting a boson star. The aim of this section is to present various sustainable orbits obtained in the presence of a rotating boson star and that cannot be observed in the Kerr metric because the star would fall below the event horizon. 

\subsection{Orbits with zero angular momentum $(l=0)$}

\begin{figure}[t]
\centering
       \includegraphics[scale=0.45]{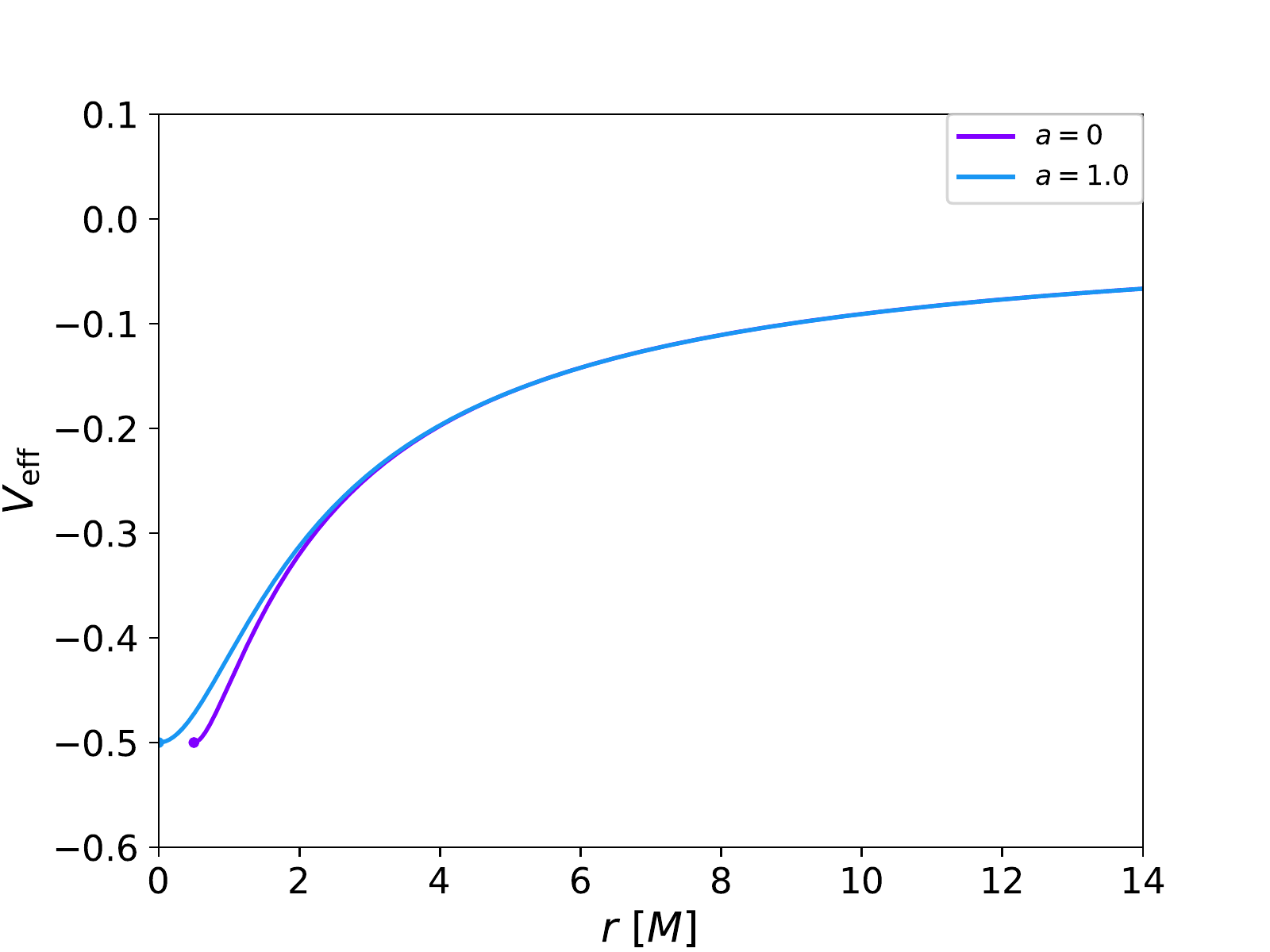}
       \quad
       \includegraphics[scale=0.45]{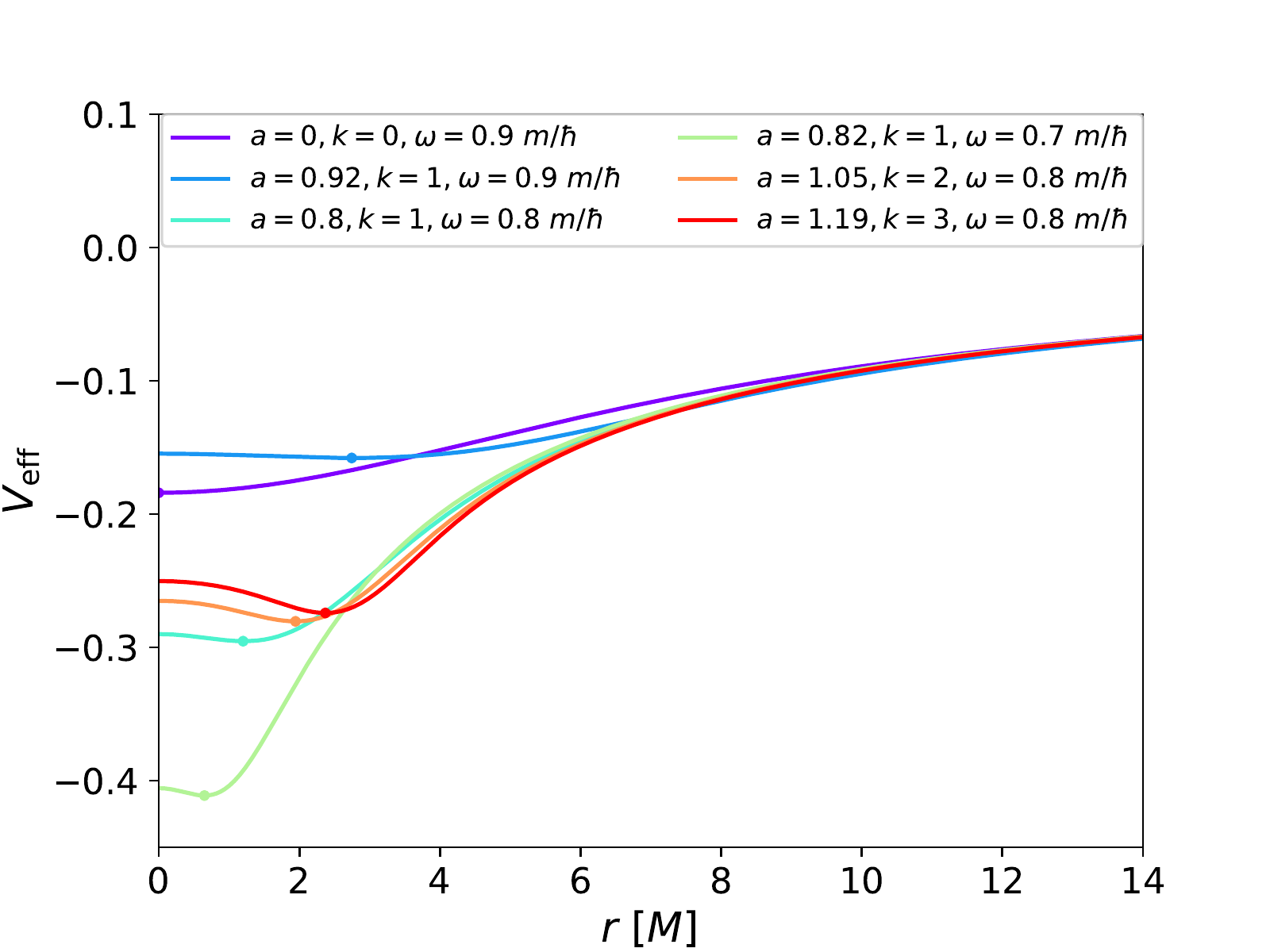}
       \caption{Effective potentials obtained for a star with $l=0$ and orbiting a black hole (left) or a boson star (right). The dots denote the minimum of the effective potential.}
       \label{fig:EPl0}
\end{figure}

\begin{figure}[t]
\centering
       \includegraphics[scale=0.3]{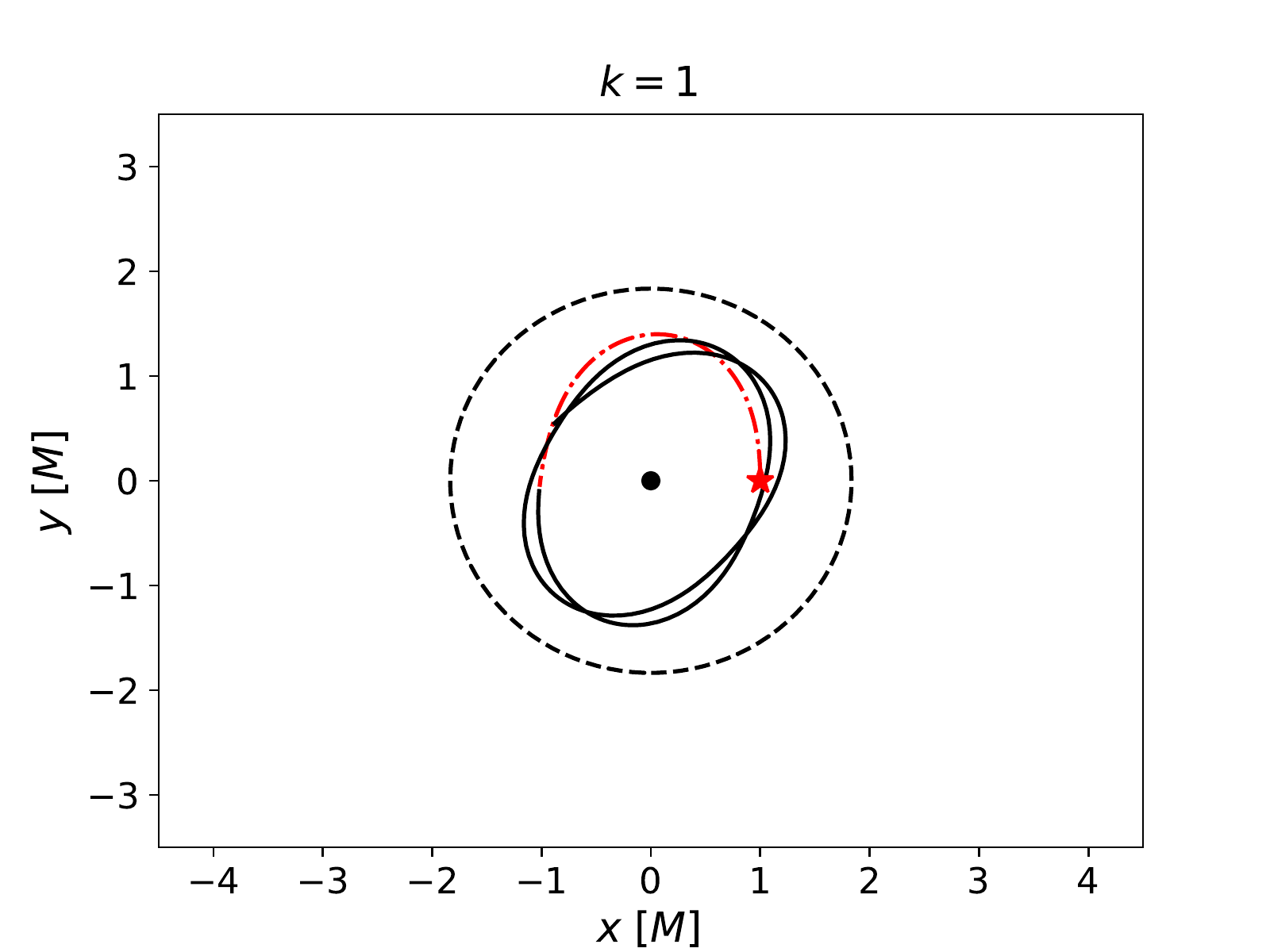}
       \includegraphics[scale=0.3]{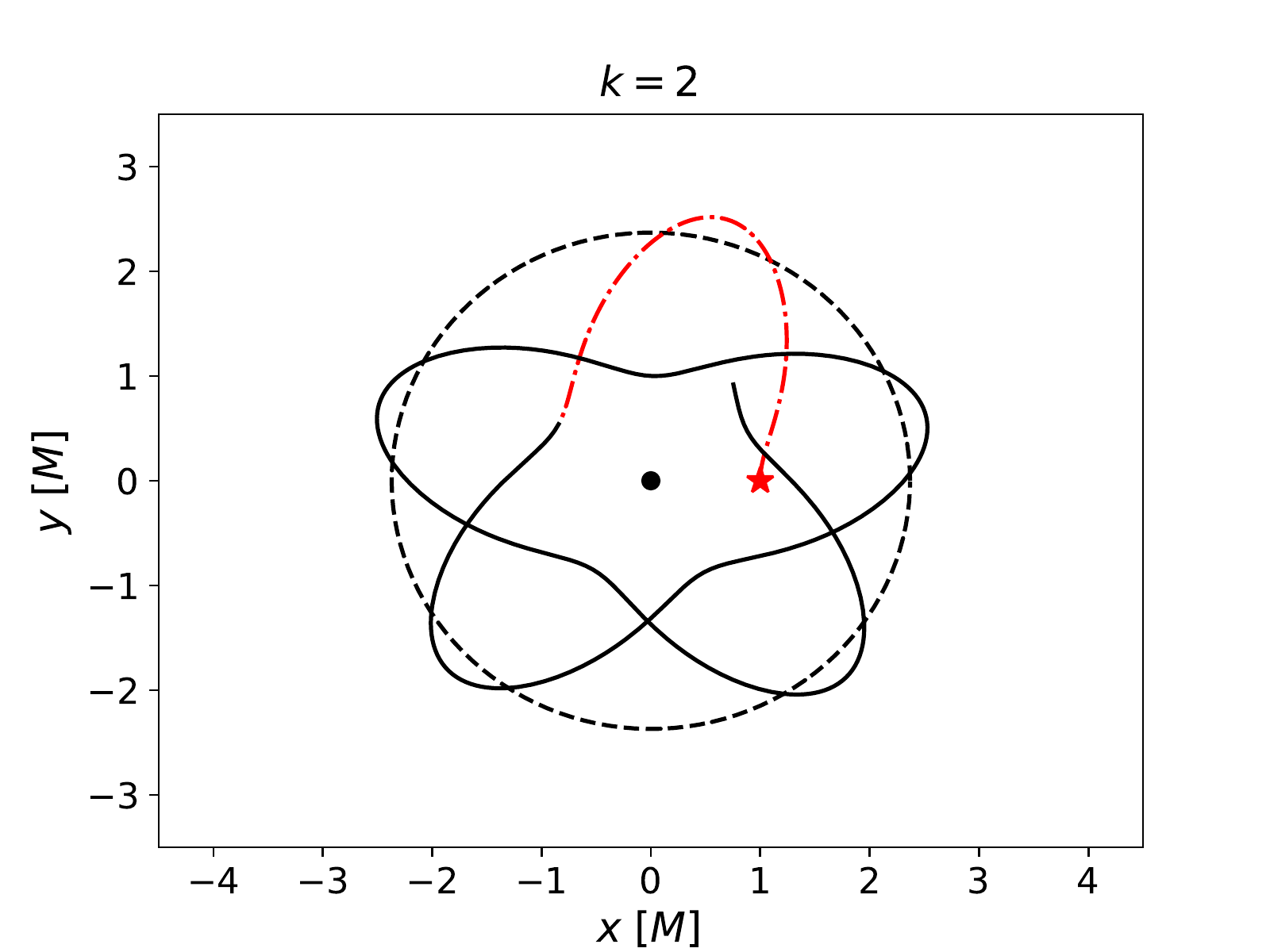}
       \includegraphics[scale=0.3]{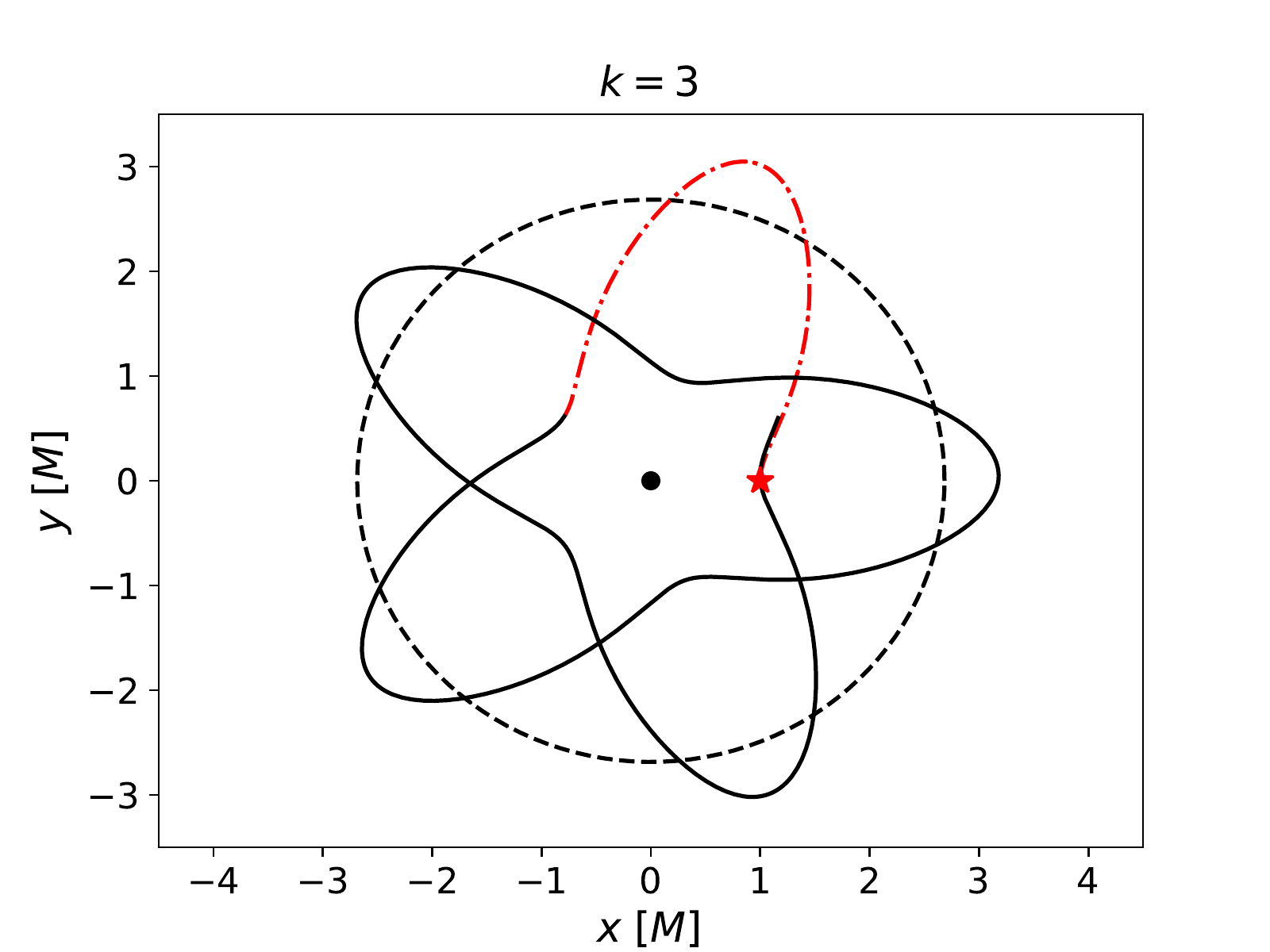}
       \quad
       \includegraphics[scale=0.3]{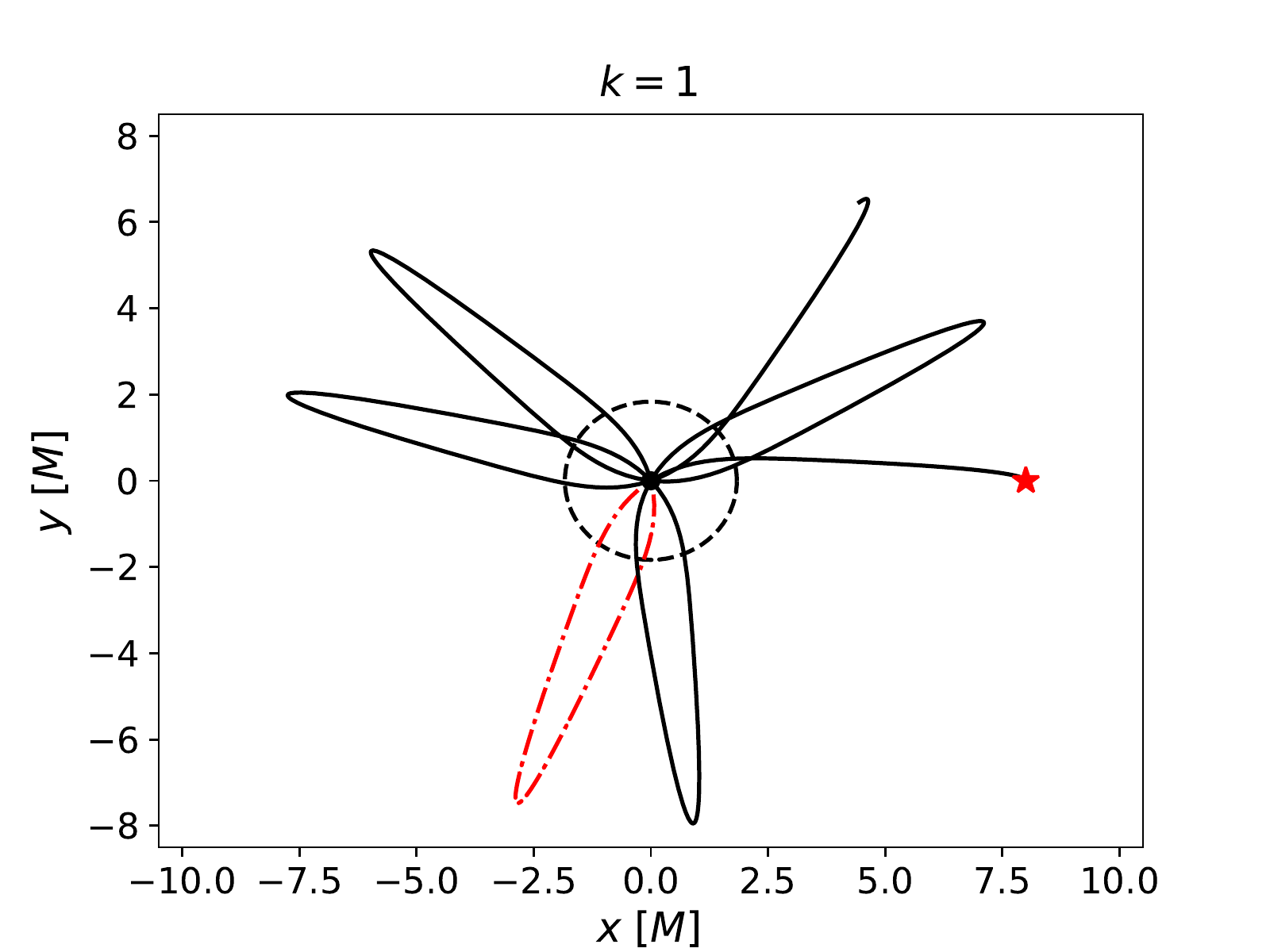}
       \includegraphics[scale=0.3]{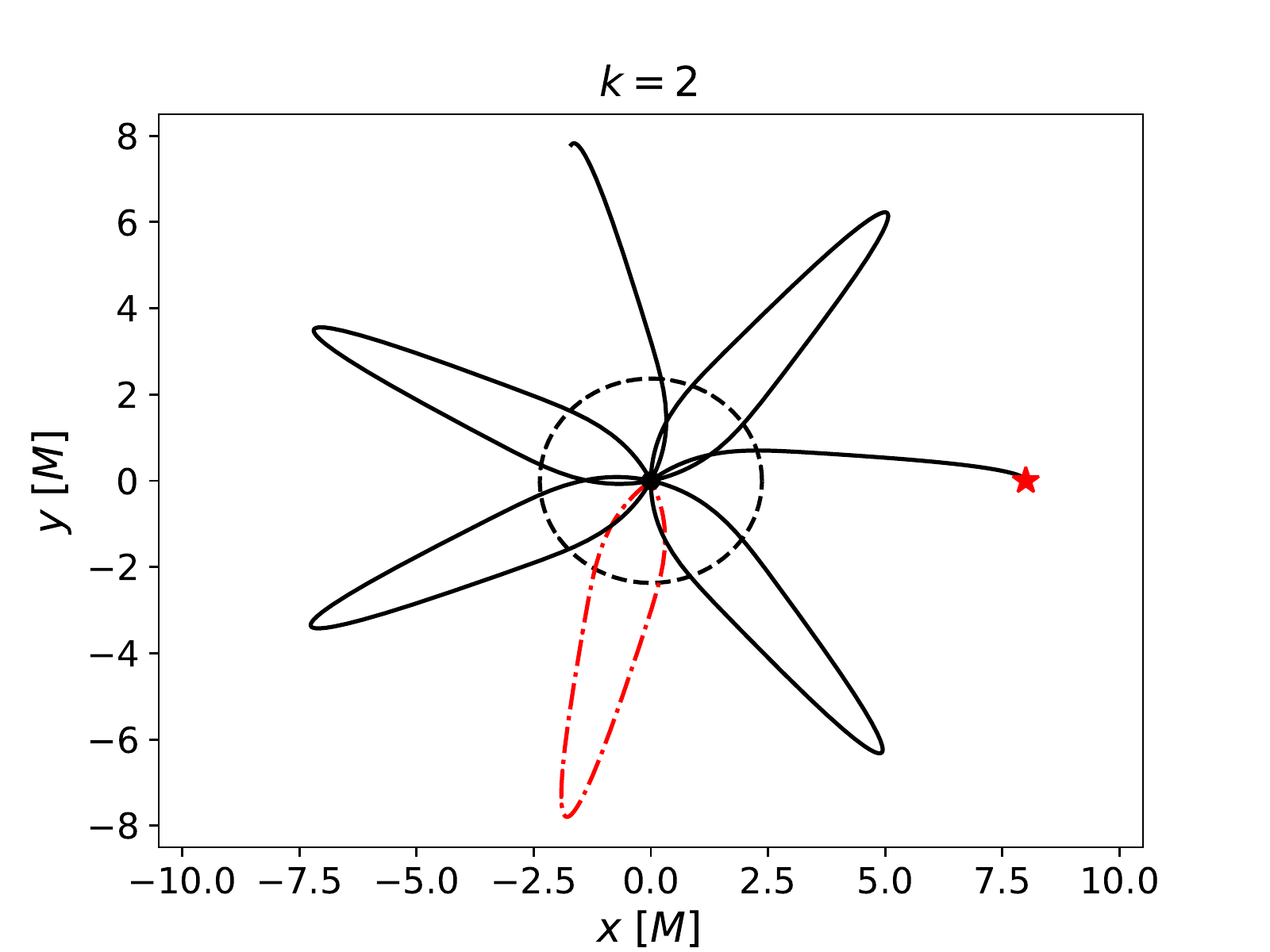}
       \includegraphics[scale=0.3]{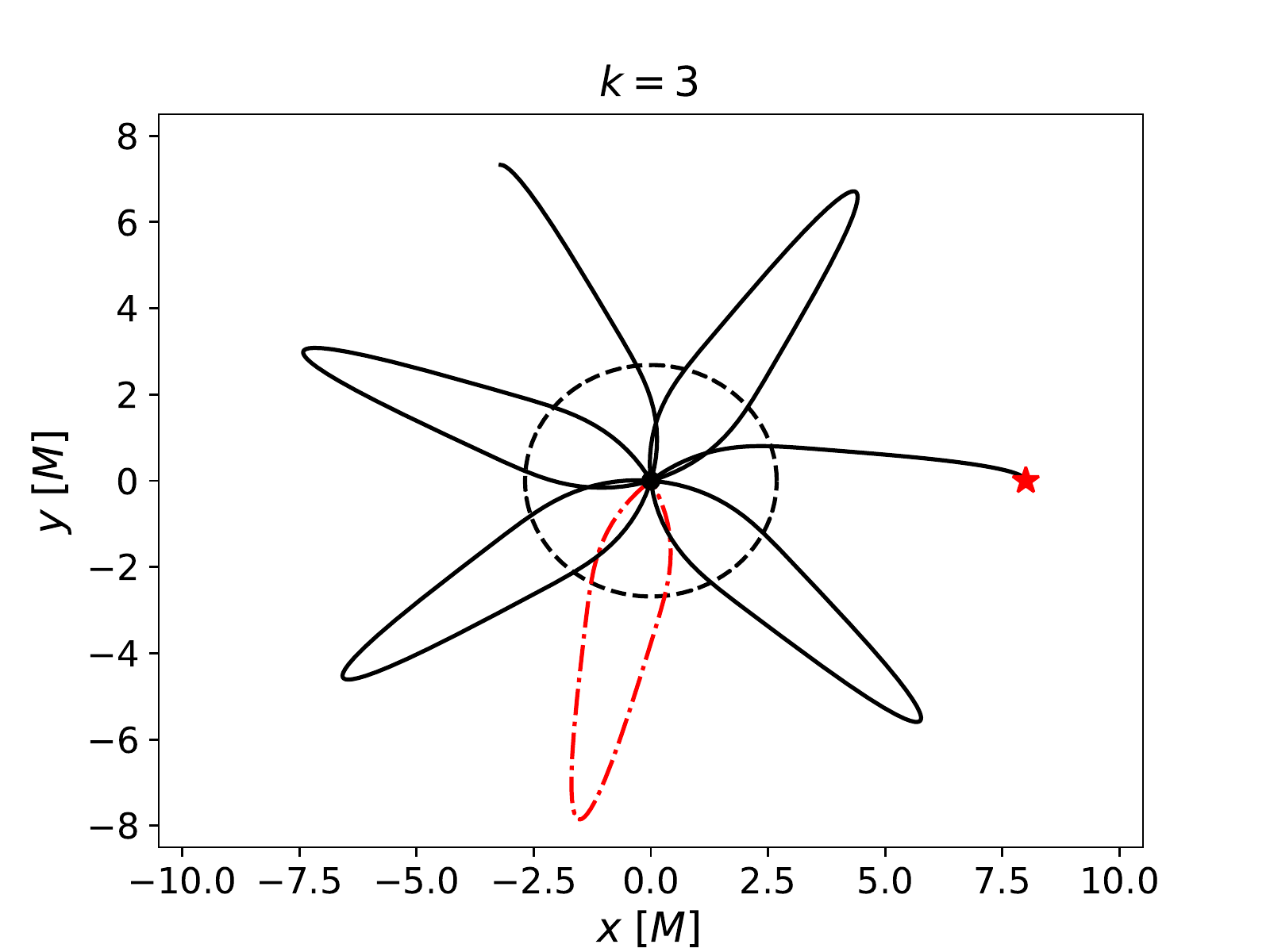}
       \quad
       \includegraphics[scale=0.3]{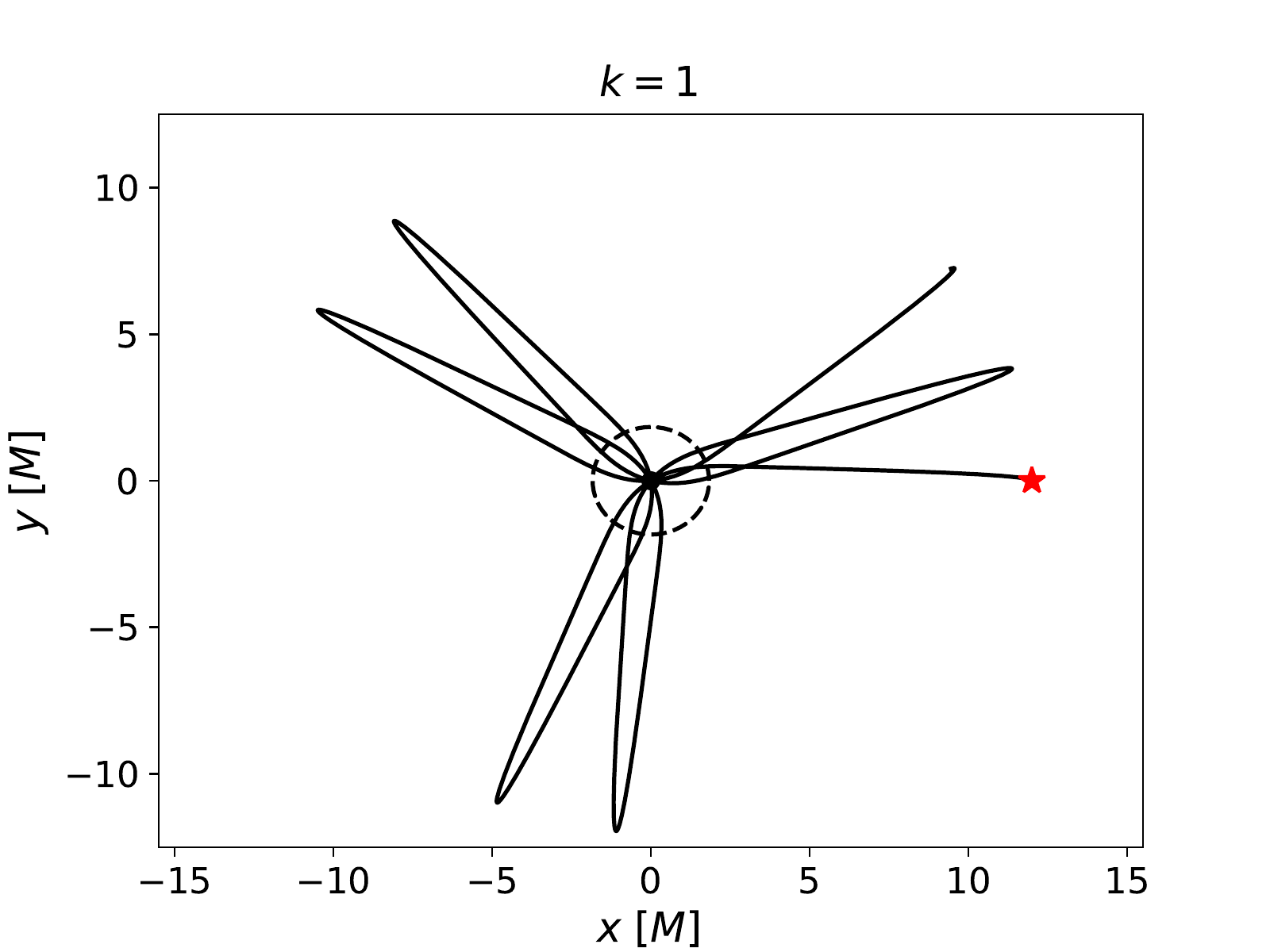}
       \includegraphics[scale=0.3]{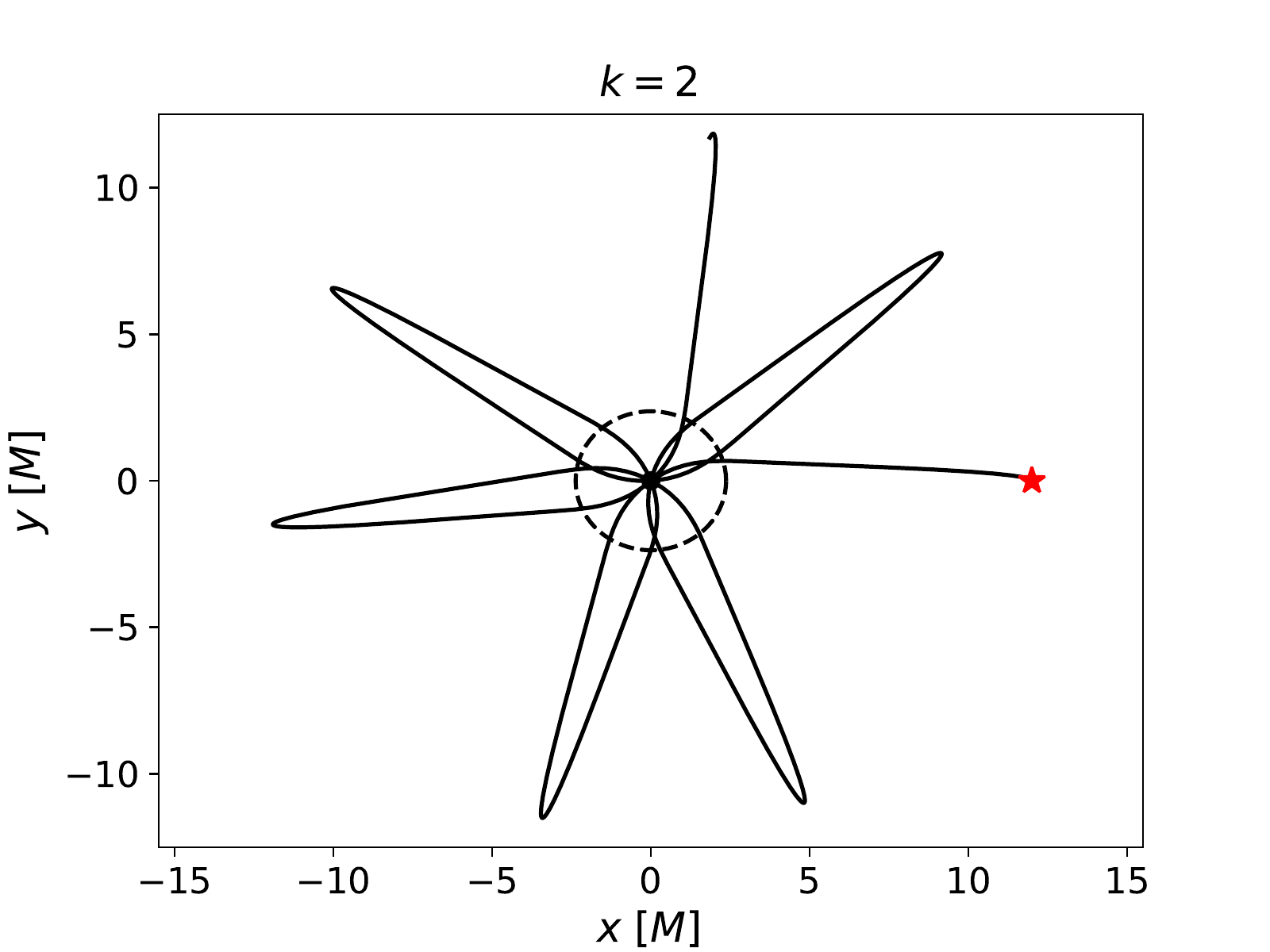}
       \includegraphics[scale=0.3]{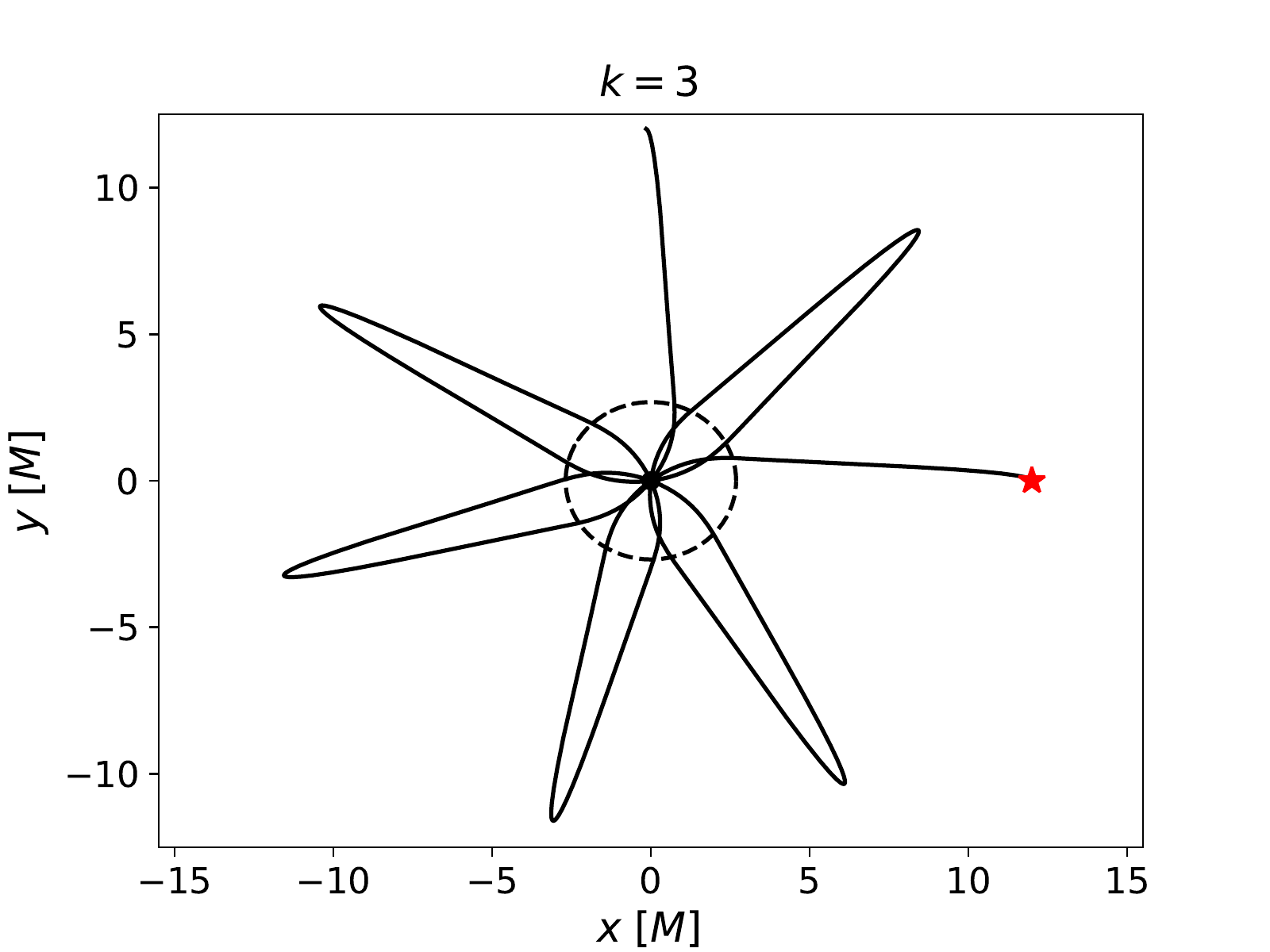}
       \caption{Orbits obtained for a star with $l=0$ and orbiting different boson stars whose frequency is fixed at $0.8~m/\hbar$. Three initial radial positions for the star are considered: $r = 1~M$ (upper plots), $r = 8~M$ (middle plots) and $r = 12~M$ (lower plots). The dashed lines correspond to the maximum of the scalar field modulus $\phi$. The black dots denote the geometrical center and the red stars denote the initial position of the star. The red dash-dotted lines illustrate what may correspond to one orbital period of the star.}
       \label{fig:Ol0}
\end{figure}

Orbits of stars with zero angular momentum are sustainable in the boson-star metric which is not the case in Kerr metrics \citep{2014PhRvD..90b4068G}. This can be shown on the right plot of Fig.~\ref{fig:EPl0} where the effective potential for rotating boson stars has the shape of a well, necessary to get a sustainable orbit. We note that this is not the case for non-rotating boson stars. However, a star cannot fall into a boson star. Thus, when this effective potential is obtained it means that the orbit of the star corresponds to a straight line: the star oscillates between two identical positions because of the symmetry in the effective potential with respect to the geometrical center of the metric. The right plot of Fig.~\ref{fig:EPl0} also shows the dependency of the effective potential to the boson star compactness parameter $\omega$: the potential well is deeper when decreasing this parameter.

\begin{figure}[t]
\centering
       \includegraphics[scale=0.45]{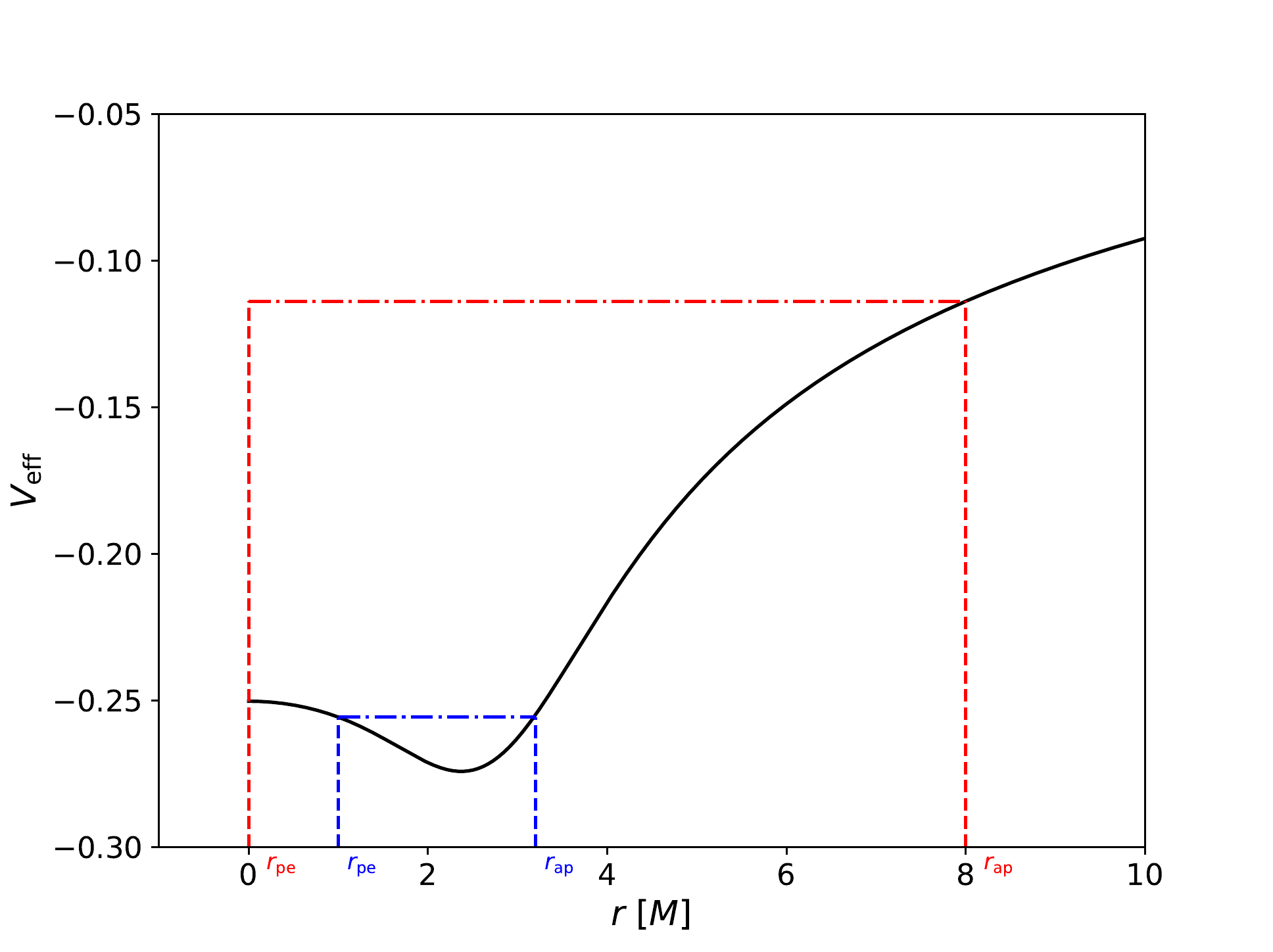}
       \caption{Effective potential obtained for a star with $l=0$ and orbiting the boson star $k=3$, $\omega=0.8~m/\hbar$. The dash-dotted lines correspond to the effective potentials of the orbit obtained with a star initialized at $r = 1~M$ (blue) and $r = 8~M$ (red). The corresponding pericenter ($r_\mathrm{pe}$) and apocenter ($r_\mathrm{ap}$) of the orbits are mentioned.}
       \label{fig:EPk3o08}
\end{figure}

Fig.~\ref{fig:Ol0} shows different orbits obtained with various rotating boson stars and considering three initial radial positions for the star. The maximum of the scalar field modulus is also presented and shows that rotating boson stars have the shape of a torus. First, we note on Fig.~\ref{fig:Ol0} that orbits generated with the initial position $r=1~M$ have different shape as those obtained with $r > 1~M$. In particular, in the former case the star always crosses the geometrical center which is not the case at $r=1~M$. For all $k$ considered and all initial positions, we find orbits very different from those that can be encountered in the Kerr metric for $l \neq 0$. In particular, on the upper plots of Fig.~\ref{fig:Ol0}, the star does not evolve through a full rotation about the geometrical center and is attracted to regions where the scalar field modulus is maximal, which gives the impression that the orbit is cut. We will call such orbits the \textit{semi orbits}. These trajectories obviously appear for energy smaller than the one obtained at the geometrical center and thus allows the star orbit to be confined in a potential well: the star oscillates between the apocenter and the pericenter (see blue lines on Fig.~\ref{fig:EPk3o08}). Such orbits can also appear for $\omega=0.9~m/\hbar$ or $\omega=0.7~m/\hbar$. On the first upper plot of Fig.~\ref{fig:Ol0}, we note that the orbit does not go through the maximum of the scalar field modulus, which is not the case for the other orbits plotted on this figure. This shows that the star can be trapped inside the torus defined by intense scalar field. This can also happen for $\omega=0.7 - 0.9~m/\hbar$. We also note that the semi orbits are obtained for apocenters located close or inside the region with maximum scalar field modulus. For instance, considering $\omega=0.8~m/\hbar$ and $k \leqslant 3$ they are formed for stars  initialized at apocenter $r < 4~M$, where the radius of the maximum scalar field modulus reaches $1.8~M$ and $2.7~M$ for $k=1$ and $k=3$, respectively.

For larger apocenter, the star passes by the geometrical center and is highly deflected. The resulting orbit looks like a petal, we will call them the \textit{petal orbits}. When the apocenter of the star increases, the petal becomes more and more pointy which is due to the fact that the component $u^\varphi$ of the star tends to zero. The star mainly has a radial motion at large distances, and closer to the boson star the component $u^\varphi$ increases rapidly giving rise to these petal orbits.

These particular orbits are also not observed in the Kerr metric since the star cannot go through the black hole. Contrary to previous orbits, these orbits are obtained for higher energies so that the star oscillates between the apocenter and the geometrical center which can be considered as the pericenter position of the orbit (see red lines on Fig.~\ref{fig:EPk3o08}). From apocenters at $8~M$, the orbits become similar for $k=2$ and $k=3$. This suggests that for all $k$ and sufficiently large apocenters, the deflection angle reaches a plateau ($\approx 260^\circ$ for $\omega=0.8~m/\hbar$). Thus, even for further apocenters from the boson star the petals exist and the deflection angle is high (this is confirmed by the simulations).

Finally, Fig.~\ref{fig:Ol0} shows that all orbits are prograde as encountered in the Kerr metric when the angular momentum of the black hole is positive. Moreover, we note a prograde relativistic shift when considering the semi orbits. However, it is retrograde for the petal orbits.

\begin{figure}[t]
\centering
       \includegraphics[scale=0.3]{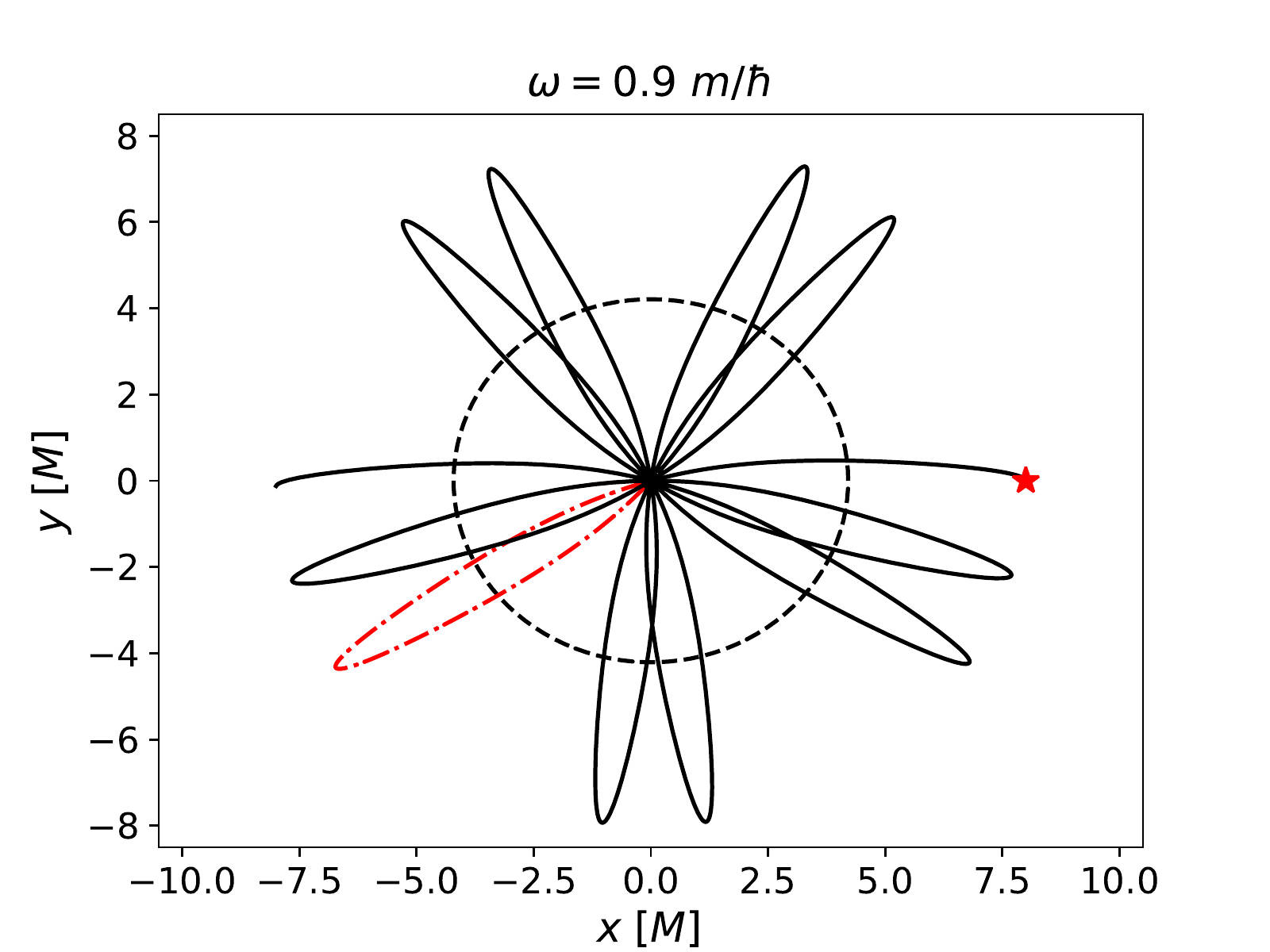}
       \includegraphics[scale=0.3]{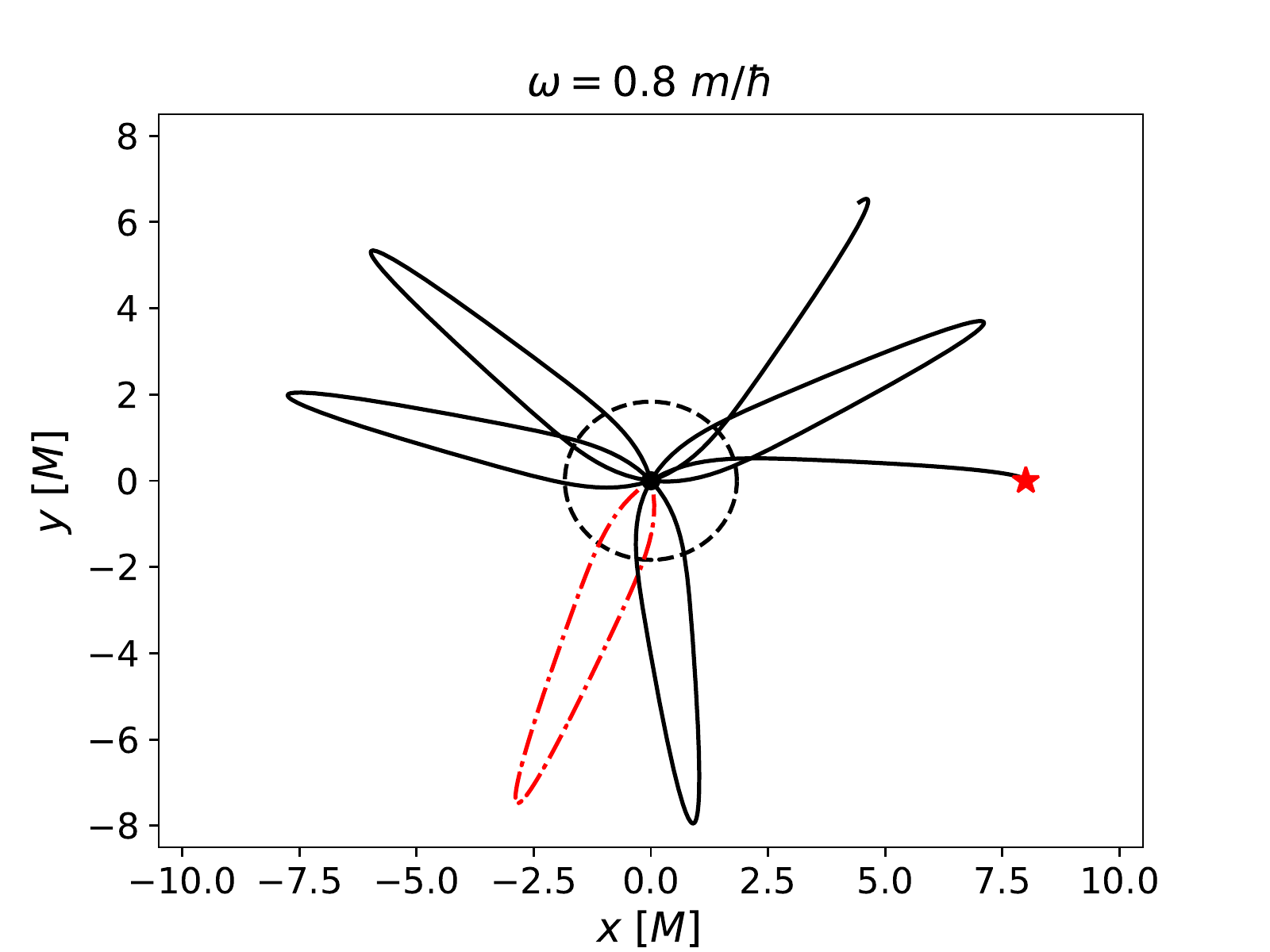}
       \includegraphics[scale=0.3]{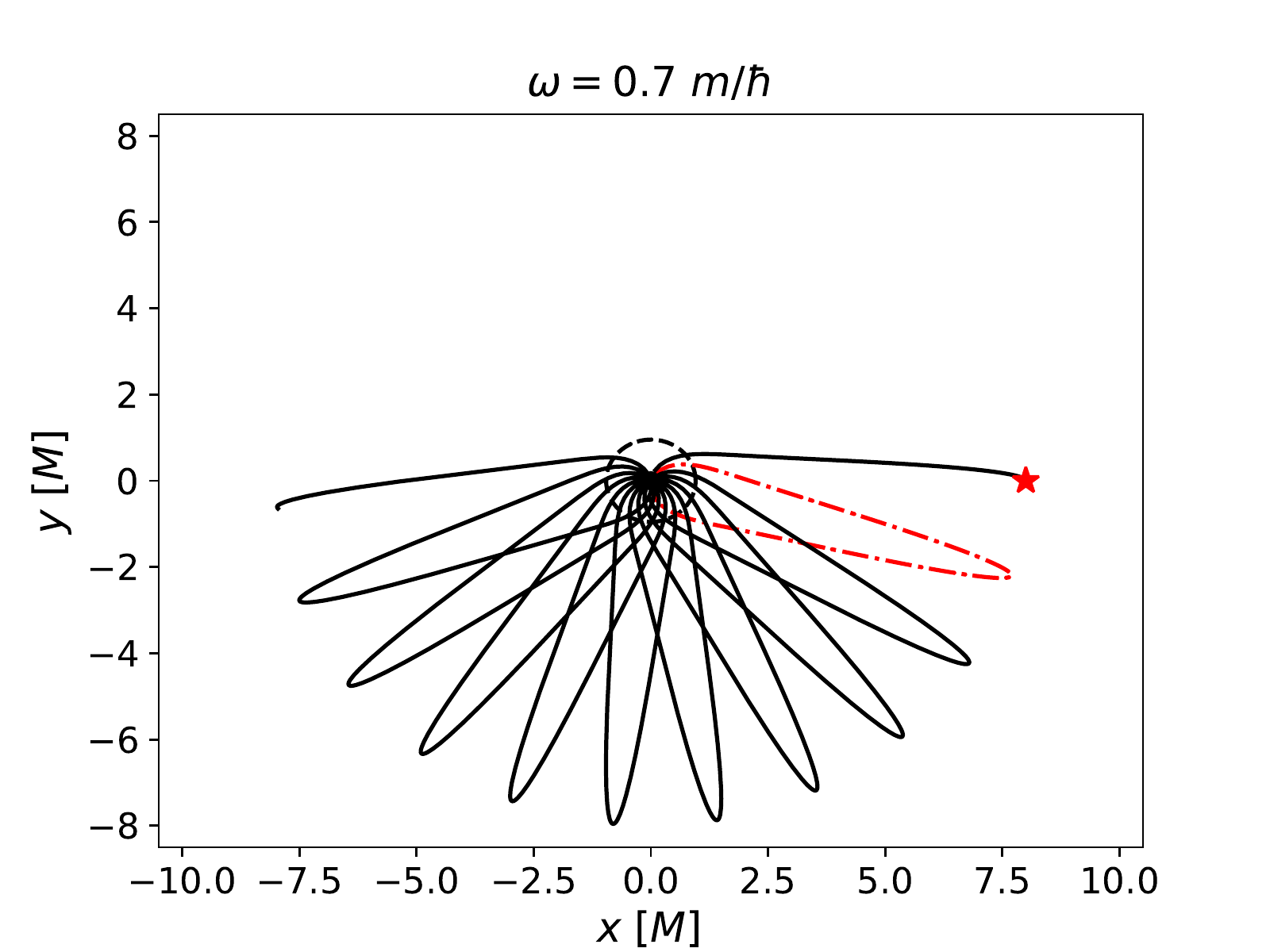}
       \caption{Orbits obtained for a star with $l=0$ and orbiting different boson stars whose azimuthal number $k$ is fixed to 1. The initial radial position of the star is at $r = 8~M$ for the three panels. The dashed lines correspond to the maximum of the scalar field modulus $\phi$. The black dots denote the geometrical center and the red stars denote the initial position of the star. The red dash-dotted lines illustrate what may correspond to one orbital period of the star.}
        \label{fig:Ol0om}
\end{figure}

Fig.~\ref{fig:Ol0om} presents the influence of the compactness parameter $\omega$ of the boson star on the orbit when considering an initial condition at $r~=~8~M$. First, the orbits remain petal orbits going through the geometrical center. Second, as mentioned in \cite{2014PhRvD..90b4068G} the parameter $\omega$ modifies the deflection angle of the star: it increases when $\omega$ decreases. This is obvious since the boson star becomes more compact (more relativistic) when this parameter decreases: the magnitude of the Lense-Thirring effect is more important. In particular, for $\omega=0.7~m/\hbar$ the deflection angle reaches a plateau at large distances of about $330^\circ$.

\subsection{Orbits with $l=1~M$}

\begin{figure}[t]
\centering
       \includegraphics[scale=0.45]{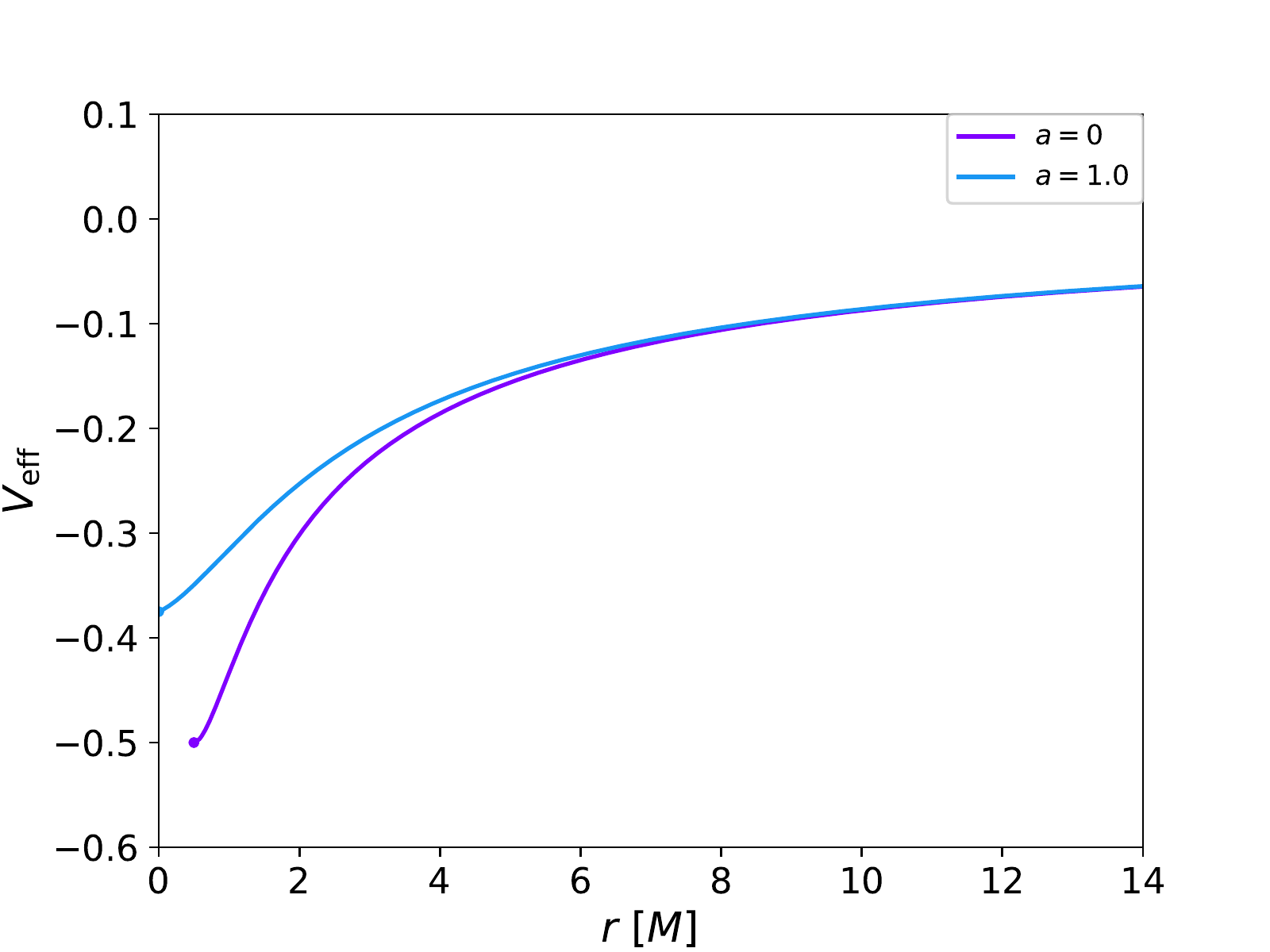}
       \quad
       \includegraphics[scale=0.45]{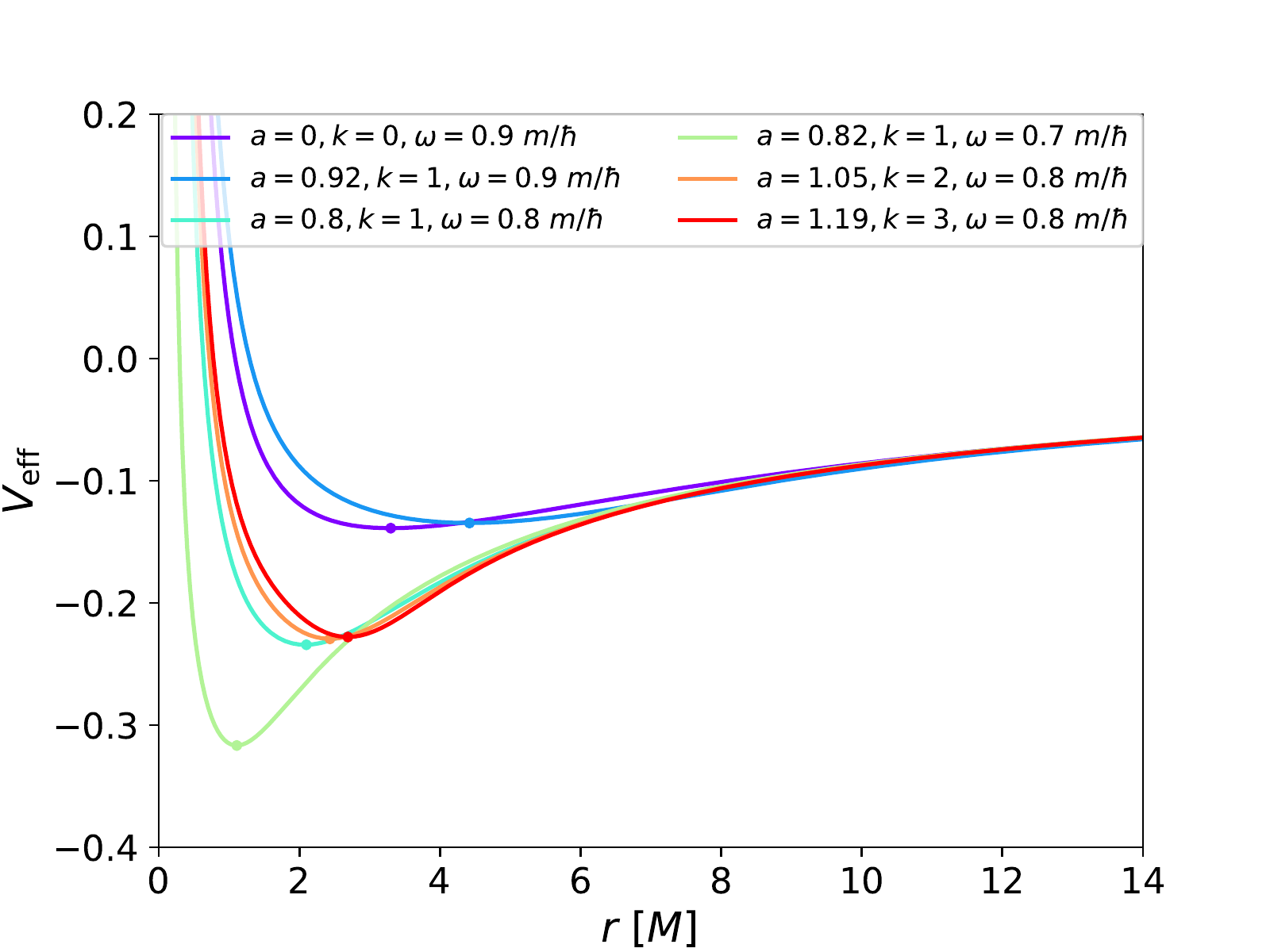}
       \caption{Effective potentials obtained for a star with $l=1~M$ and orbiting a black hole (left) or a boson star (right). The dots denote the minimum of the effective potential.}
       \label{fig:EPl1}
\end{figure}

As for zero angular momentum, the star falls into the black hole when $l~=~1~M$ (see the left panel of Fig.~\ref{fig:EPl1}). However, as mentioned before this is never the case in boson-star metrics \citep{2013PhRvD..88d4025D,2014PhRvD..90b4068G}. On the right panel of Fig.~\ref{fig:EPl1} we can notice that contrary to $l=0$, the effective potentials go to infinity at small radii. This means that contrary to the petal orbits, the orbits obtained with such effective potentials will not passe through the geometrical center. 

\begin{figure}[t]
\centering
       \includegraphics[scale=0.3]{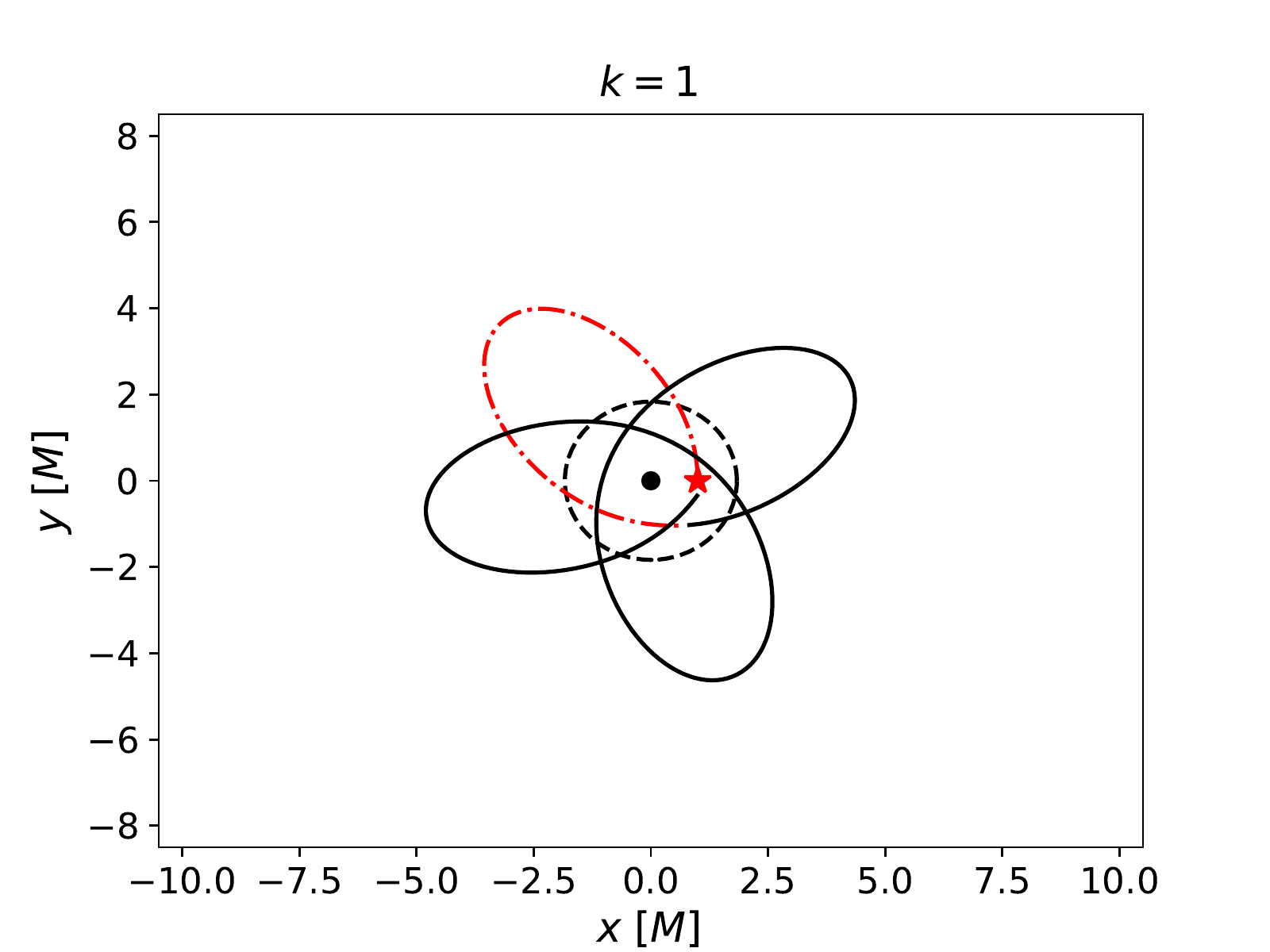}
       \includegraphics[scale=0.3]{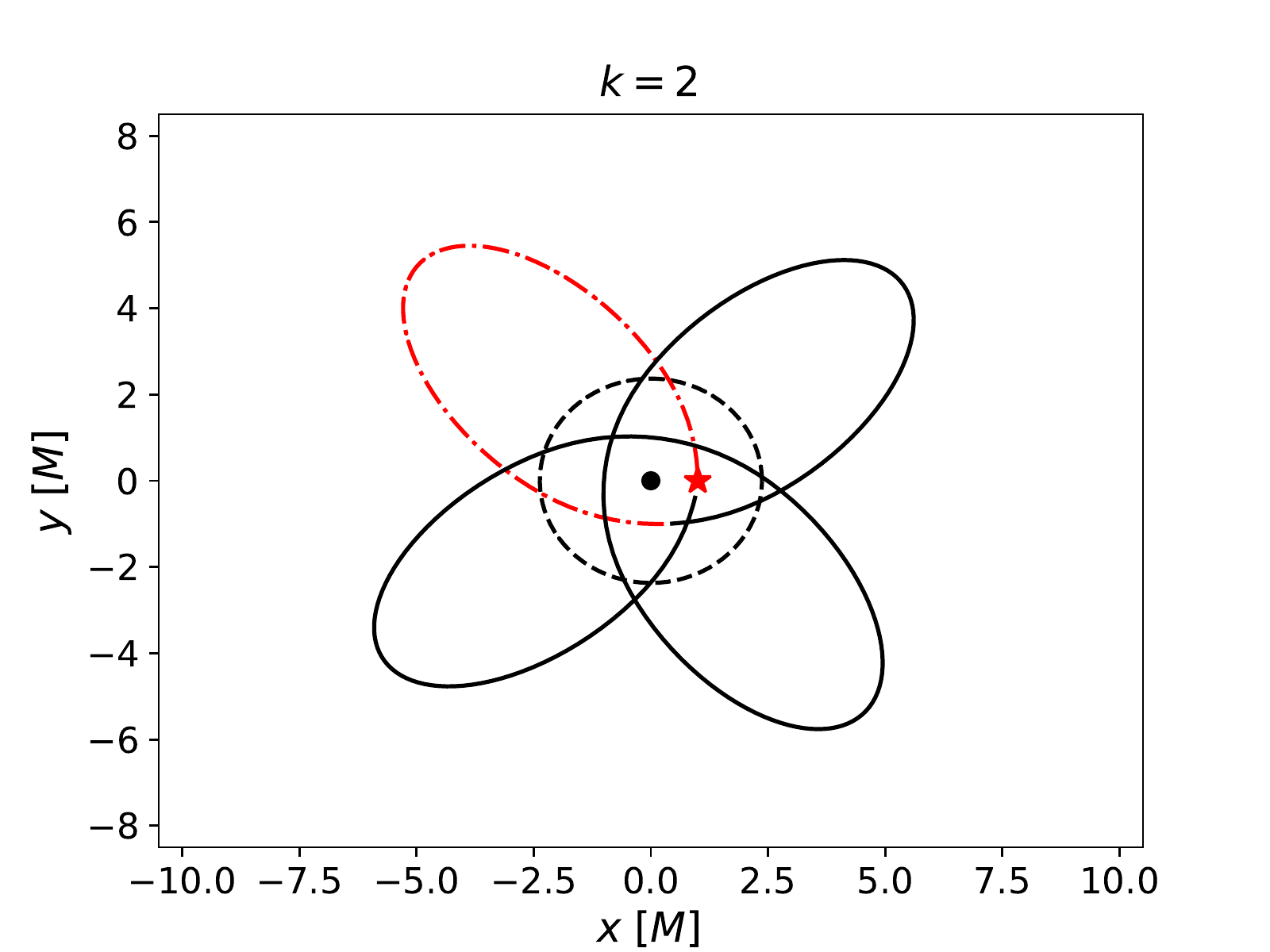}
       \includegraphics[scale=0.3]{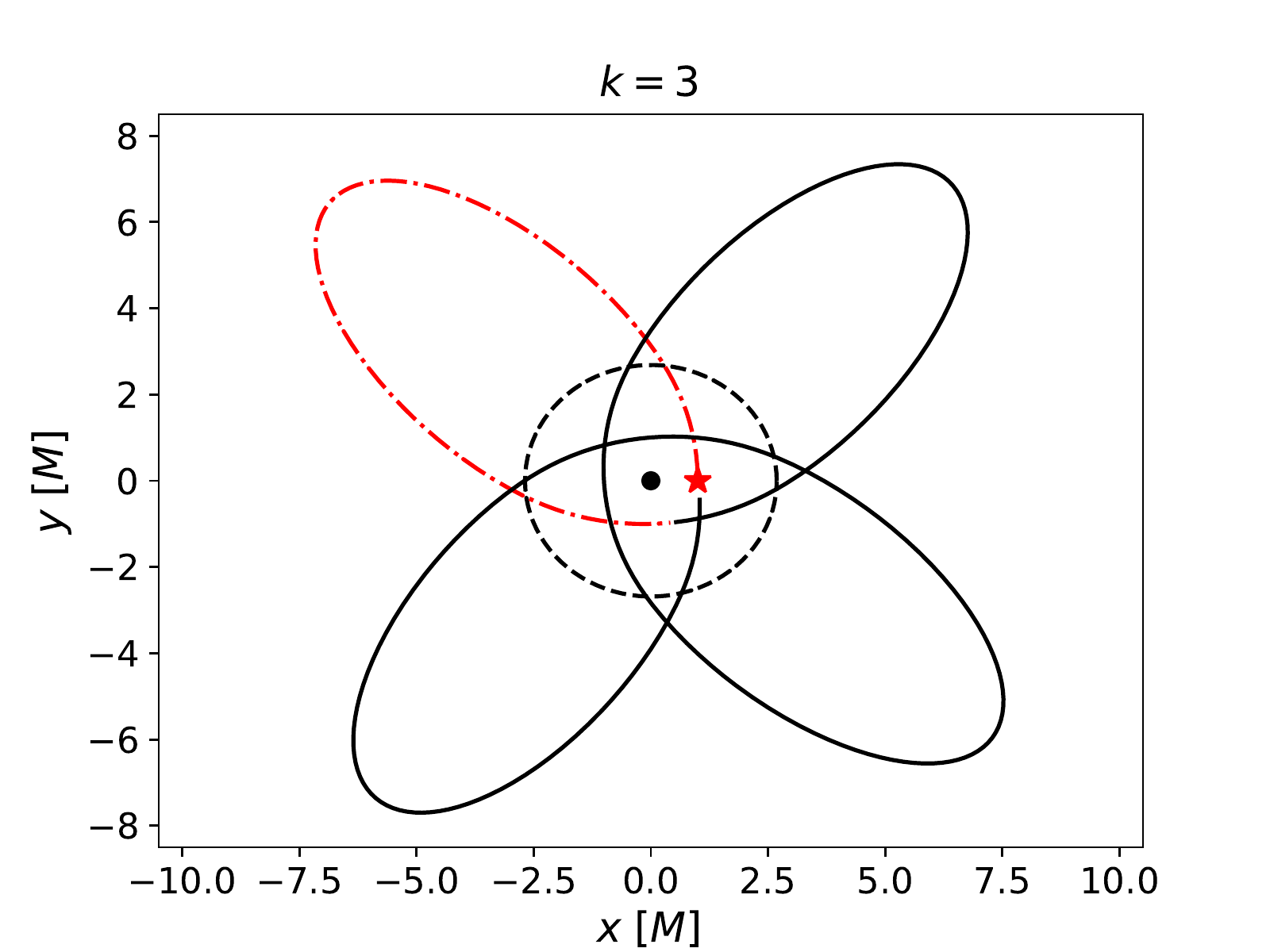}
       \quad
       \includegraphics[scale=0.3]{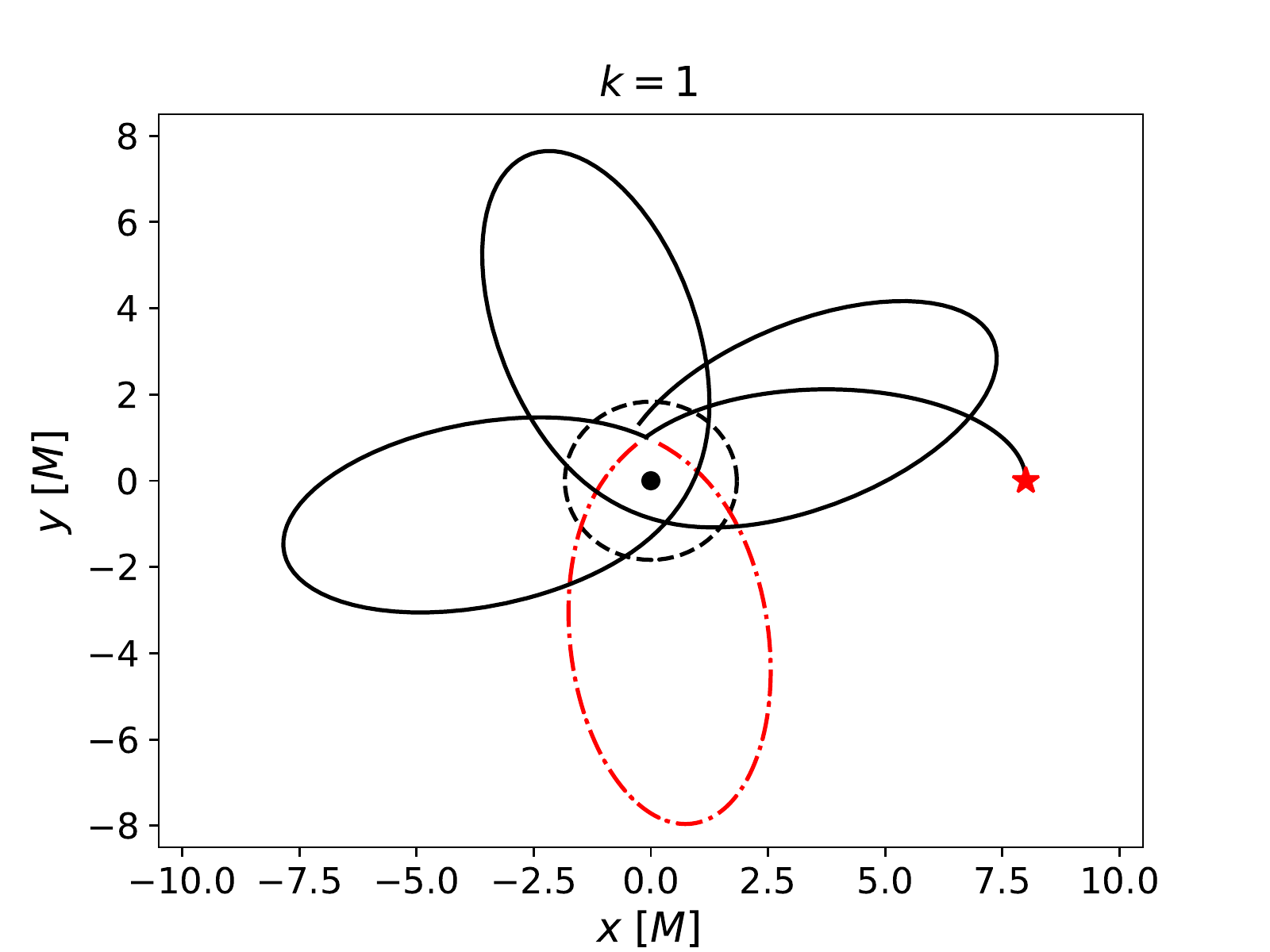}
       \includegraphics[scale=0.3]{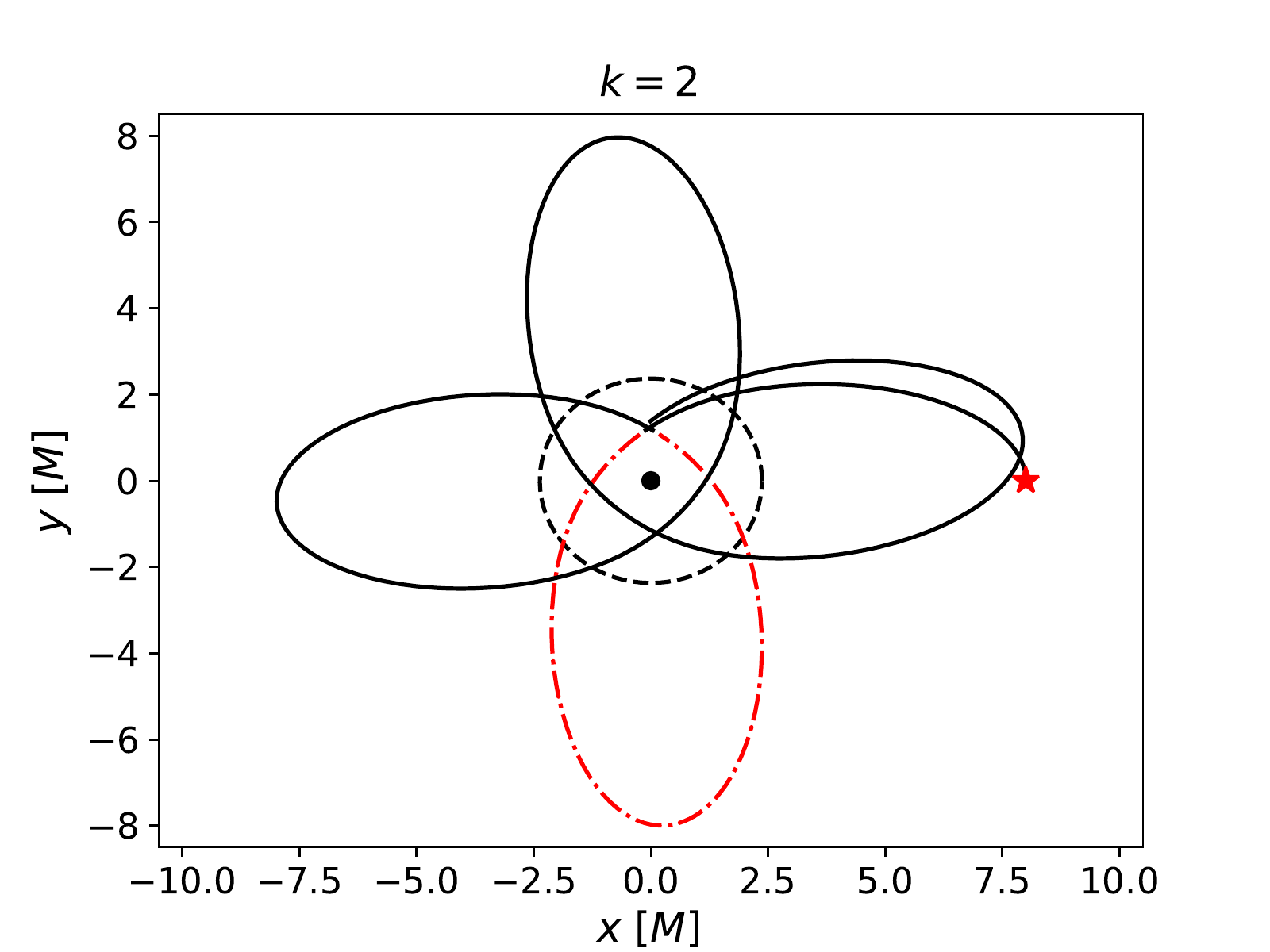}
       \includegraphics[scale=0.3]{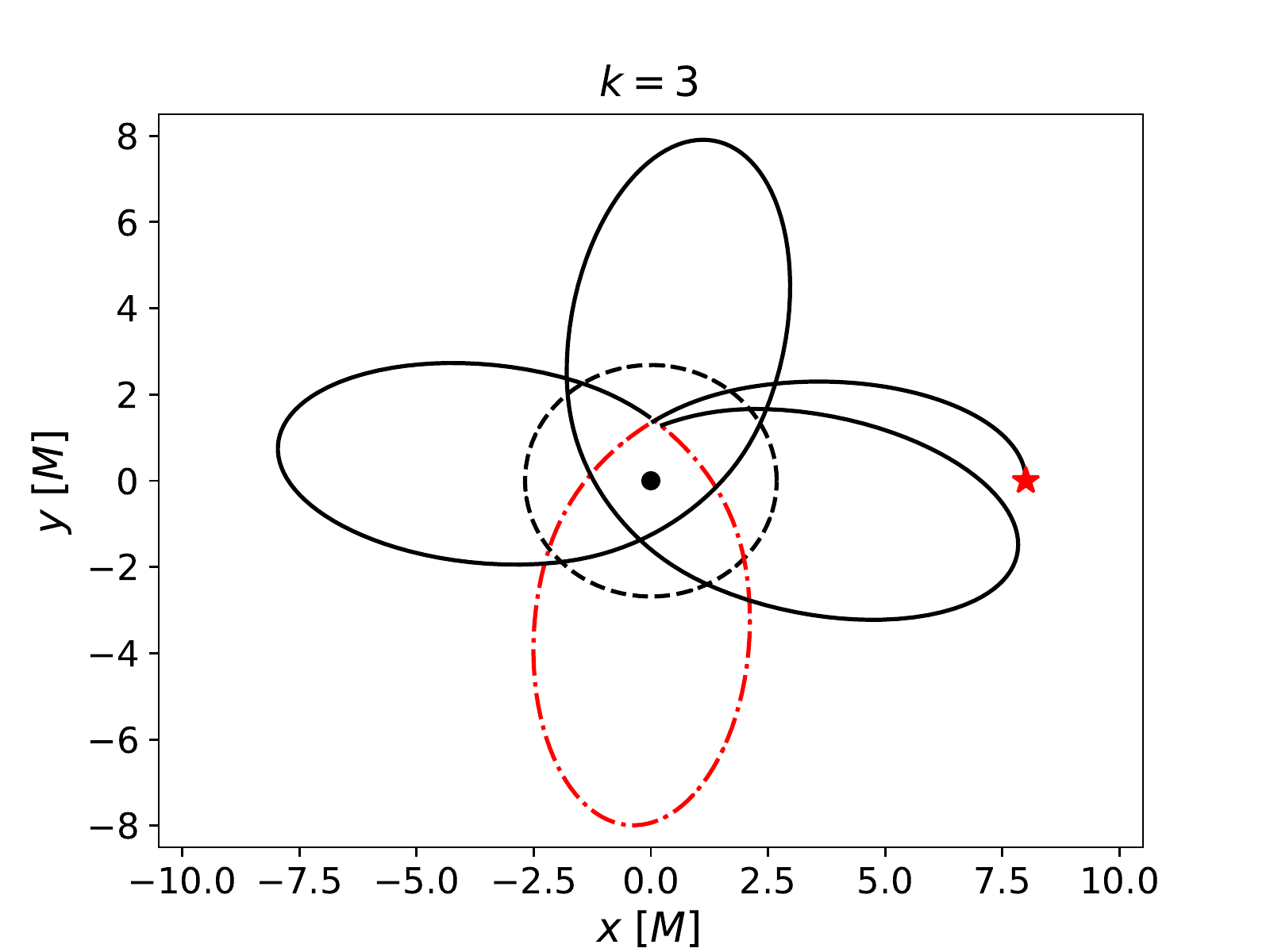}
       \quad
       \includegraphics[scale=0.3]{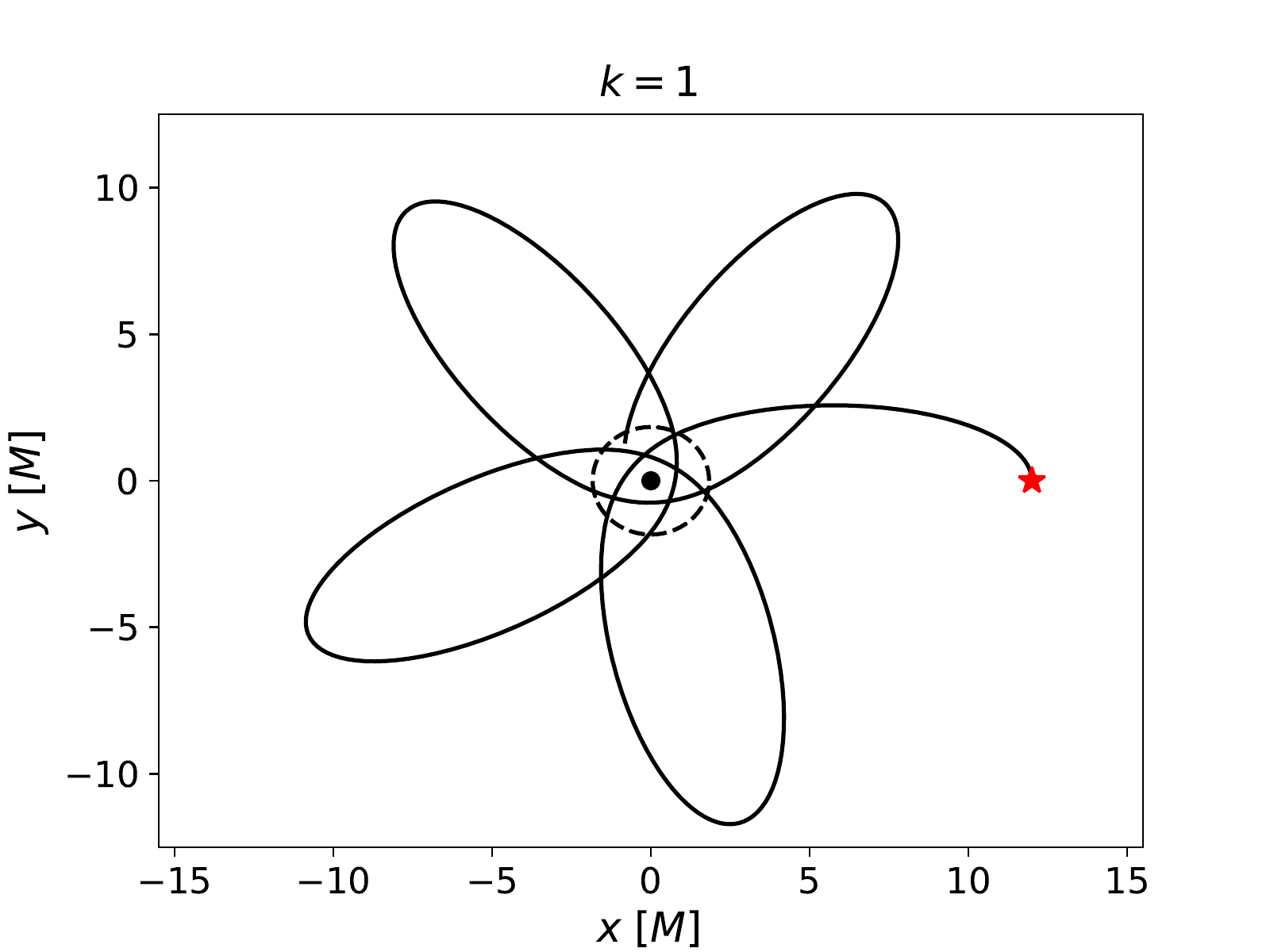}
       \includegraphics[scale=0.3]{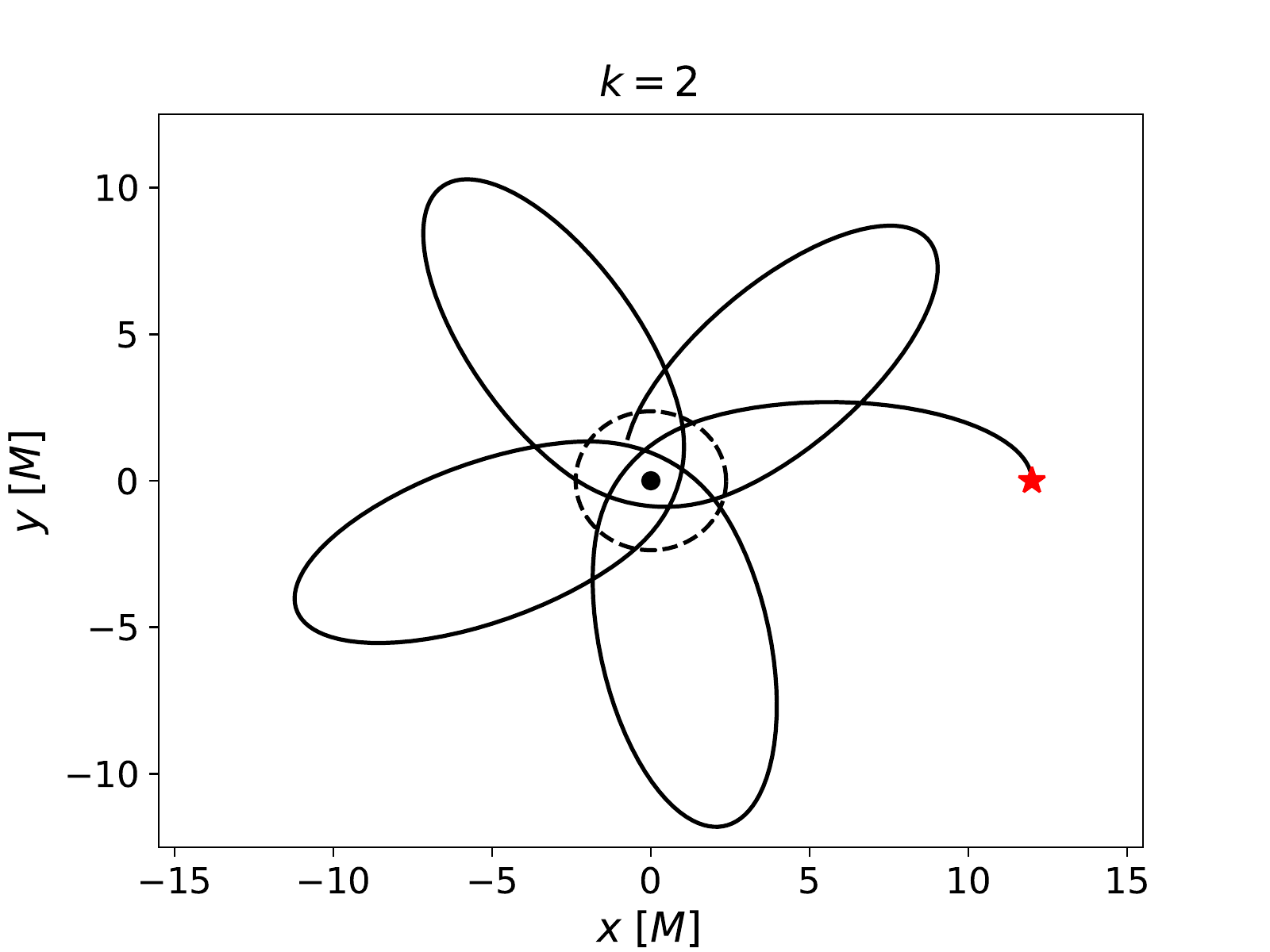}
       \includegraphics[scale=0.3]{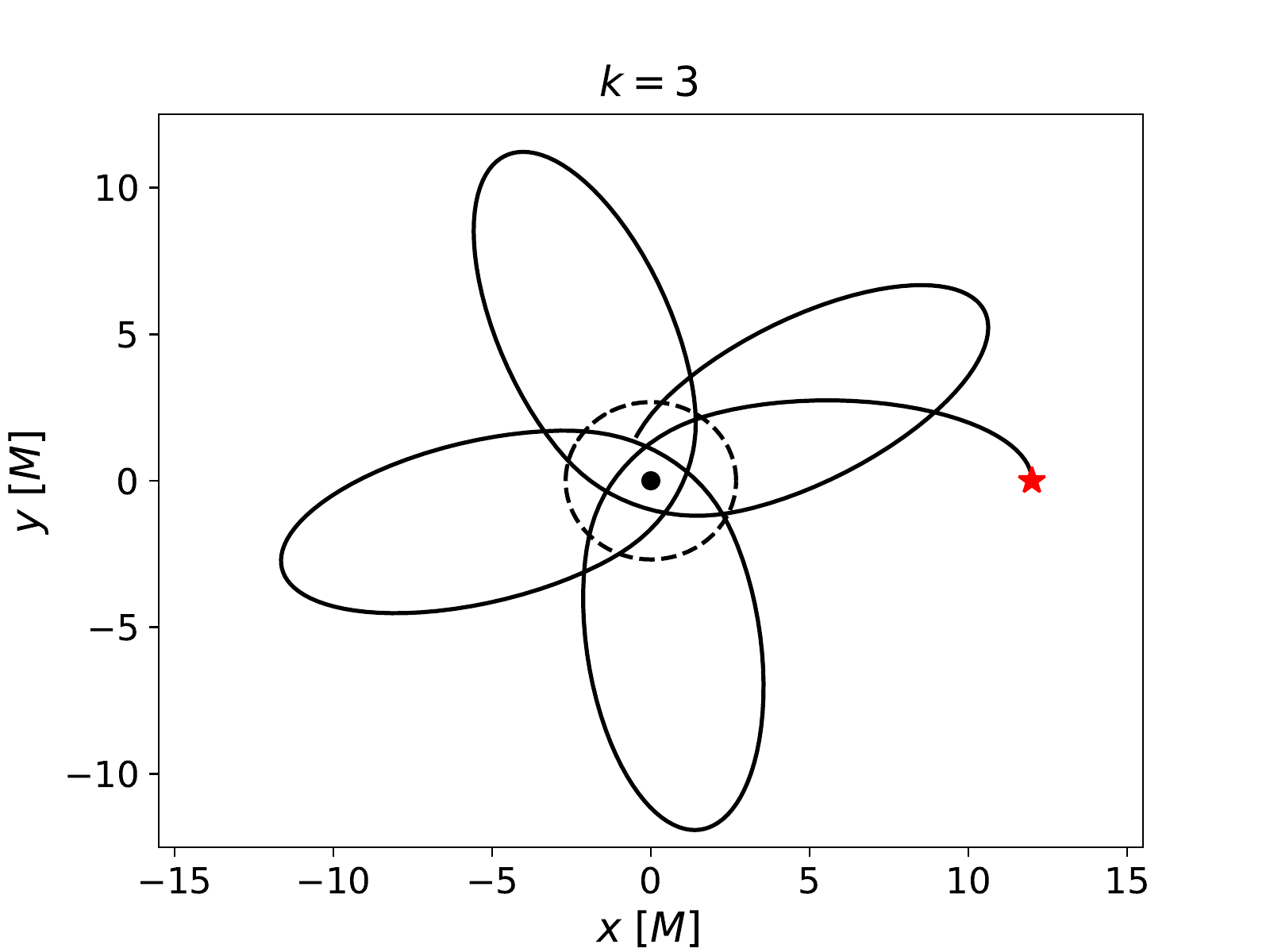}
       \caption{Orbits obtained for a star with $l=1~M$ and orbiting different boson stars whose frequency is fixed at $0.8~m/\hbar$. Three initial radial positions for the star are considered: $r = 1~M$ (upper plots), $r = 8~M$ (middle plots) and $r = 12~M$ (lower plots). The dashed lines correspond to the maximum of the scalar field modulus $\phi$. The black dots denote the geometrical center and the red stars denote the initial position of the star. The red dash-dotted lines illustrate what may correspond to one orbital period of the star.}
       \label{fig:Ol1}
\end{figure}

Fig.~\ref{fig:Ol1} is similar to Fig.~\ref{fig:Ol0} but is obtained considering a star with the angular momentum $l=1~M$. The found orbits have shapes close to those that can be obtained with a Kerr black hole. As encountered for petal orbits, a retrograde relativistic shift is observed. Besides, the orbits become similar for all $k$ when the apocenter reaches a certain distance and the deflection angle plateaued (at $\approx 290^\circ$ in this case).

\begin{figure}[t]
\centering
       \includegraphics[scale=0.5]{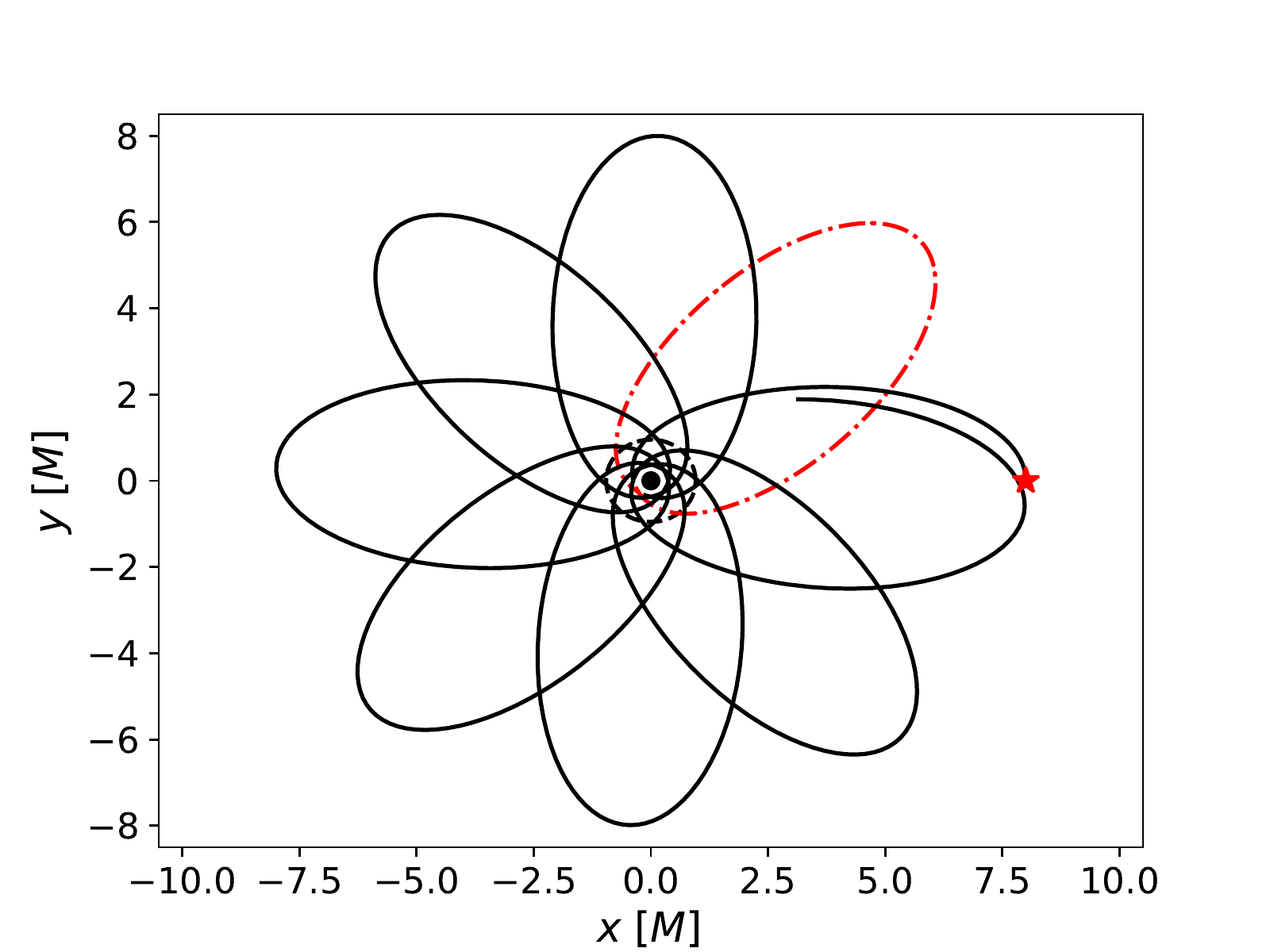}
       \caption{Orbit obtained for a star with $l=1~M$ and orbiting the boson star $k=1$, $\omega=0.7~m/\hbar$. The star is launched from the initial position $r = 8~M$. The dashed lines correspond to the maximum of the scalar field modulus $\phi$. The black dot denotes the geometrical center and the red star denotes the initial position of the star. The dash-dotted line shows an one-period orbit of the star.}
       \label{fig:Ol1om}
\end{figure}

When considering more relativistic boson stars (e.g. $\omega = 0.7~m/\hbar$), the shape of the orbit is also familiar to those that can be found in the Kerr metric, we do not have exotic orbits as previously (see Fig.~\ref{fig:Ol1om}). However, the star passes very close to the center ($r \approx 0.4~M$).

\subsection{Orbits of stars initially at rest}

\begin{figure}[t]
\centering
       \includegraphics[scale=0.5]{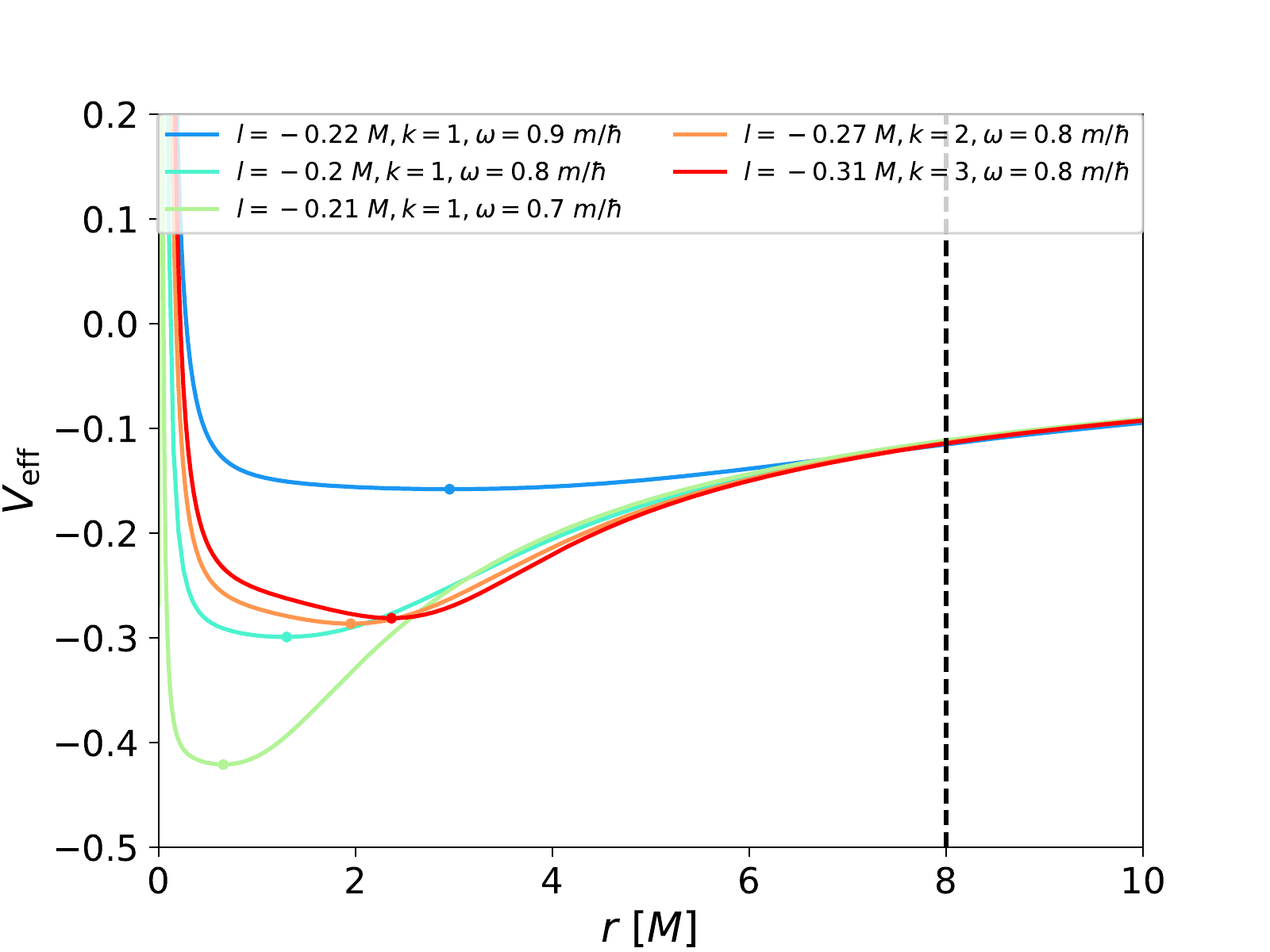}
       \caption{Effective potentials obtained for particular angular momentums allowing to get a star initialized at rest at $r=8~M$ from the center (dash line). The dots denote the minimum of the effective potential.}
       \label{fig:EPrest}
\end{figure}

We now consider a star initially at rest in its frame meaning that $u^\varphi$ is null in addition to $u^r$ and $u^\theta$. Note that such orbits have already been studied by \cite{2014PhRvD..90b4068G}. In the Kerr metric, the formation of sustainable orbits with such initial condition is obviously not possible since the star falls into the black hole. By using the last equality of the system~\eref{eq:CIBS}, we can deduce the angular momentum of such orbits 
\begin{equation}
l = - \varepsilon \frac{g_{t \varphi}}{g_{tt}}.
\end{equation}
Since the energy $\varepsilon$ is positive \citep{2014PhRvD..90b4068G}, and $g_{t \varphi}$ and $g_{tt}$ are negative, the angular momentum $l$ is always negative. Thus, orbits of stars initially at rest and orbiting a rotating boson star have negative angular momentums. To prove their existence, we plot on Fig.~\ref{fig:EPrest} the effective potentials obtained for the angular momentums allowing to satisfy the condition $u^i = 0$ for a star initialized at $r = 8~M$ in each boson-star metric considered. Indeed, for a given boson star a particular angular momentum is needed to verify the condition $u^i= 0$. We can see on Fig.~\ref{fig:EPrest} that at $8~M$ (dash line), i.e where the star is at rest in the different metrics, sustainable orbits exist since they oscillate between $8~M$ (the apocenter) and $r < 1~M$ (the pericenter) for each boson star. 

\begin{figure}[t]
\centering
       \includegraphics[scale=0.3]{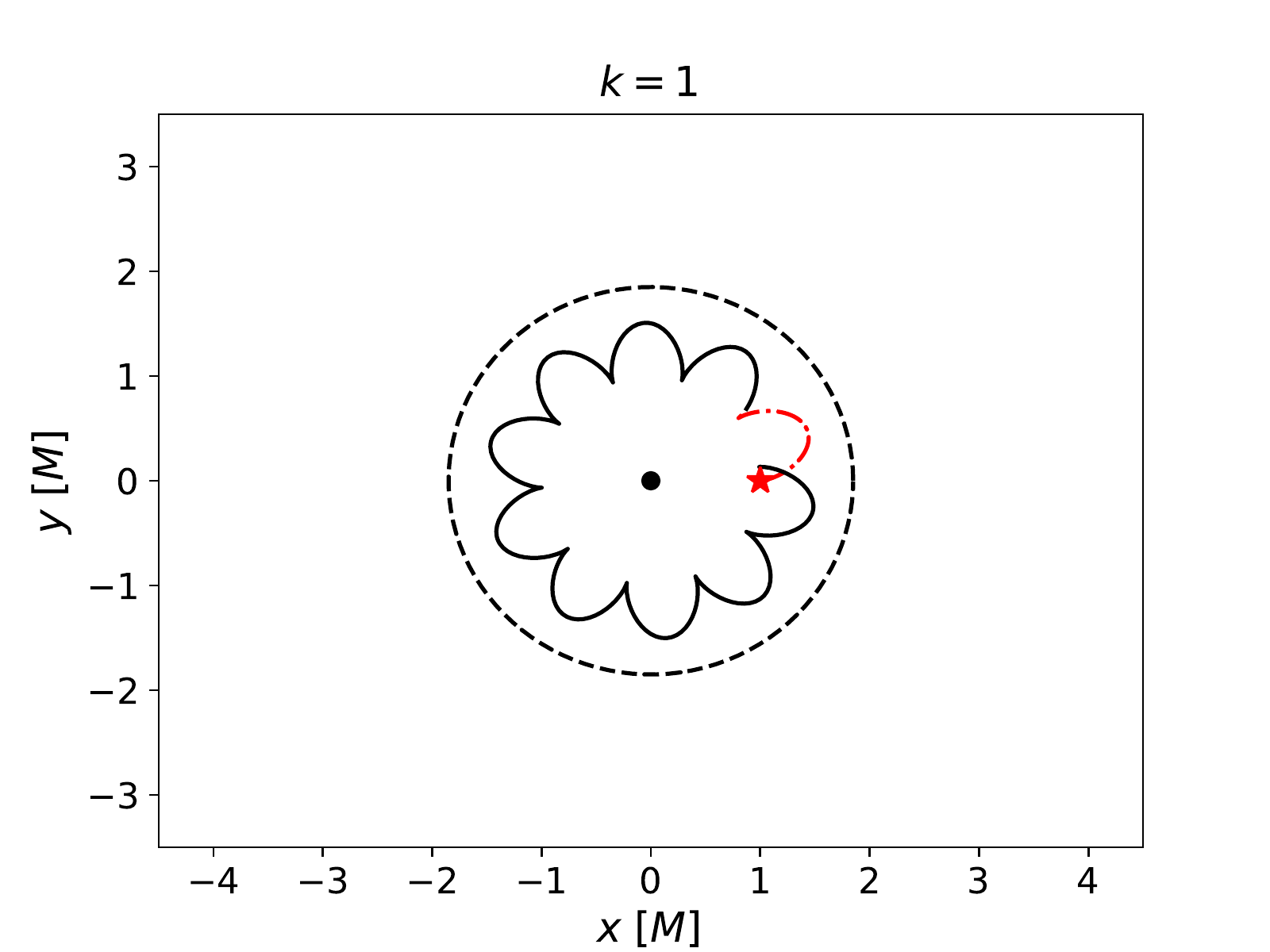}
       \includegraphics[scale=0.3]{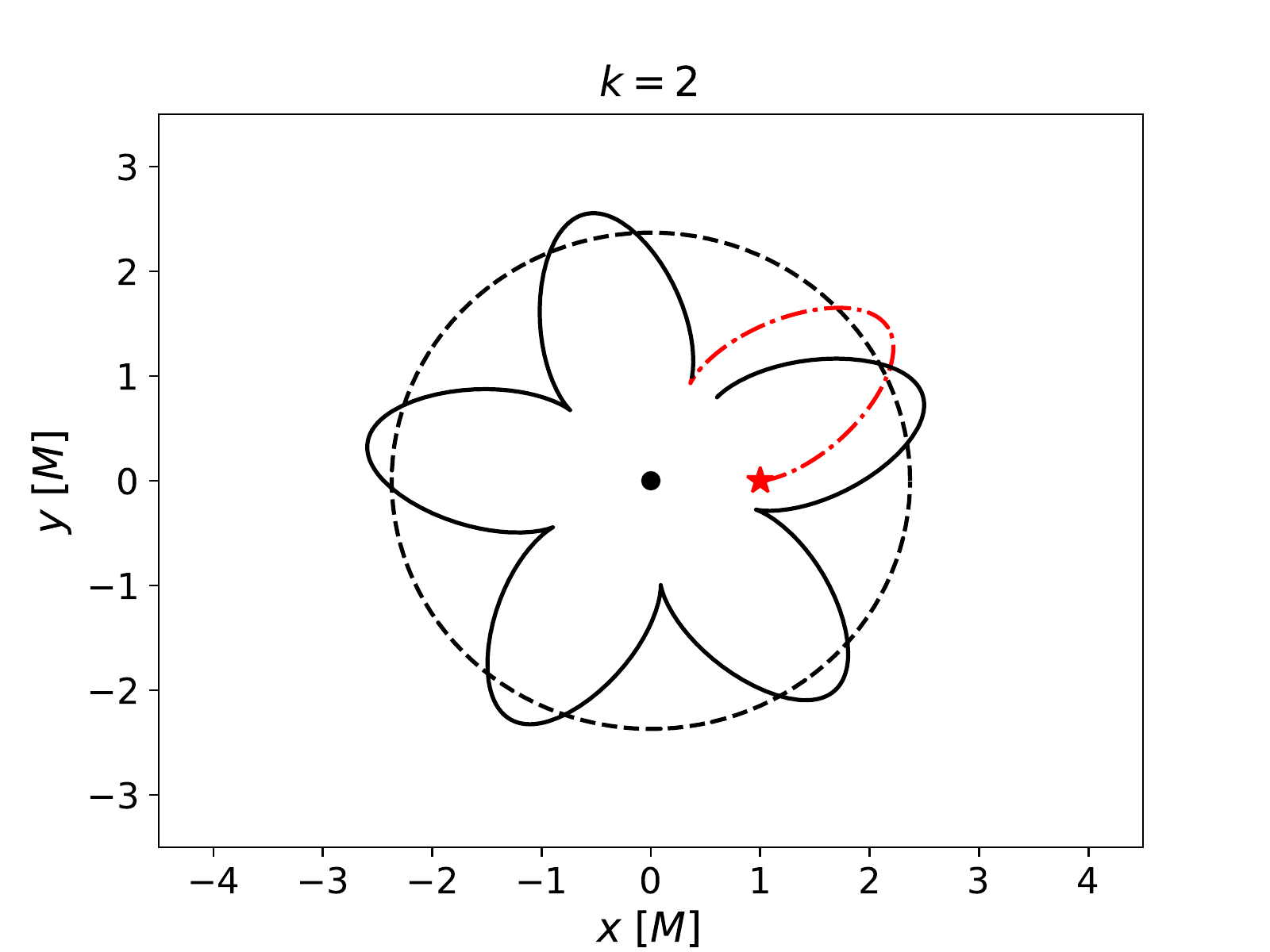}
       \includegraphics[scale=0.3]{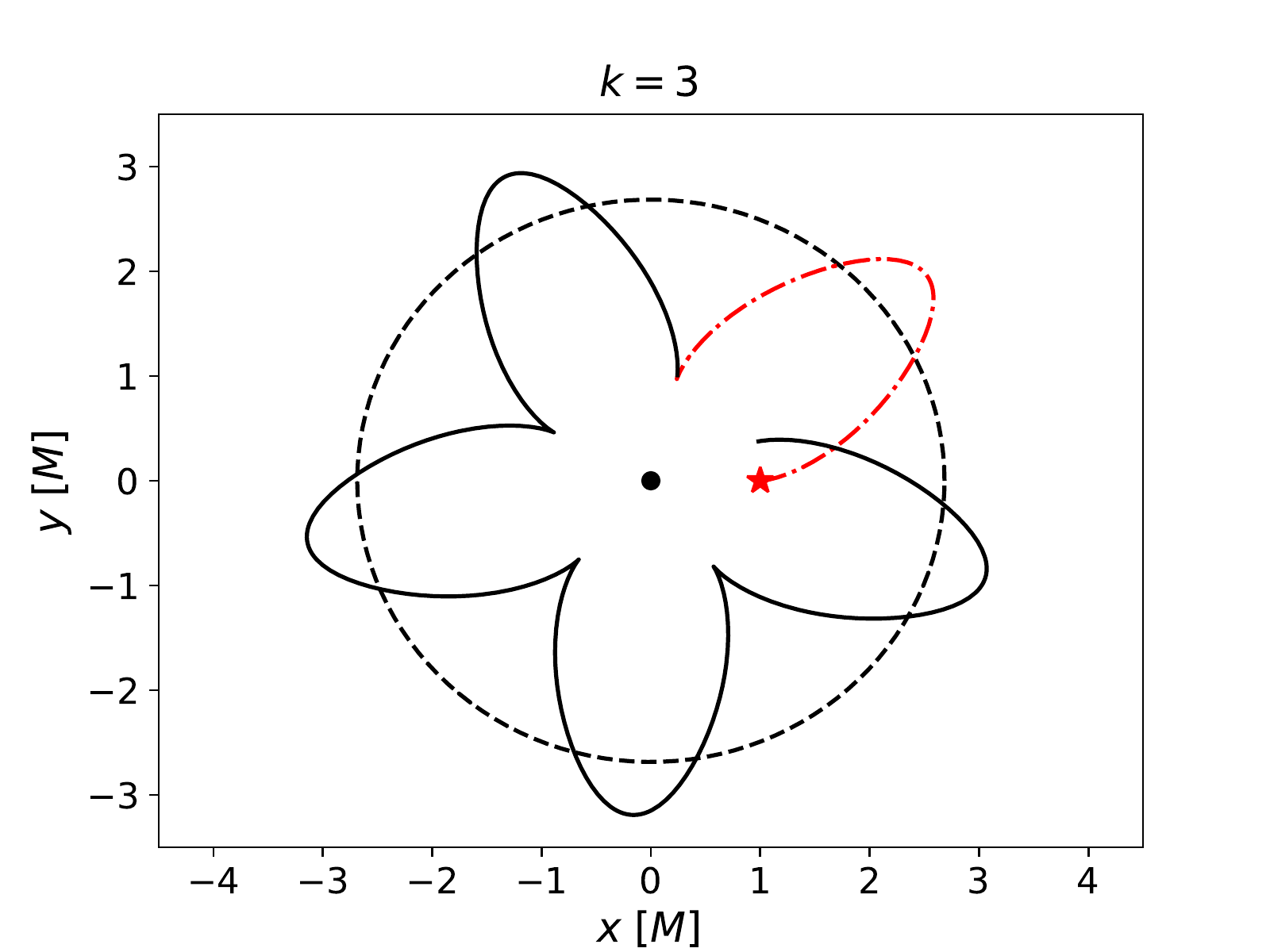}
       \quad
       \includegraphics[scale=0.3]{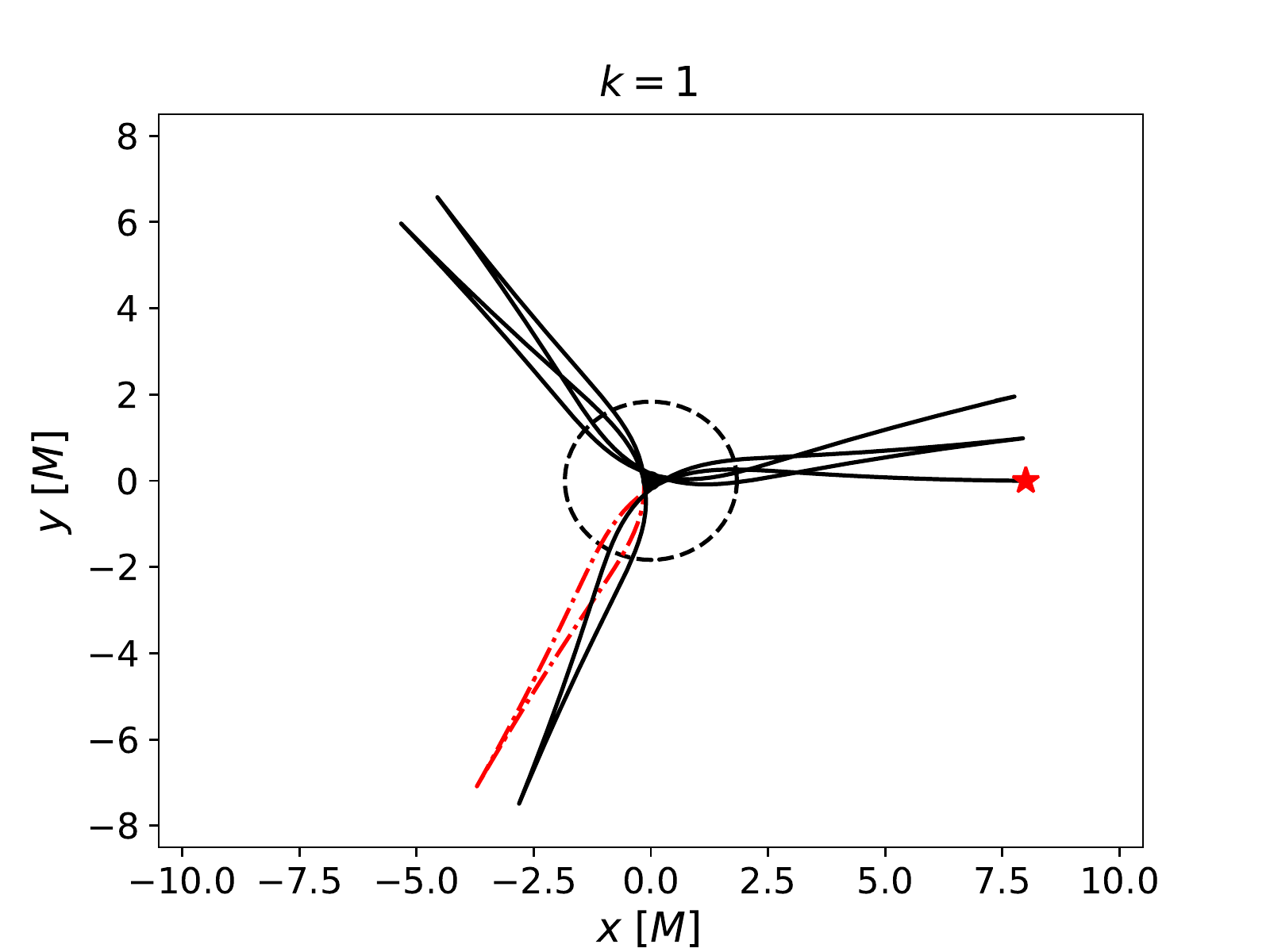}
       \includegraphics[scale=0.3]{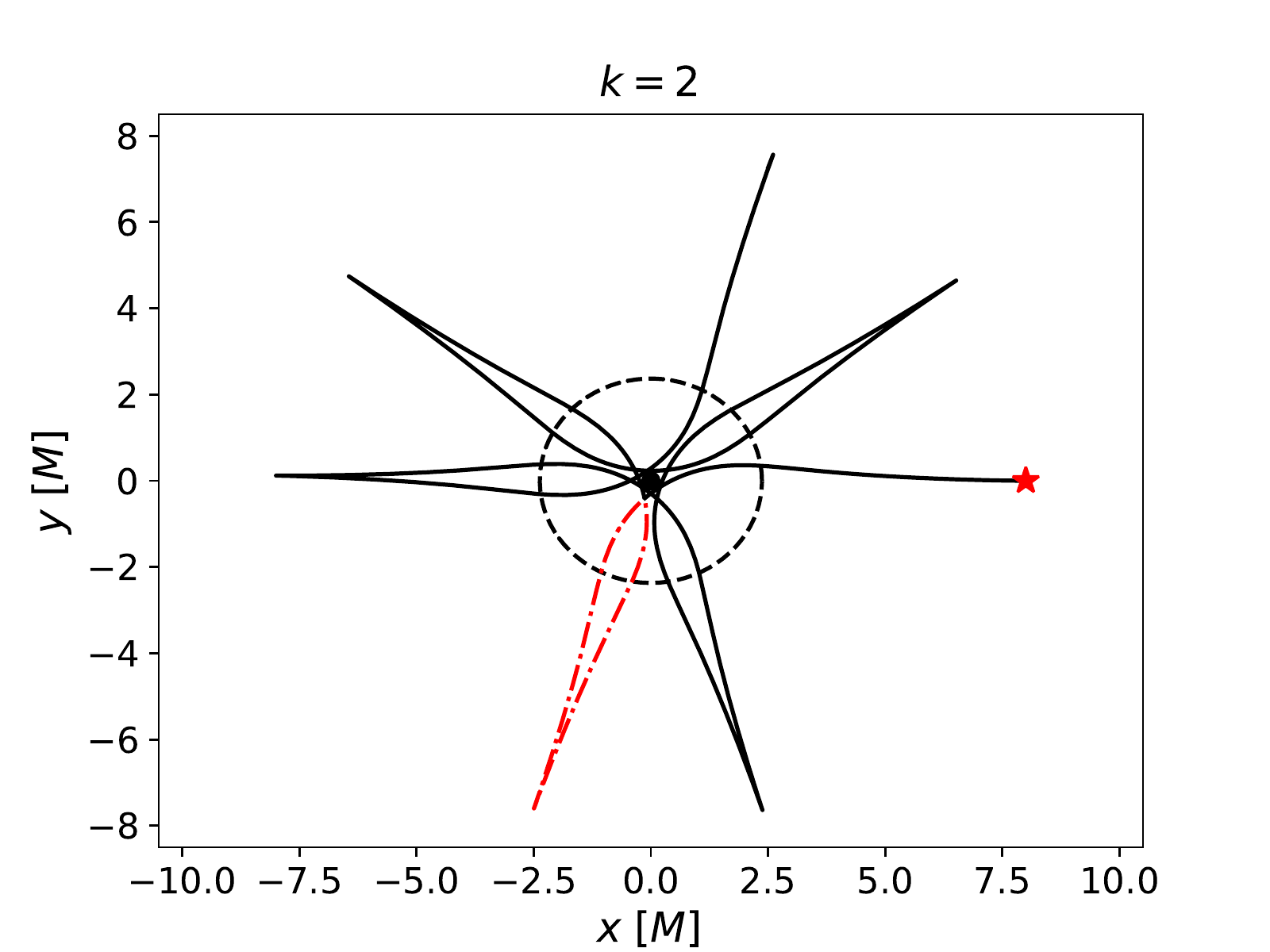}
       \includegraphics[scale=0.3]{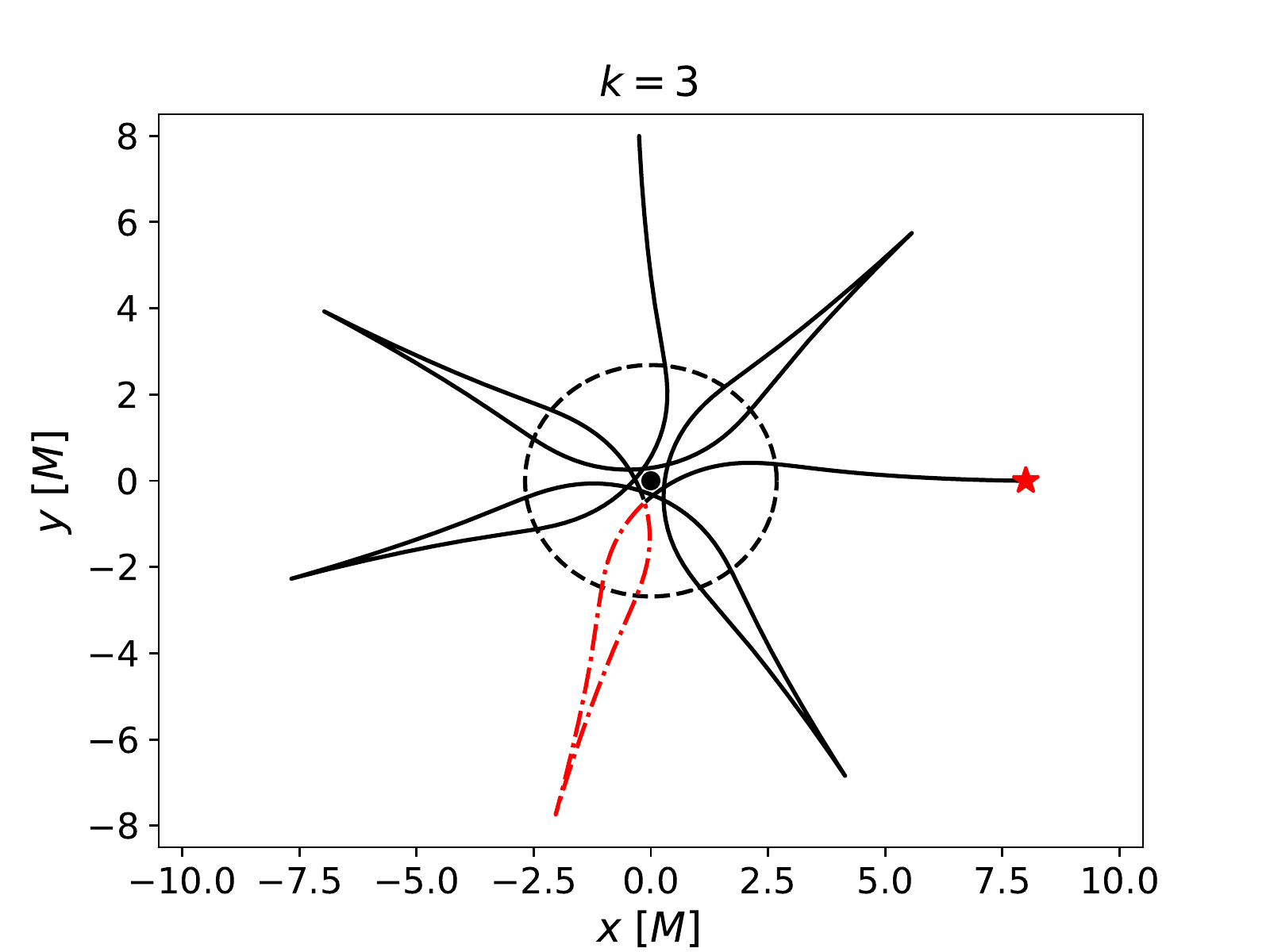}
       \quad
       \includegraphics[scale=0.3]{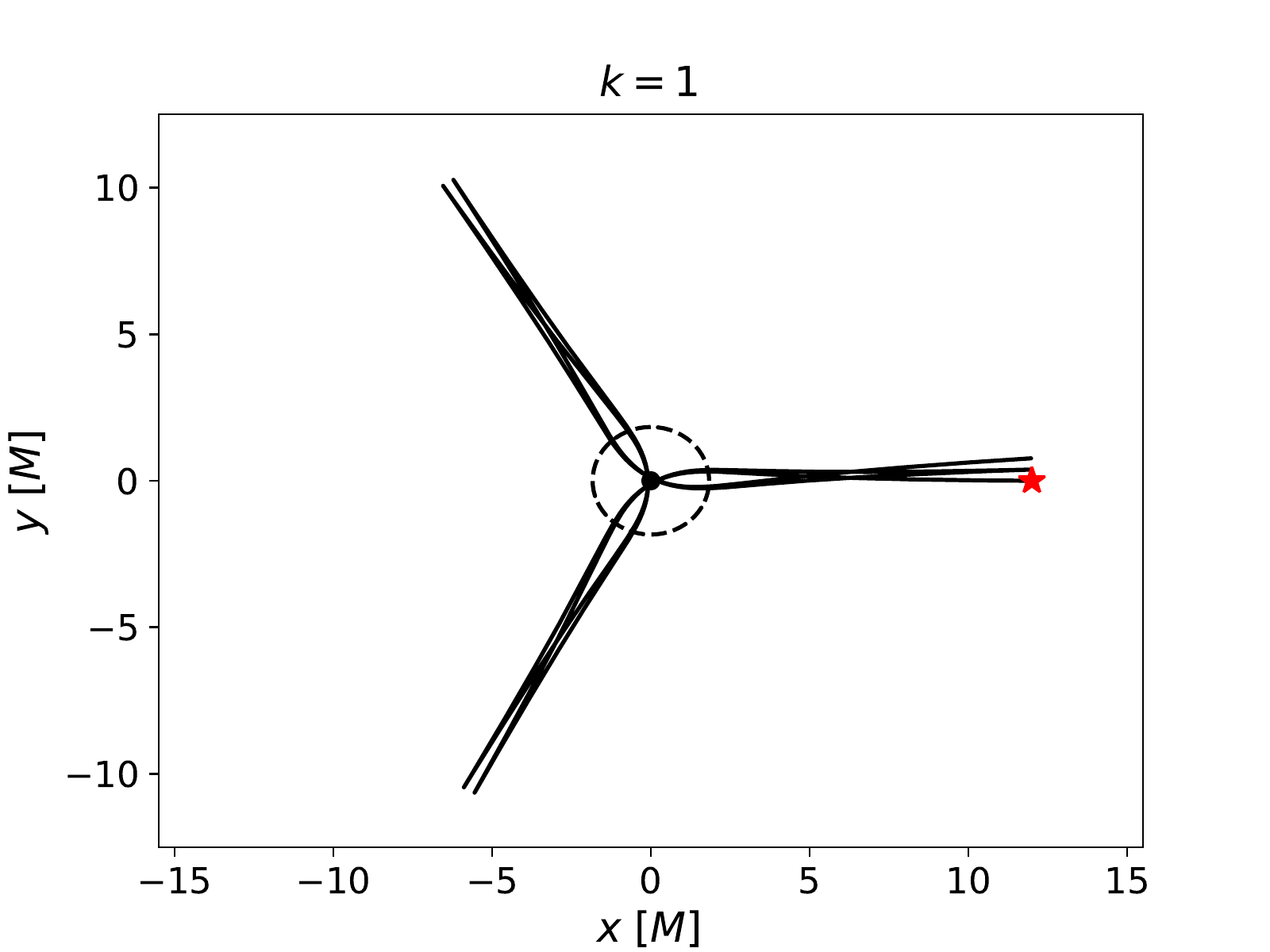}
       \includegraphics[scale=0.3]{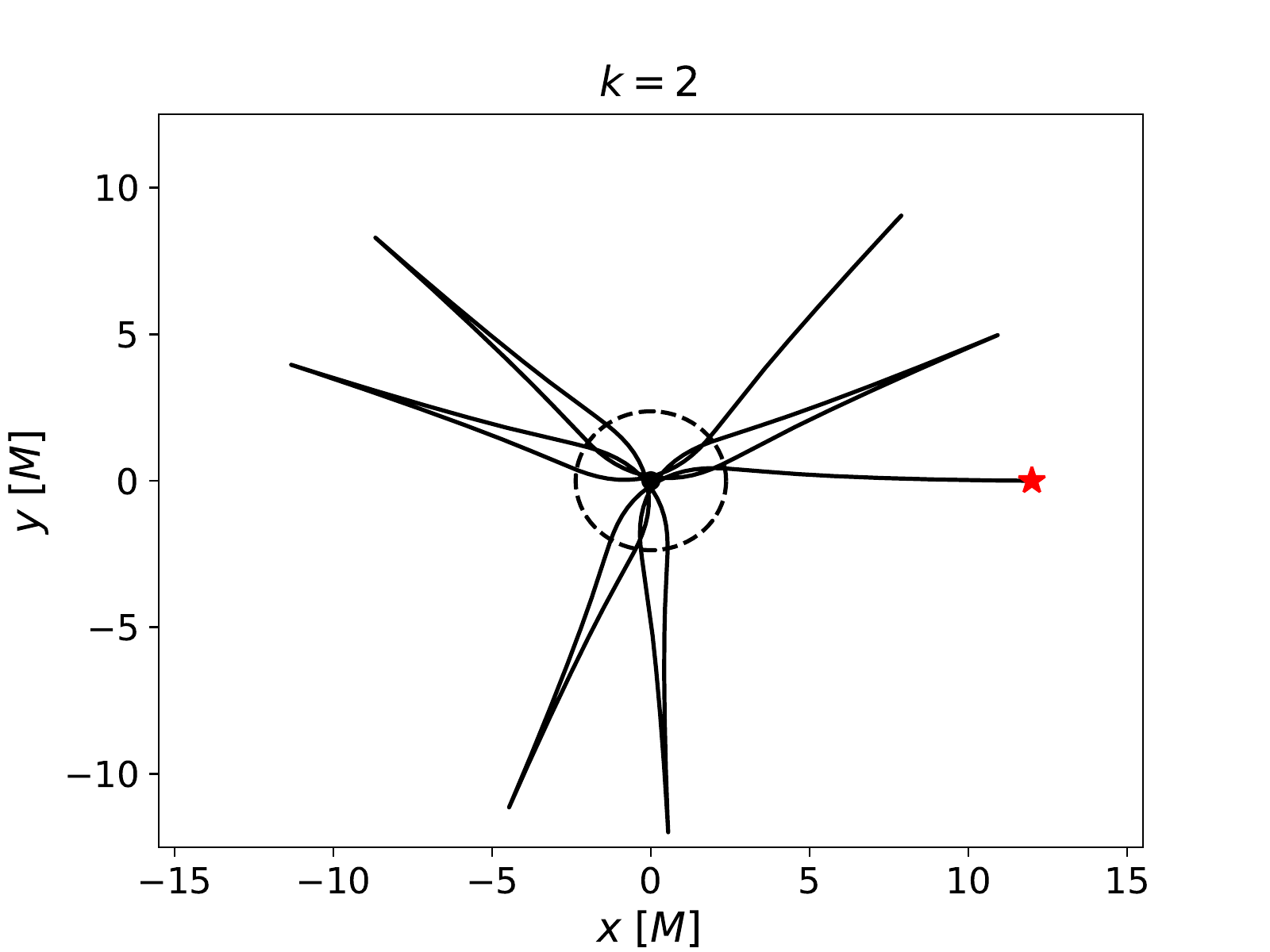}
       \includegraphics[scale=0.3]{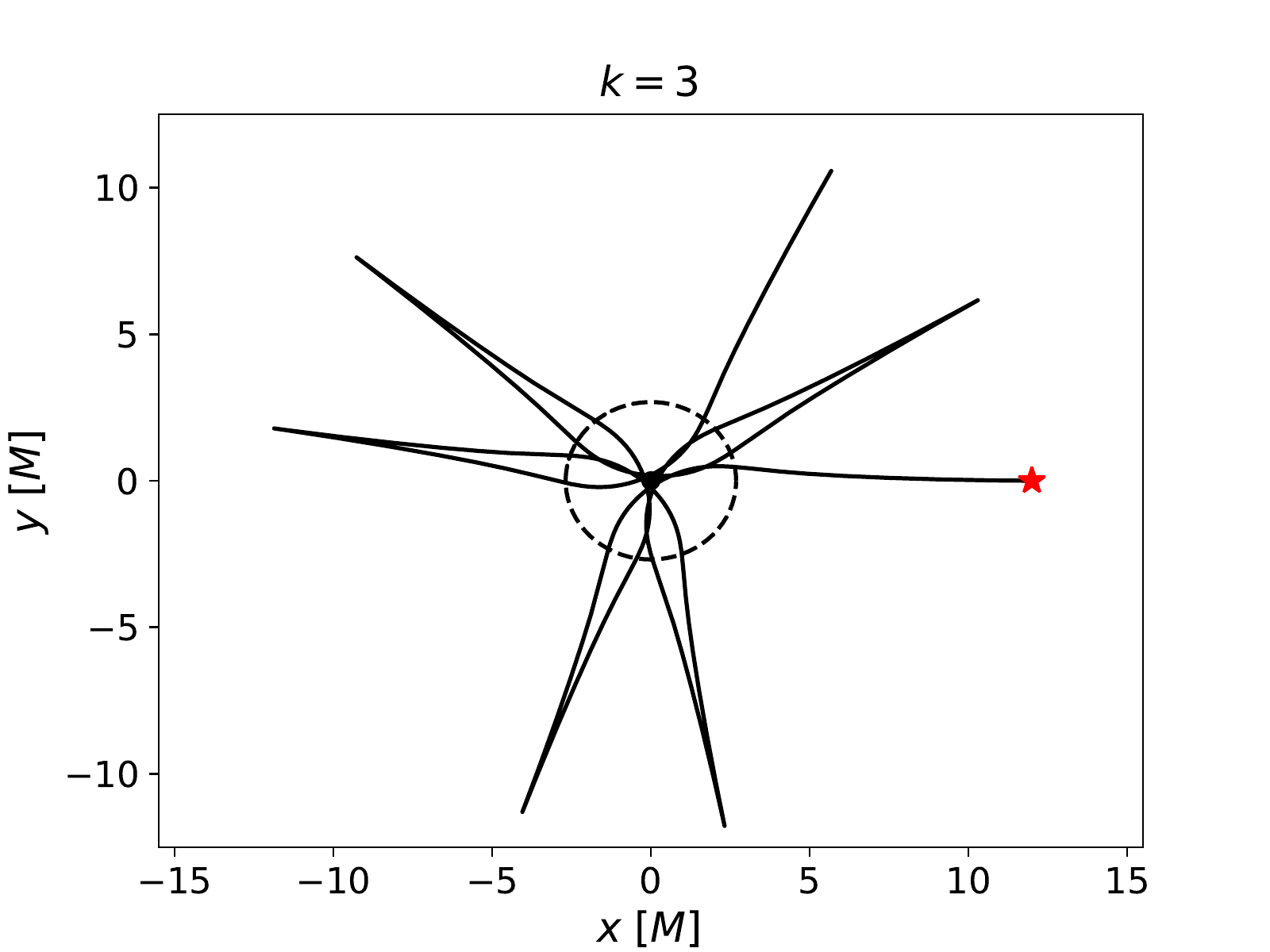}
       \caption{Orbits obtained considering a star initially at rest and orbiting different boson stars whose frequency is fixed at $0.8~m/\hbar$. Three initial radial positions for the star are considered: $r = 1~M$ (upper plots), $r = 8~M$ (middle plots) and $r = 12~M$ (lower plots). The dashed lines correspond to the maximum of the scalar field modulus $\phi$. The black dots denote the geometrical center and the red stars denote the initial position of the star. The red dash-dotted lines illustrate what may correspond to one orbital period of the star.}       \label{fig:Olrest}
\end{figure}

\begin{figure}[t]
\centering
       \includegraphics[scale=0.5]{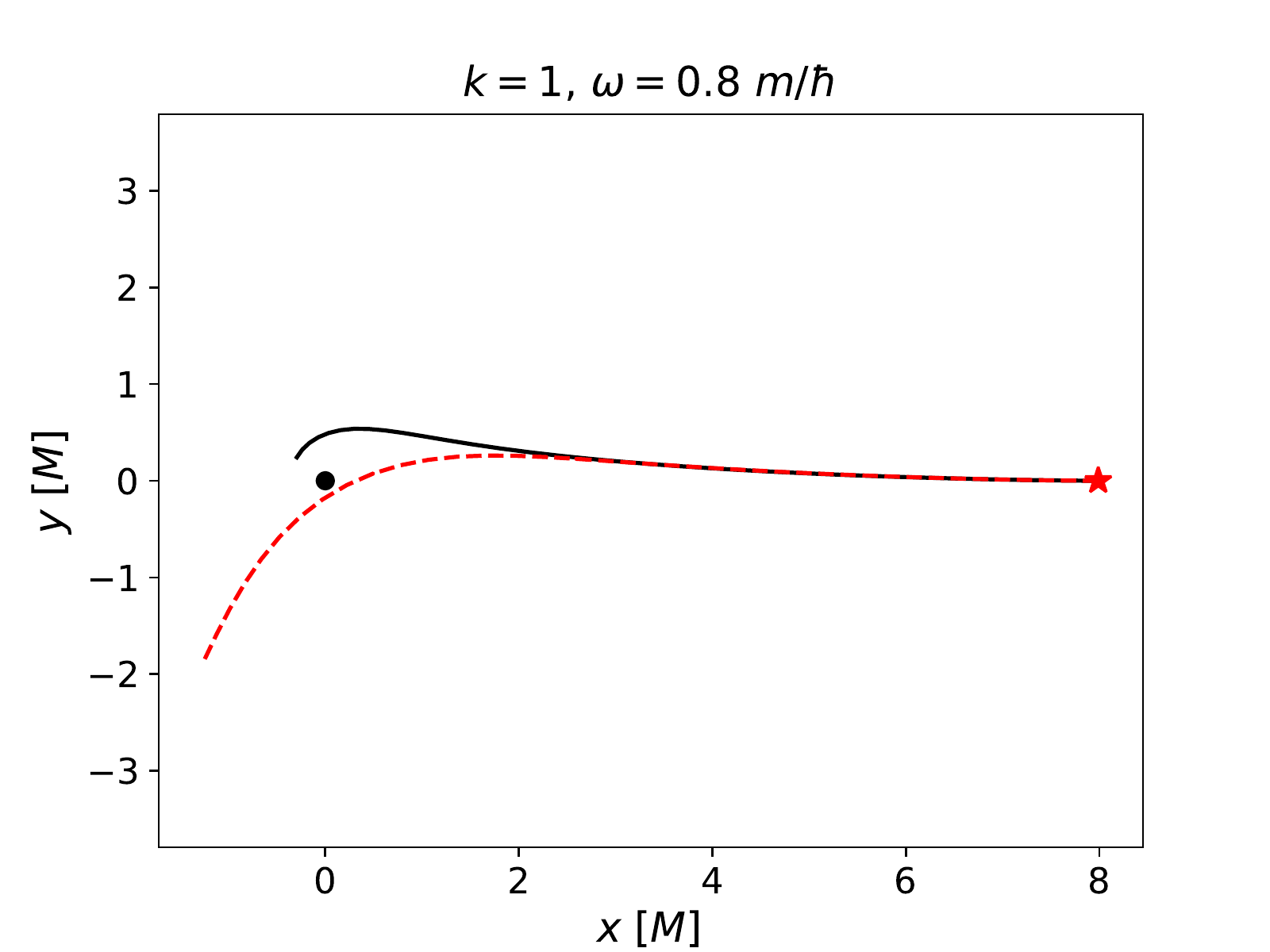}
       \caption{Orbits obtained considering a star initially at rest and orbiting a Kerr black hole at $a=0.802$ (black solid line) and a boson star (red dash line) with $k=1$, $\omega=0.8~m/\hbar$ ($a=0.802$). The initial condition of the star is taken at $r_\mathrm{ap}=8$ $M$. The black dot denotes the geometrical center and the red star denotes the initial position of the star in both metrics. The orbit obtained with the boson star is a portion of the orbit visible on the left middle plot of Fig.~\ref{fig:Olrest}.}
       \label{fig:OlrestBH}
\end{figure}

Fig.~\ref{fig:Olrest} shows the orbits when considering a star initialized at rest. As found by \cite{2014PhRvD..90b4068G}, we observe the formation of \textit{pointy petal orbits}. Contrary to what it is claim in their study, these particular orbits are not formed for zero angular momentums but negative. We note that when the star has zero three-velocity at pericenter the orbit appears like semi orbits. The pointy petal orbits are thus always formed when the three-velocity of the star is null at apocenter. Regarding the relativistic shift, it is prograde for the semi orbits and retrograde for the pointy petal orbits. Contrary to all previous orbits, the star evolves in the retrograde sense with respect to the geometrical center for the pointy petal orbits. On Fig.~\ref{fig:OlrestBH} is plotted the trajectories of a star orbiting a rotating boson star and a Kerr black hole for $a=0.802$. This figure confirms the fact that the star evolution differs when it gets closer to the geometrical center. Moreover, we can see that before falling into the black hole, the motion of the star is very close to the one obtained with the boson star.

The pointy petal orbits are also formed for $\omega=0.7~m/\hbar$ and $\omega=0.9~m/\hbar$, and the deflection angle increases when $\omega$ decreases. Even far from the compact object such orbits exist and the deflection angle plateaued for instance at $\approx~250^\circ$ for $\omega=0.8~m/\hbar$ and all considered azimuthal numbers.

\section{Comparisons between timelike geodesics obtained in the Kerr and boson-star metrics}
\label{CompBlackHoleBosonStar}

In this section, we focus on the comparison between orbits of stars obtained in both the Kerr and the boson-star metrics.

\subsection{Method}
\label{Method}

In order to compare timelike geodesics generated in each metric, we consider similar initial coordinates of the star in both spacetimes. First, we choose to fix the position and the velocity of the star in the Kerr metric. These initial coordinates are generated at pericenter or apocenter and in the equatorial plane. Thus, the initial three-position of the star in the Kerr metric is $(r,\pi/2,0)$, and its initial velocity is given by the system \eref{eq:CIBS} where the energy of the star and its initial radial position in the Kerr metric are fixed (see the beginning of Sect.~\ref{TLBosonStar}). The corresponding initial position and velocity of the star in the boson-star metric are equal to those obtained in the Kerr metric since they are both expressed in the quasi-isotropic system. Note that the energy and the angular momentum of the star in the boson-star metric can be deduced from its initial coordinates by
\begin{eqnarray}
\label{eq:ELVBS}
\varepsilon =  - g_{tt} u^t - g_{t\varphi} u^\varphi, \nonumber \\ 
l = g_{\varphi t} u^t + g_{\varphi\varphi} u^\varphi. 
\end{eqnarray}
In what follows, energy and angular momentum of the star in the Black Hole and Boson-Star metrics will be indexed by BH and BS, respectively.

\subsection{Energy}

\begin{figure}[t]
\centering
      \includegraphics[scale=0.38]{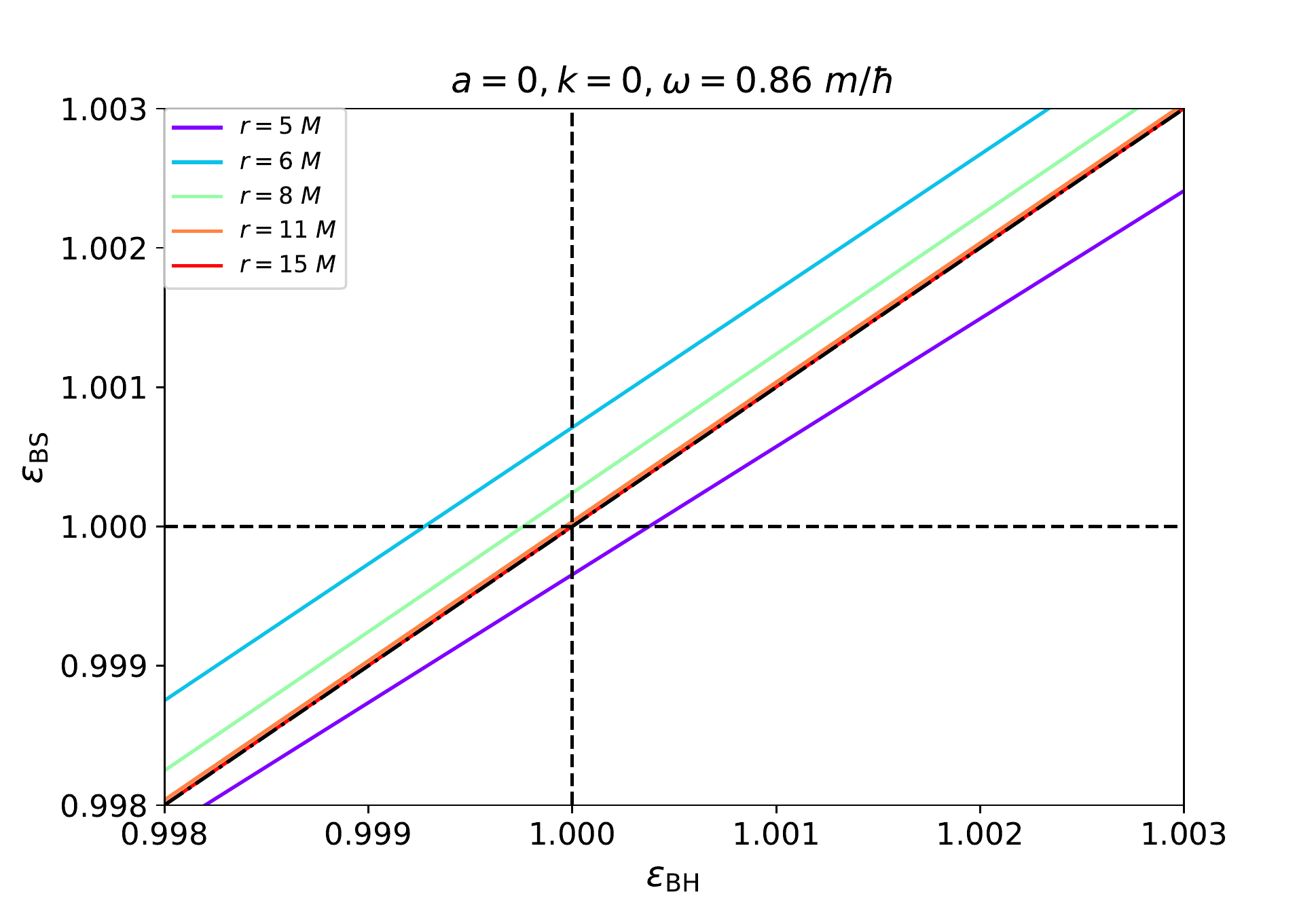}
      \quad
      \includegraphics[scale=0.38]{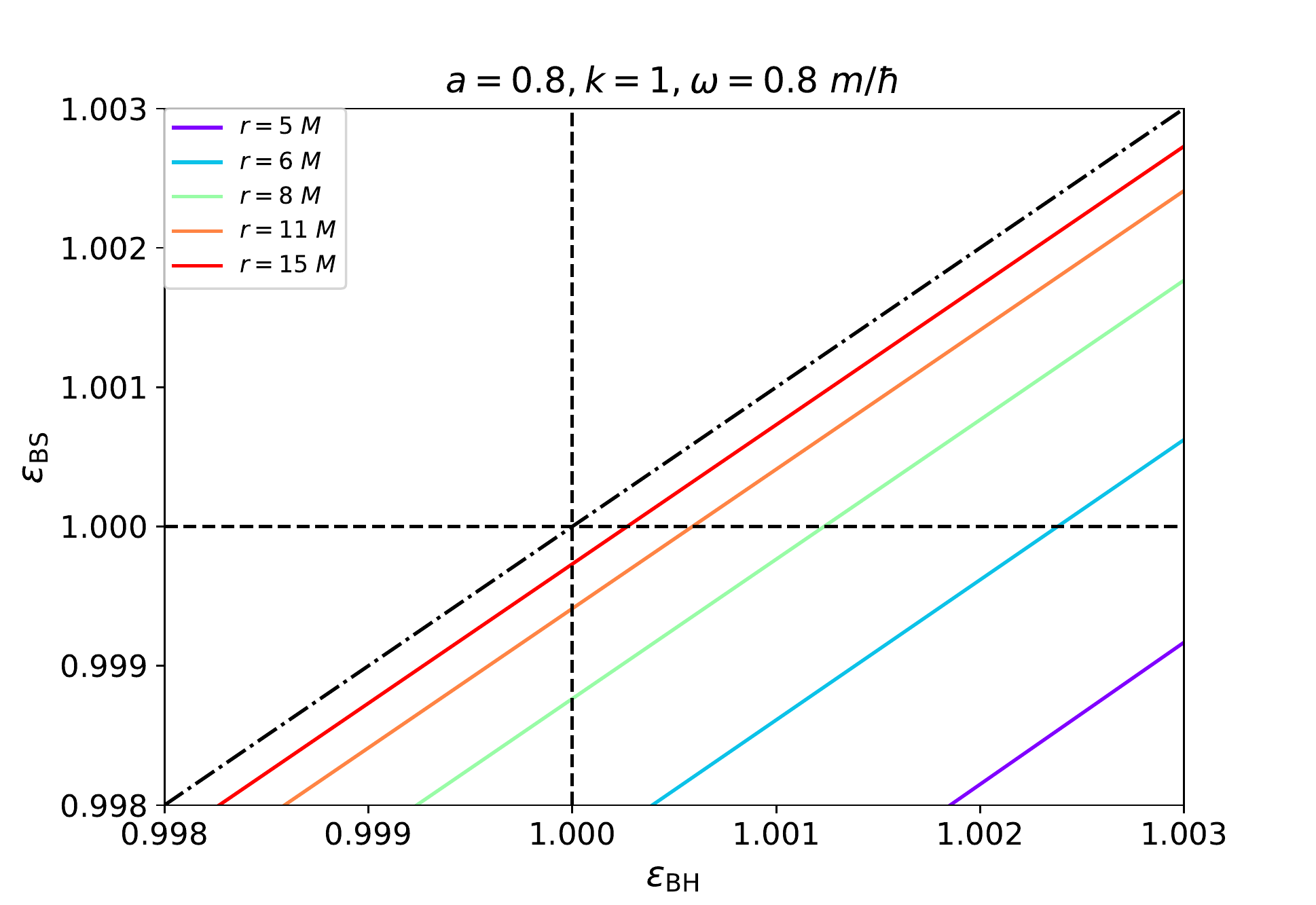}
      \quad
      \includegraphics[scale=0.38]{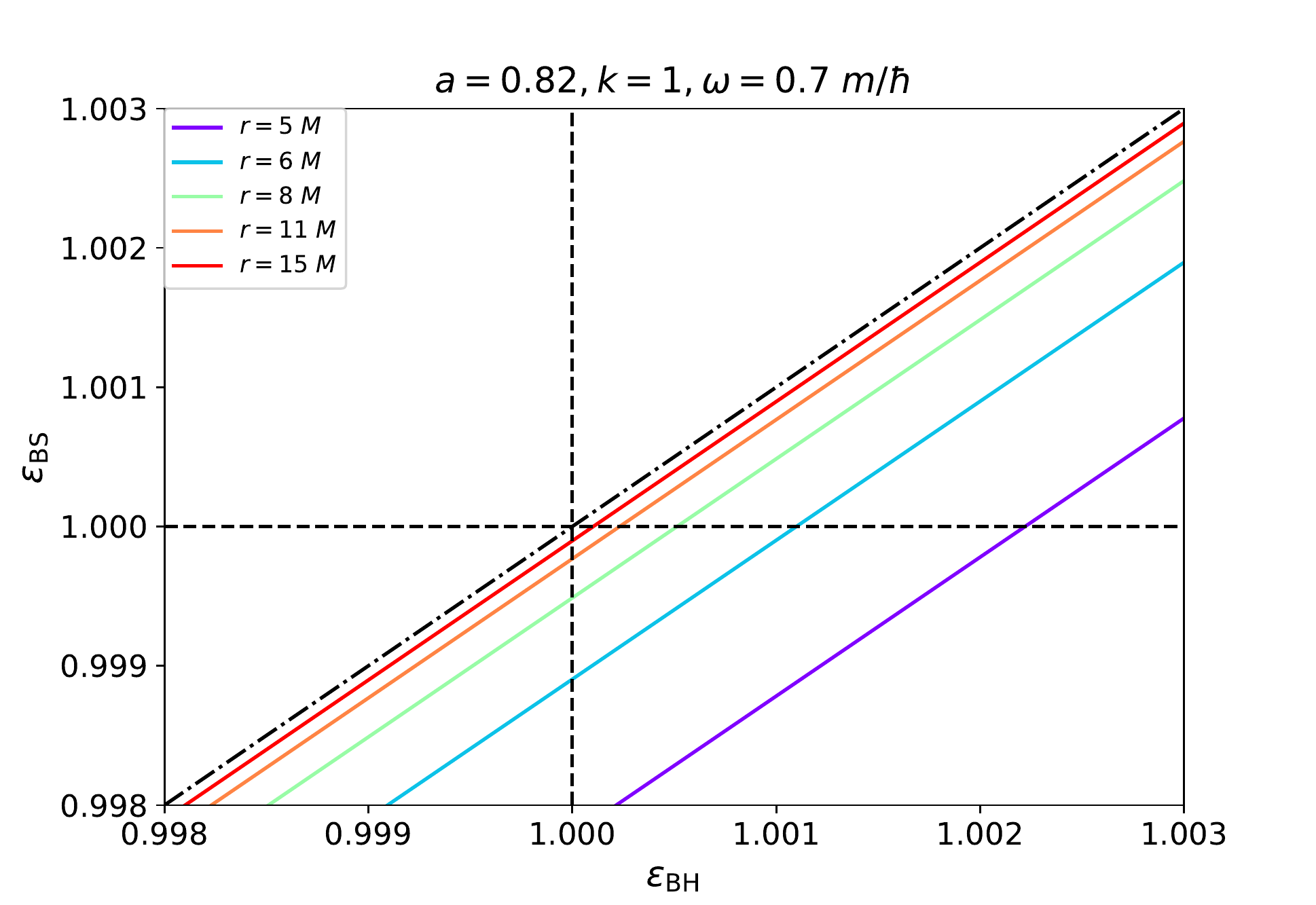}
      \quad
      \includegraphics[scale=0.38]{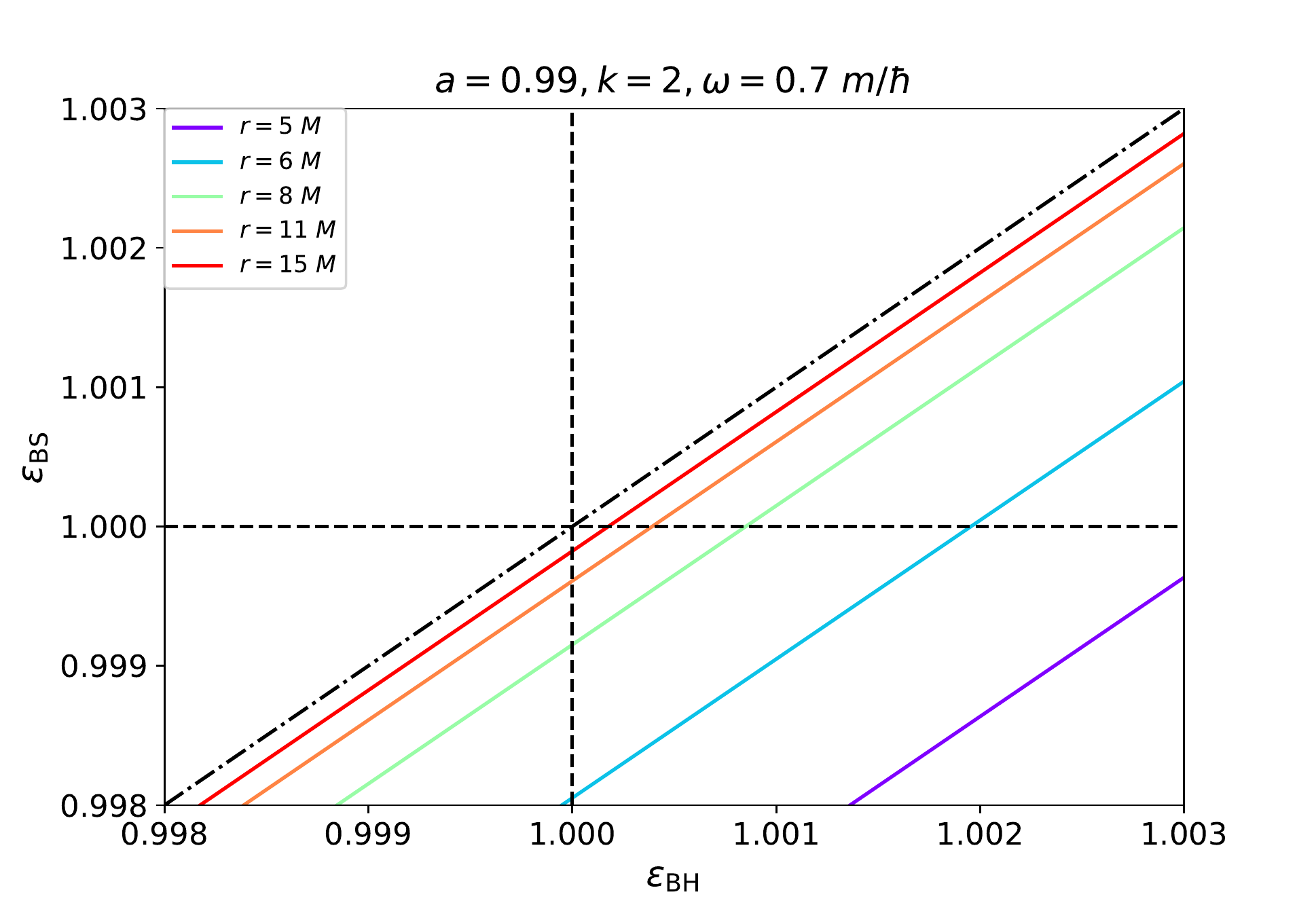}
   \caption{Energies of the star obtained in various boson-star metrics and when considering different initial radial positions and energies of this star in the black hole metric. The dash-dotted lines correspond to the case where energies of the star in both metrics are equal.}
   	\label{fig:EAM}
\end{figure} 

Before studying the trajectory of the star, we compare its energies $\varepsilon_\mathrm{BH}$ and $\varepsilon_\mathrm{BS}$. We remind that the latter depends on the initial coordinates of the star chosen in the Kerr metric. Indeed, by fixing $\varepsilon_\mathrm{BH}$ and the initial radial position we are able to get the initial coordinates of the star in the boson-star metric and thus its energy $\varepsilon_\mathrm{BS}$, given by the first equation of \eref{eq:ELVBS}.

Fig.~\ref{fig:EAM} gives the energy $\varepsilon_\mathrm{BS}$ obtained when considering initial position and velocity generated with $\varepsilon_\mathrm{BH}$ varying between 0.998 and 1.003, and various initial distances $r$. We mention that all angular momentums considered here (and obtained by solving the equation~\eref{eq:LBH} for the Kerr black hole and by using the second equality of equation~\eref{eq:ELVBS} for the boson star) implies the existence of sustainable orbits in both metrics. In particular, in the Kerr metric the star does not fall into the black hole.

First, we note on Fig.~\ref{fig:EAM} that the energy $\varepsilon_\mathrm{BS}$ tends to $\varepsilon_\mathrm{BH}$ when the initial radial position of the star is taken further from the compact object. For non-rotating objects, the convergence between both metrics appears faster (i.e for small initial distances) than with rotating objects. Besides, when the frequency $\omega$ decreases at a given azimuthal number $k$, the deviation slightly decreases. This is due to the fact that when $\omega$ tends to $m/\hbar$, the boson star is less relativistic meaning that it is not as compact as a black hole. Thus, there is more similarity between both metrics when the frequency is small. Considering the two lower plots of Fig.~\ref{fig:EAM}, we also note that when $k$ increases the deviation increases which is due to the increase of the Lense-Thirring effect. 

Focusing now on the upper left plot of Fig.~\ref{fig:EAM}, we note that at $r = 5~M$ a part of the unbound orbits in the non-rotating black hole metric ($\varepsilon_\mathrm{BH} \geqslant 1$) are bound in the non-rotating boson-star metric ($\varepsilon_\mathrm{BS} < 1$), and for larger distances the opposite is observed: orbits are unbound in the non-rotating boson-star metric when they are bound in the non-rotating black hole one. For rotating compact objects we find bound orbits in the boson-star metric when they are unbound in the Kerr metric. Note that for the frequencies considered here, we never find the opposite case as previously. These results show that even considering the same initial coordinates of the star in both metrics, we can obtain different types of orbits. \\

In following subsections we compare the evolution of the star in both metrics considering identical initial coordinates. To facilitate the comparison between timelike geodesics, we consider initial position and velocity in the Kerr metric allowing to recover similar types (bound or unbound) of orbits in both metrics.

\subsection{Bound orbits: $\varepsilon<1$}

\begin{figure}[t]
\centering
       \includegraphics[scale=0.8]{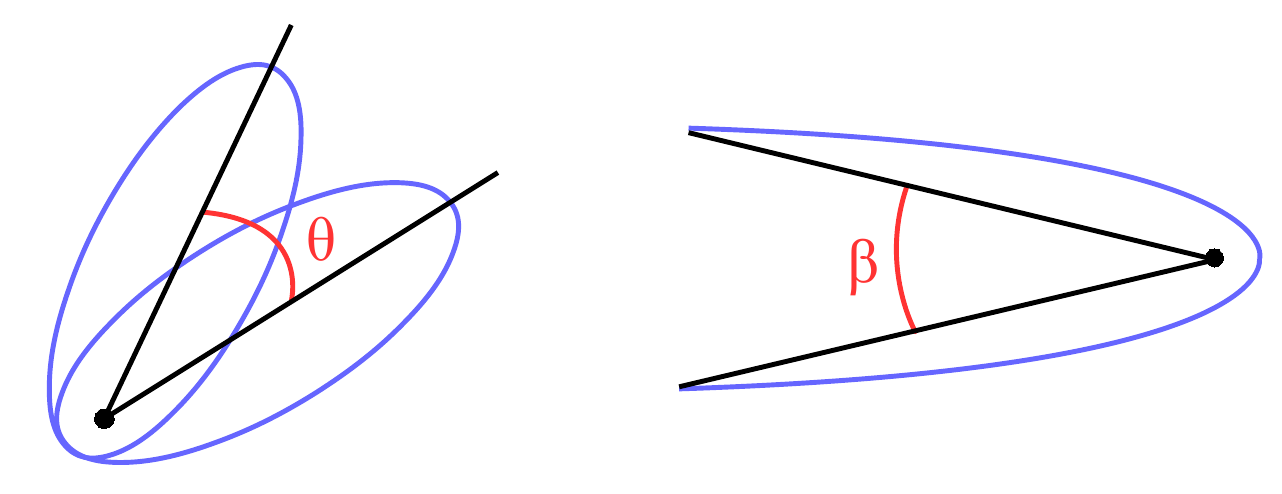}
       \caption{Angles measured to compare bound (left) or unbound (right) orbits in both Kerr and boson-star metrics. The black dot denotes the geometrical center of the metrics.}
       \label{fig:LNLO}
\end{figure} 

Contrary to the previous subsection, we generate the initial coordinates of the star in the Kerr metric by fixing its angular momentum (and its radial initial position) instead of its energy, because we want to investigate the influence of $l_\mathrm{BH}$ on the difference between orbits computed in each metric. Note that the energy $\varepsilon_\mathrm{BH}$ is obtained by solving the equation~\eref{eq:LBH}. We also mention that in this subsection we only consider stars initialized at apocenter.

To compare the orbits of the star obtained in Kerr and boson-star metrics, we use the relativistic precession of the orbit whose measure corresponds to the angle between two consecutive apocenter passages, and which is directly linked to the deflection angle of the star (see the left illustration of Fig.~\ref{fig:LNLO}). The relativistic shift takes into account the pericenter advance effect and the Lense-Thirring effect corresponding to two relevant precessions used to test general relativity. 

\begin{figure}[t]
\centering
       \includegraphics[scale=0.45]{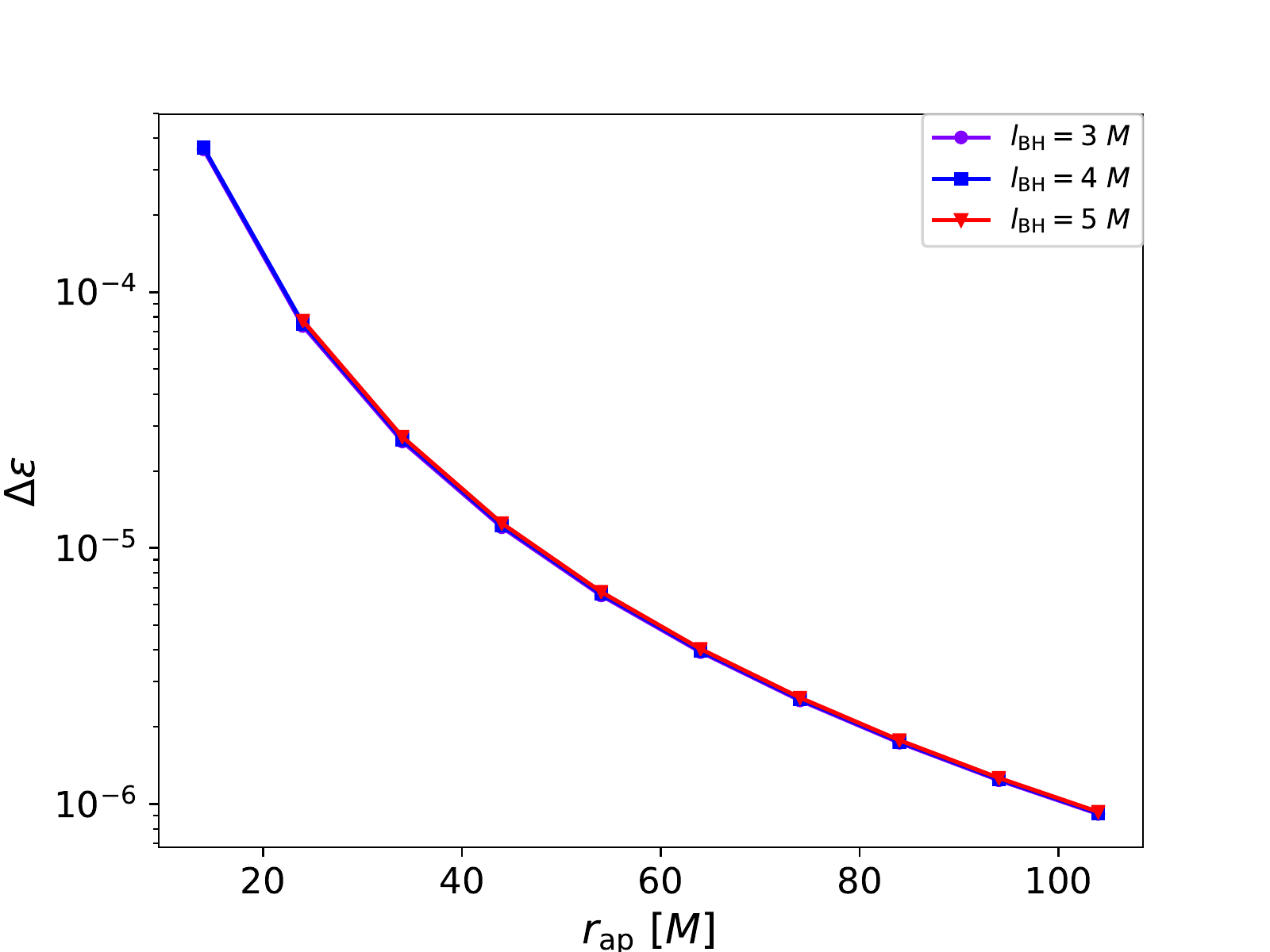}
       \quad
       \includegraphics[scale=0.45]{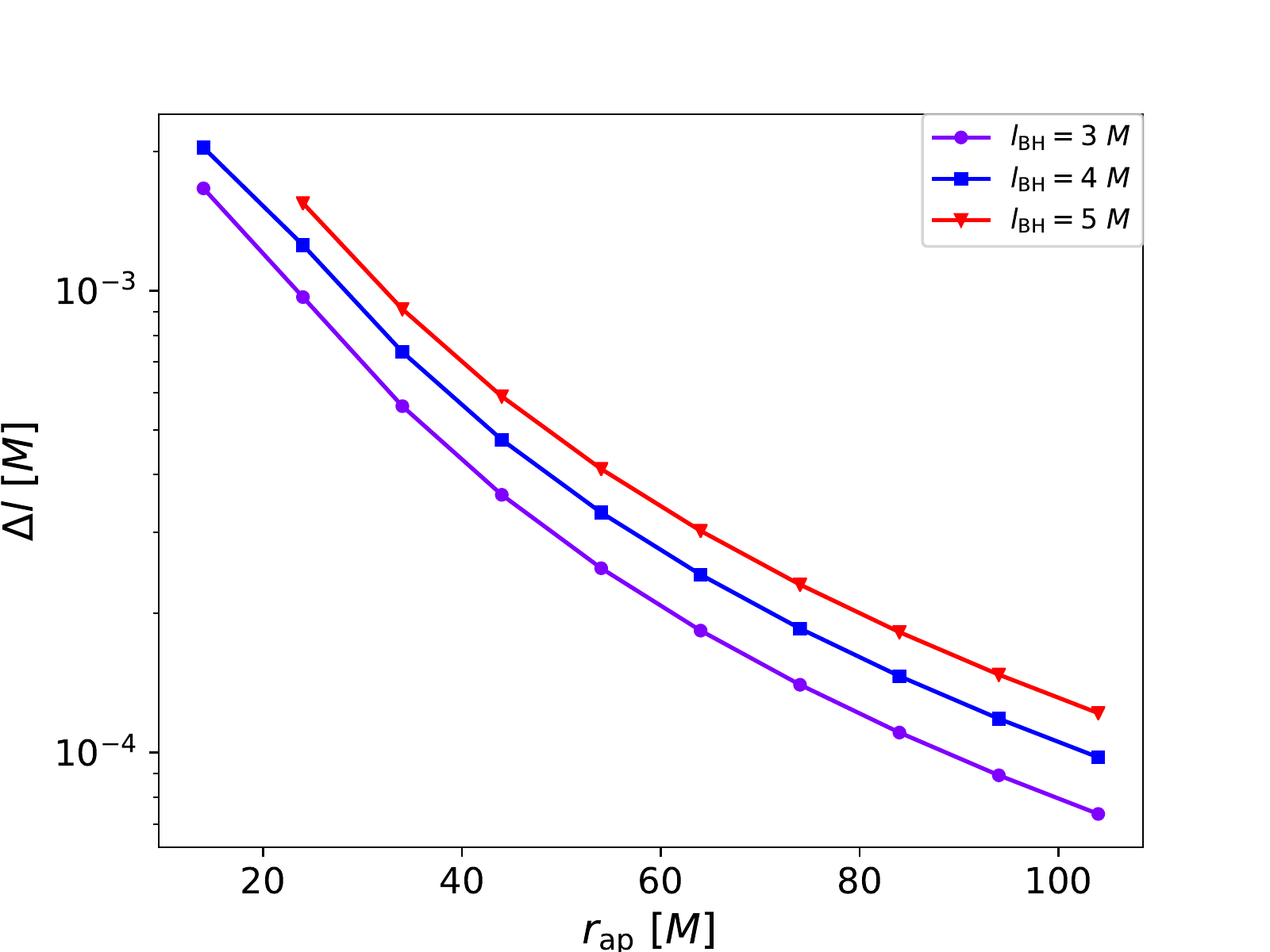}
       \quad
       \includegraphics[scale=0.45]{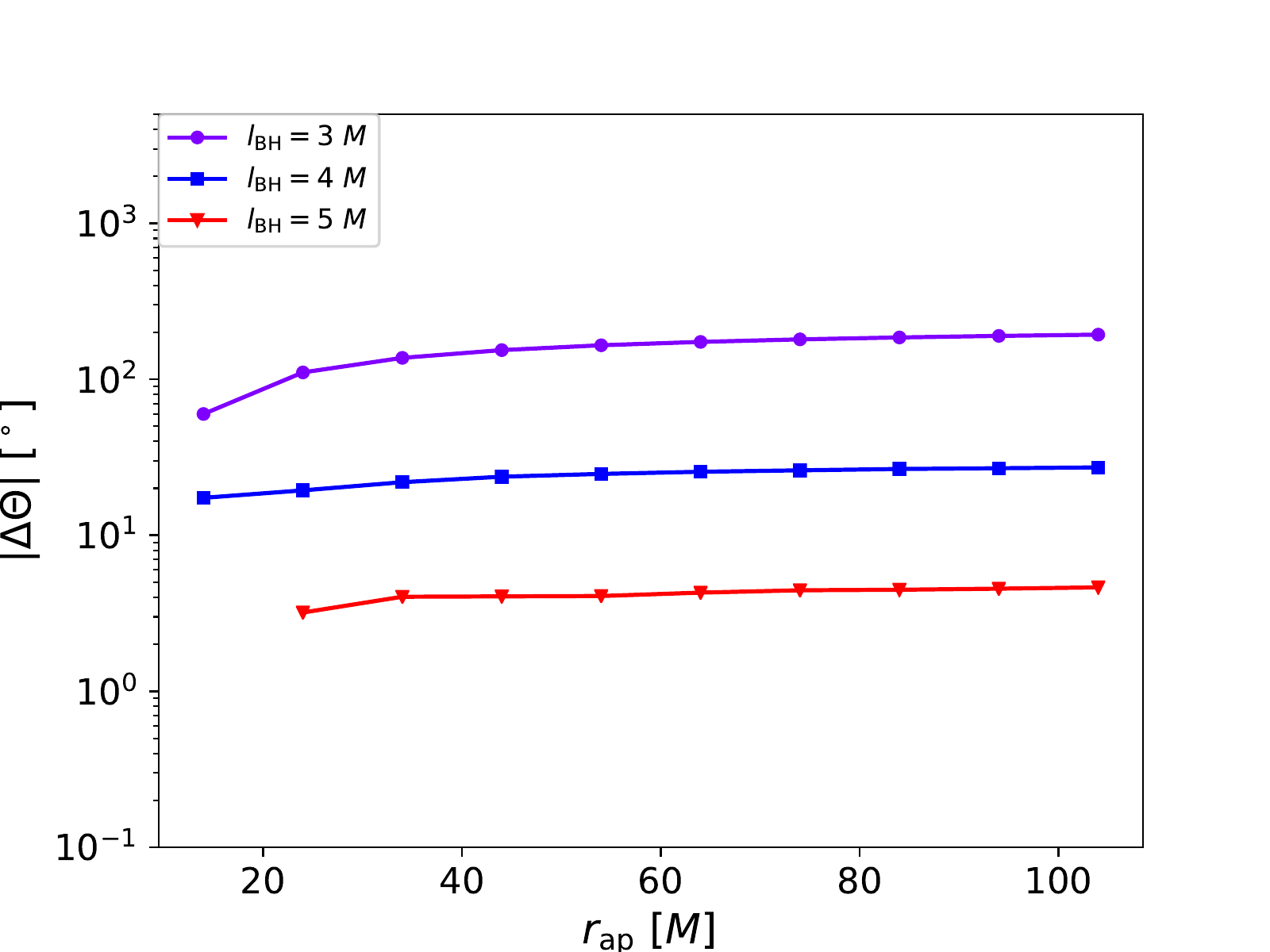}
       \caption{Comparison between timelike geodesics computed in the Kerr metric with $a=0.802$ and the boson-star metric with $k=1$ and $\omega=0.8~m/\hbar$ ($a=0.802$). \textit{Upper left:} energy differences. \textit{Upper right:} angular momentum differences. \textit{Lower:} relativistic shift differences. The types of curves denote the different angular momentums of the star considered to generate its initial coordinates in the Kerr metric. All curves are plotted versus the initial radial coordinate of the star (at apocenter) taken in both metrics.}
       \label{fig:LOK1W08}
\end{figure} 

\begin{figure}[t]
\centering
       \includegraphics[scale=0.47]{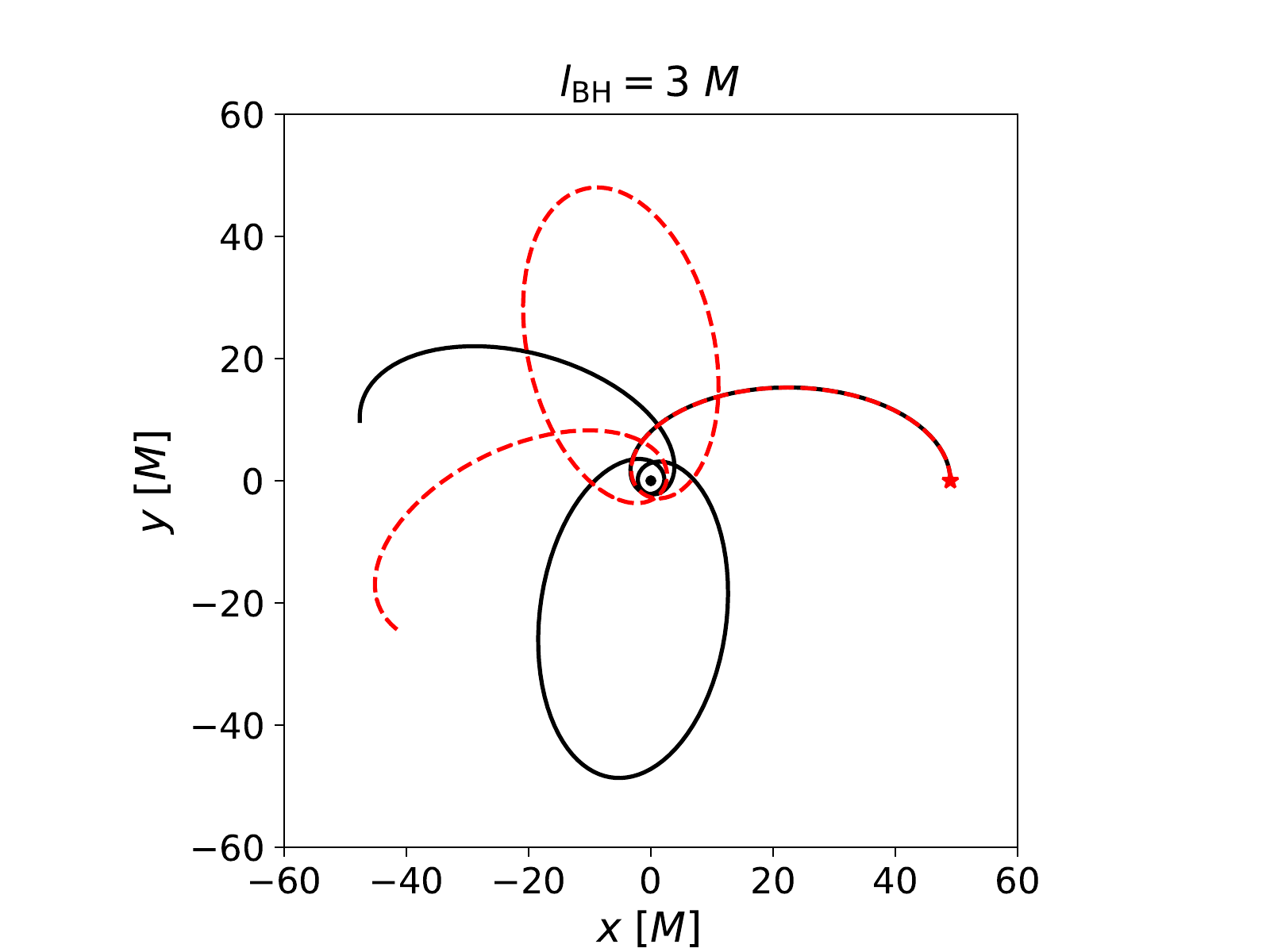}
       \includegraphics[scale=0.47]{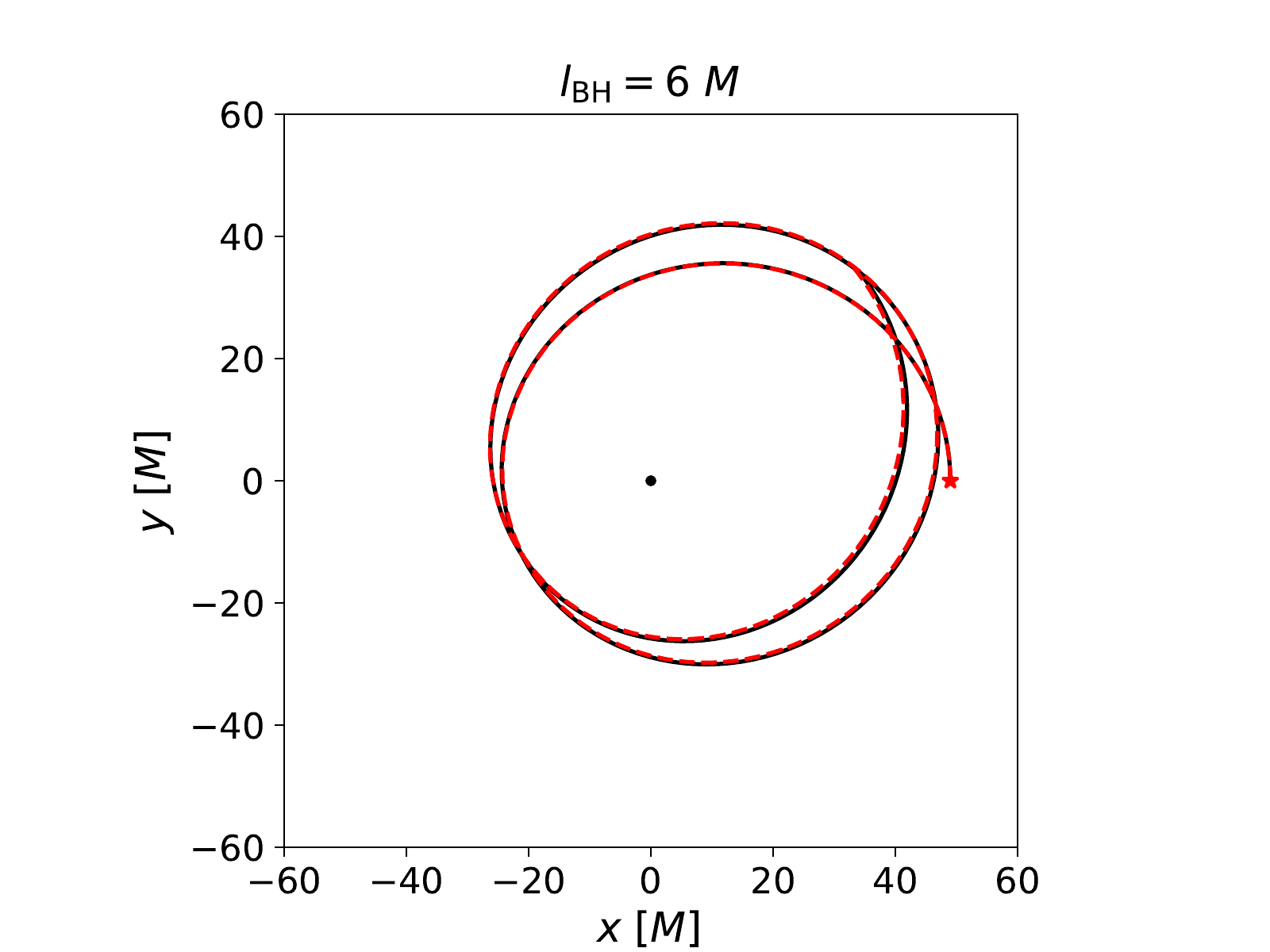}
       \caption{Orbits of stars obtained in the Kerr metric with $a=0.802$ (black solid lines) and the boson-star metric with $k=1$, $\omega=0.8~m/\hbar$ ($a=0.802$, dash red lines). The initial position of the star in both metrics is at $r=50~M$, marked by the red star. The initial four-velocity of the star on the left and right panels is generated considering two different $l_\mathrm{BH}$.}
       \label{fig:LOK1W08_2}
\end{figure}

Fig.~\ref{fig:LOK1W08} presents the comparison between parameters of the orbit (relativistic shift, energy and angular momentum) evaluated with rotating black hole and boson star at $a~=~0.8$, versus the initial radial position of the star corresponding to the apocenter. The two upper plots of Fig.~\ref{fig:LOK1W08} show that for all apocenters, the angular momentum and the energy of the star in both metrics are nearly similar. However, we can see on the lower plot of Fig.~\ref{fig:LOK1W08} that the orbits are very different since the difference between relativistic shifts can be very high. This appears in particular for small angular momentums $l_\mathrm{BH}$ where we find for instance $\Delta \Theta~\approx~200^\circ$ at $r_\mathrm{ap}~=~100~M$ and $l_\mathrm{BH}~=~3~M$. 
For higher angular momentums, the difference $\Delta \Theta$ decreases and tends to zero meaning that the orbits in both metrics become similar. These different behaviors are explained by the fact that the pericenter of the star increases when the angular momentum $l_\mathrm{BH}$ increases. Thus, when $l_\mathrm{BH}$ is small the star is more affected by the strong gravitational field generated by the compact object, which modifies its trajectory differently in the two spacetimes. This results in orbits shifted differently. Fig.~\ref{fig:LOK1W08_2} shows the orbits computed in each metric considering two initial four-velocities for the star and obtained by using two different angular momentums $l_\mathrm{BH}$. For both initial coordinates, the star is initialized at $50~M$ from the compact object. For $l_\mathrm{BH} = 3~M$, the orbits are similar for the first dates but differ when orbiting in the vicinity of the object. In this case, the minimal distance of the star from the latter is of about $2~M$. We note also that the evolution of the relativistic shift is different: it is prograde in the boson-star metric and retrograde in the Kerr metric. This is due to the fact that the deflection angle of the star near the black hole is much higher ($\approx~260^\circ$) than with the boson star ($\approx~100^\circ$). For $l_\mathrm{BH} = 6~M$, the orbits are very close and their minimal radial distances to the compact object is of about $24~M$.

\begin{figure}[t]
\centering
       \includegraphics[scale=0.45]{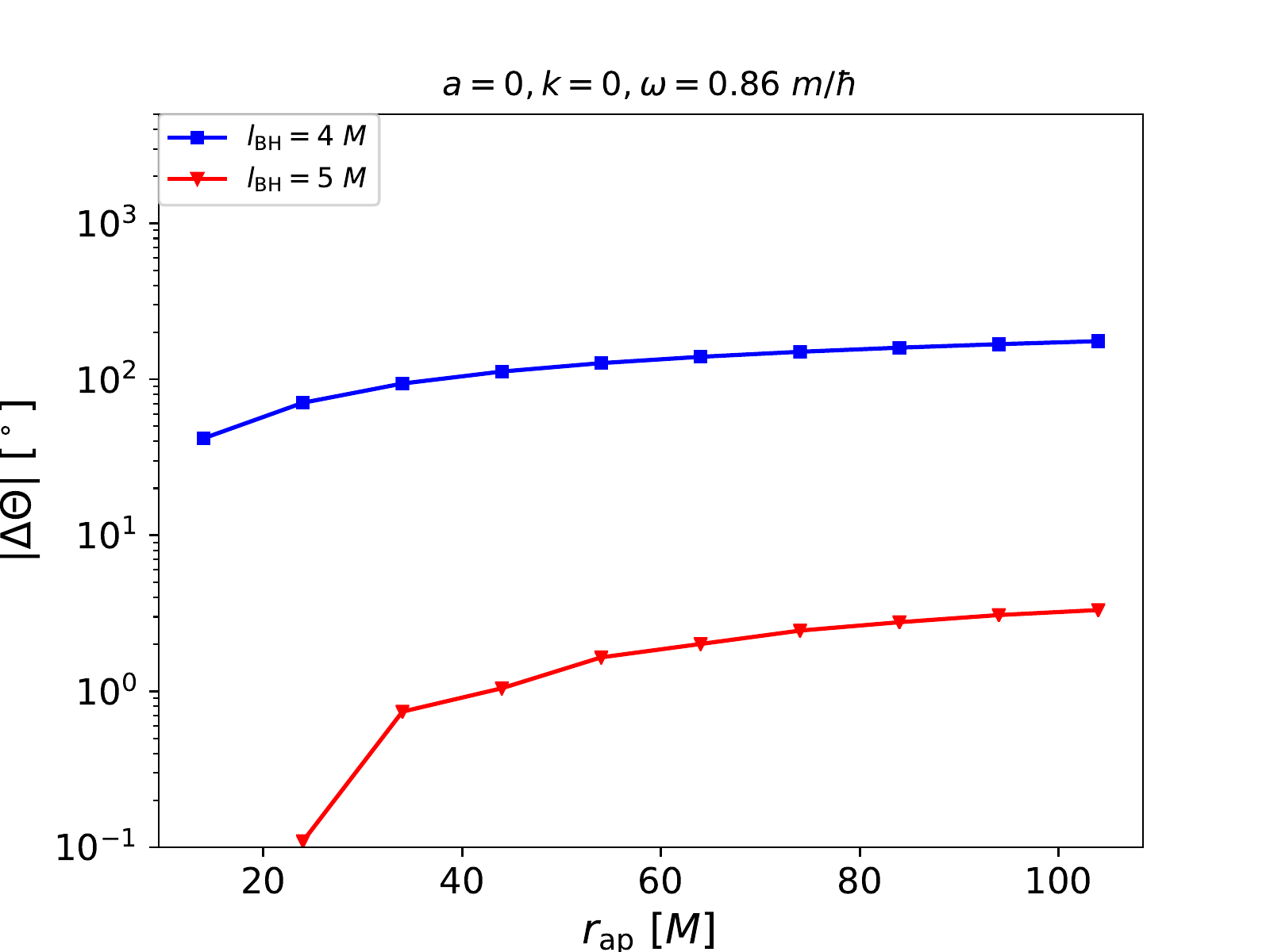}
       \quad
       \includegraphics[scale=0.45]{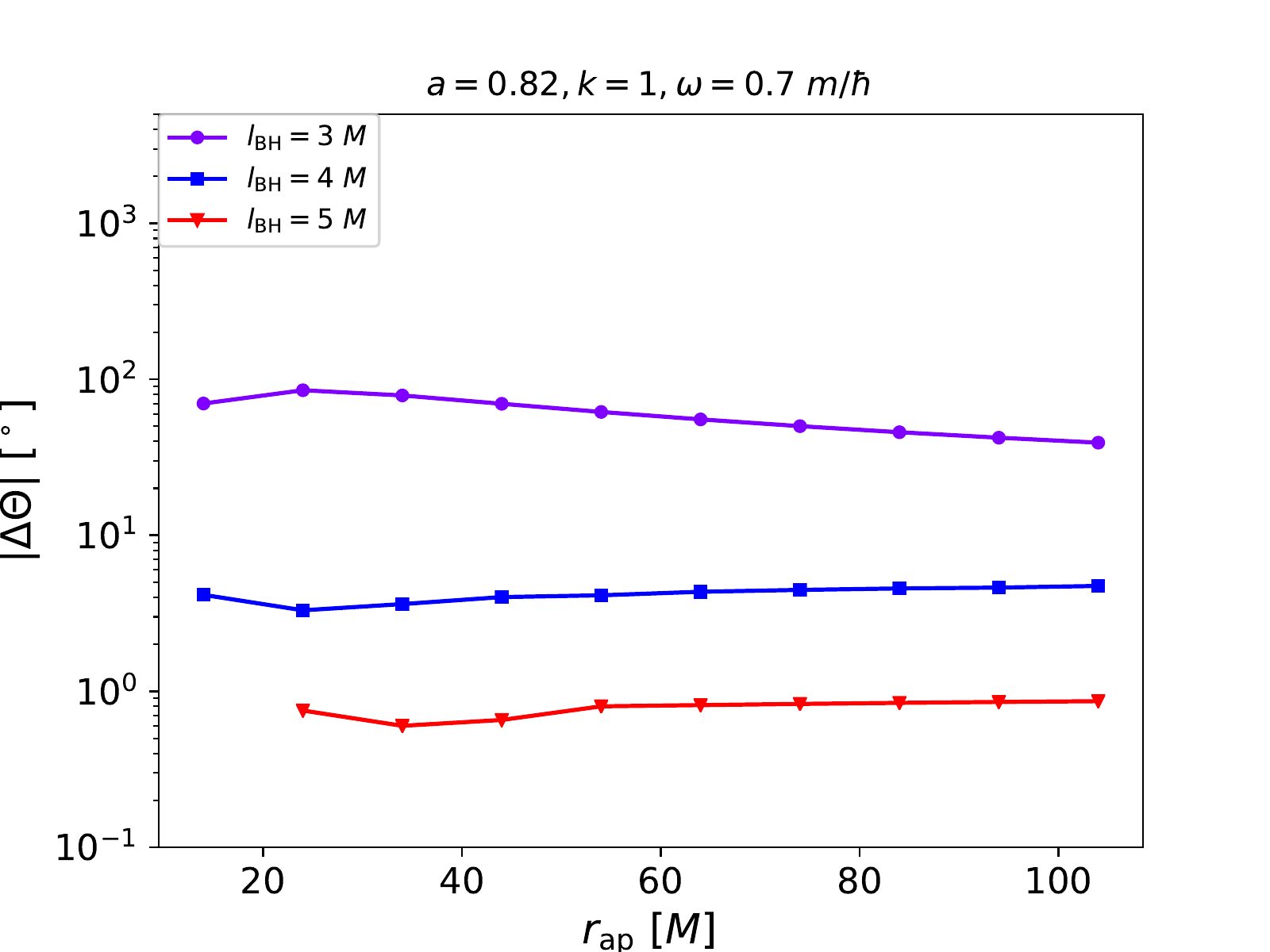}
       \quad
       \includegraphics[scale=0.45]{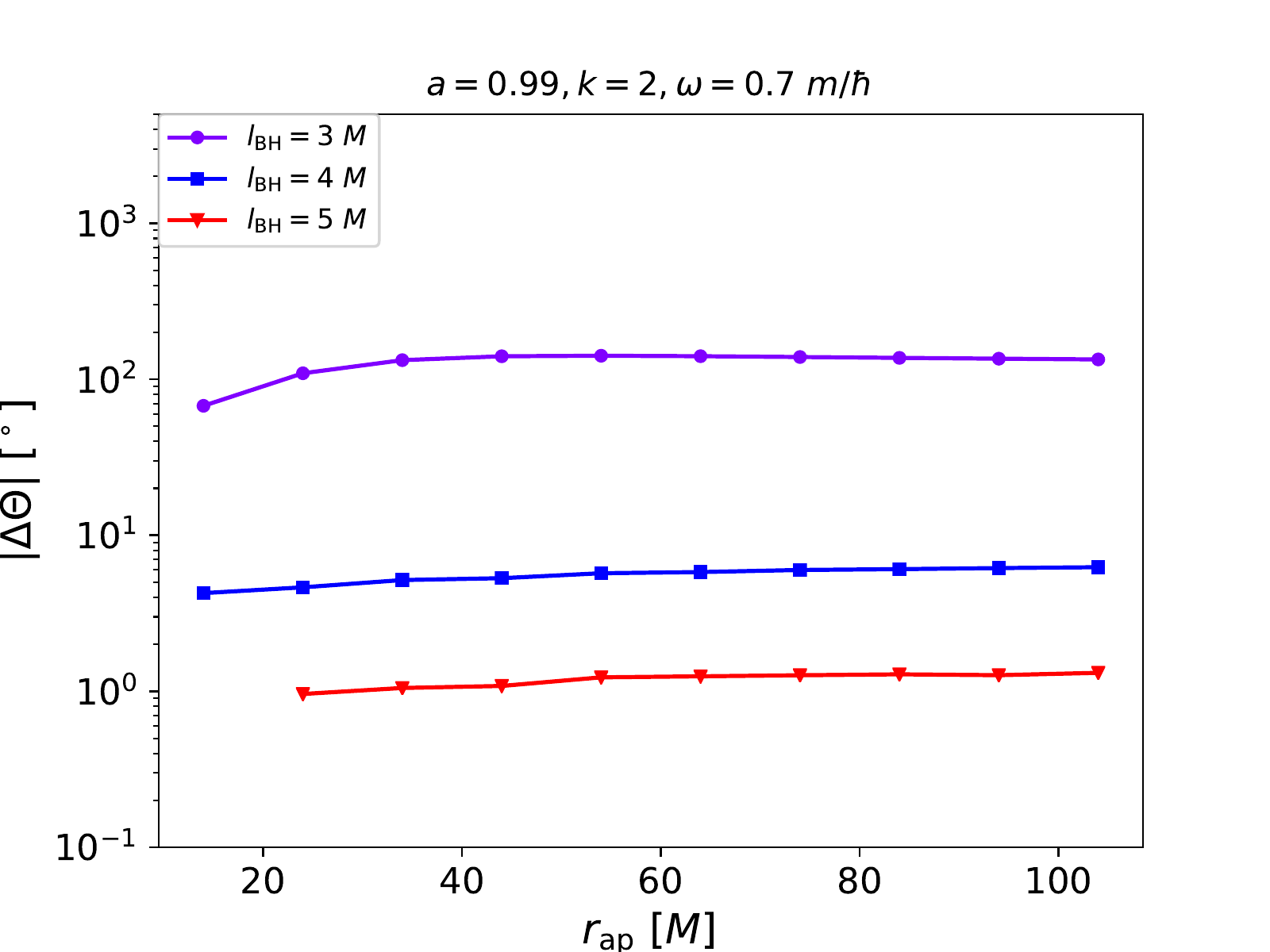}
       \caption{Comparison of relativistic shifts of orbits obtained in a black hole and a boson-star metric.}
       \label{fig:LO}
\end{figure}

Fig.~\ref{fig:LO} presents the difference between relativistic shifts of orbits obtained in various black hole and boson-star metrics. We mention that for all cases, the differences between energies or angular momentums of the star in both metrics are weak ($\Delta \varepsilon < 10^{-2}$ and $\Delta l < 0.1~M$). The upper left plot of Fig.~\ref{fig:LO} is obtained for non-rotating compact objects. The relativistic shift is thus only due to the pericenter advance. The case at $l_\mathrm{BH}~=~3~M$ is not presented on this plot since the star falls into the black hole for such angular momentum (the minimal value of $l_\mathrm{BH}$ allowed to get an orbit in this spacetime is of about $3.5~M$). On this plot, we note that the difference $\Delta \Theta$ at $l_\mathrm{BH}~=~4~M$ is as high as the one obtained for rotating compact objects at $l_\mathrm{BH}~=~3~M$ (e.g. at $a=0.802$, see the lower plot of Fig.~\ref{fig:LOK1W08}). In other words, this means that $\Delta \Theta$ can be high for larger pericenter when considering non-rotating objects. In particular, we find $\Delta \Theta \approx 170^\circ$ for both orbits with $r_\mathrm{pe} = 2~M$ in metrics where $a=0.802$ and orbits with $r_\mathrm{pe} = 4~M$ in metrics where $a=0$. The important offset obtained in the latter case can be linked to the fact that the limit of the event horizon of a non-rotating black hole is larger than a rotating one. Thus, the strong gravitational field appears for larger distances from the geometrical center, leading to orbits strongly deviated from those found in the boson-star metric.

The two other plots of Fig.~\ref{fig:LO} are obtained considering a boson star with $\omega~=~0.7~m/\hbar$ and two different azimuthal numbers. In both cases, we find high differences when decreasing the angular momentum of the star. Even if the boson star tends to an object as compact as a black hole, the timelike geodesics can highly differ. 

\begin{figure}[t]
\centering
       \includegraphics[scale=0.45]{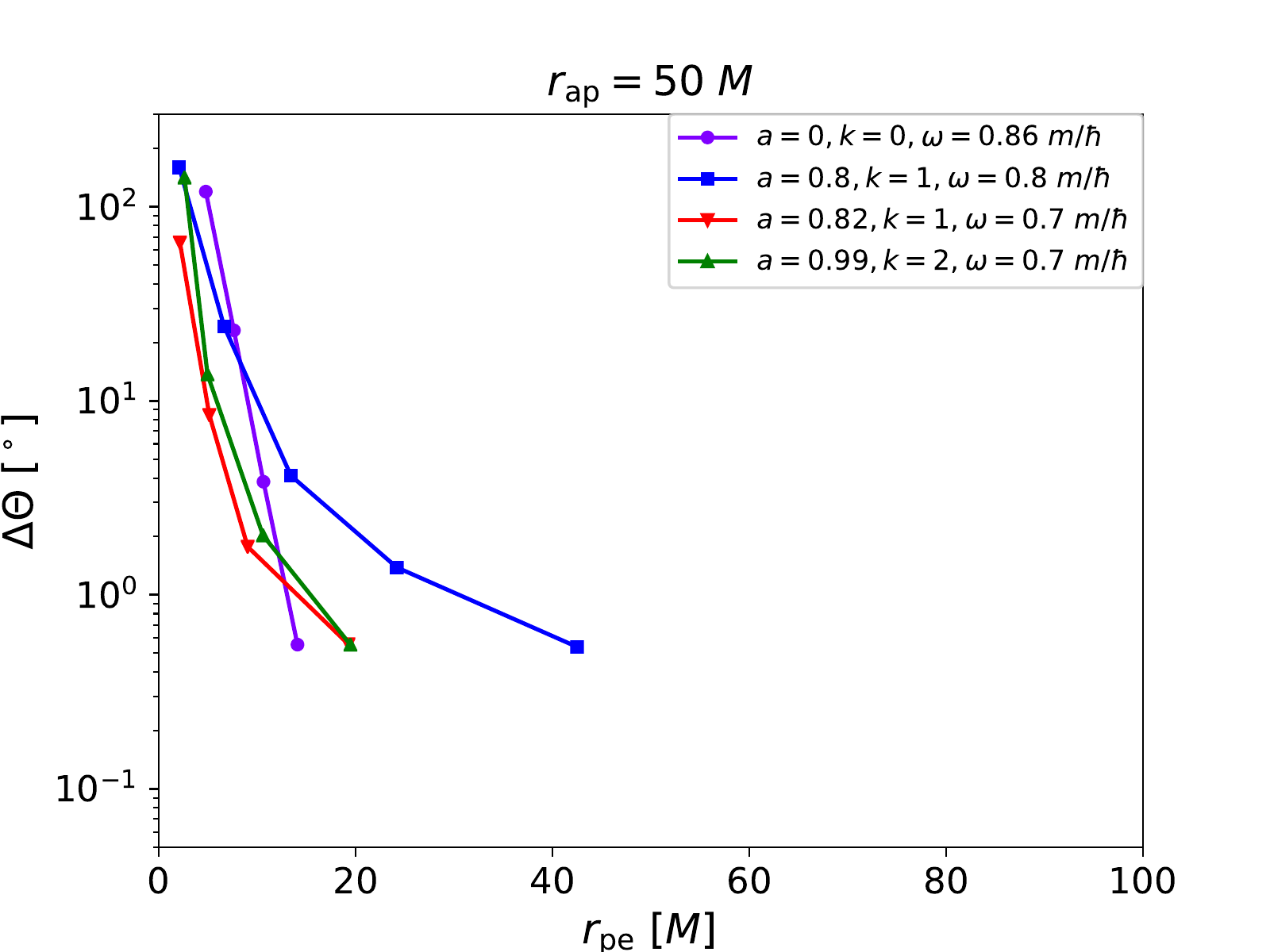}
       \includegraphics[scale=0.45]{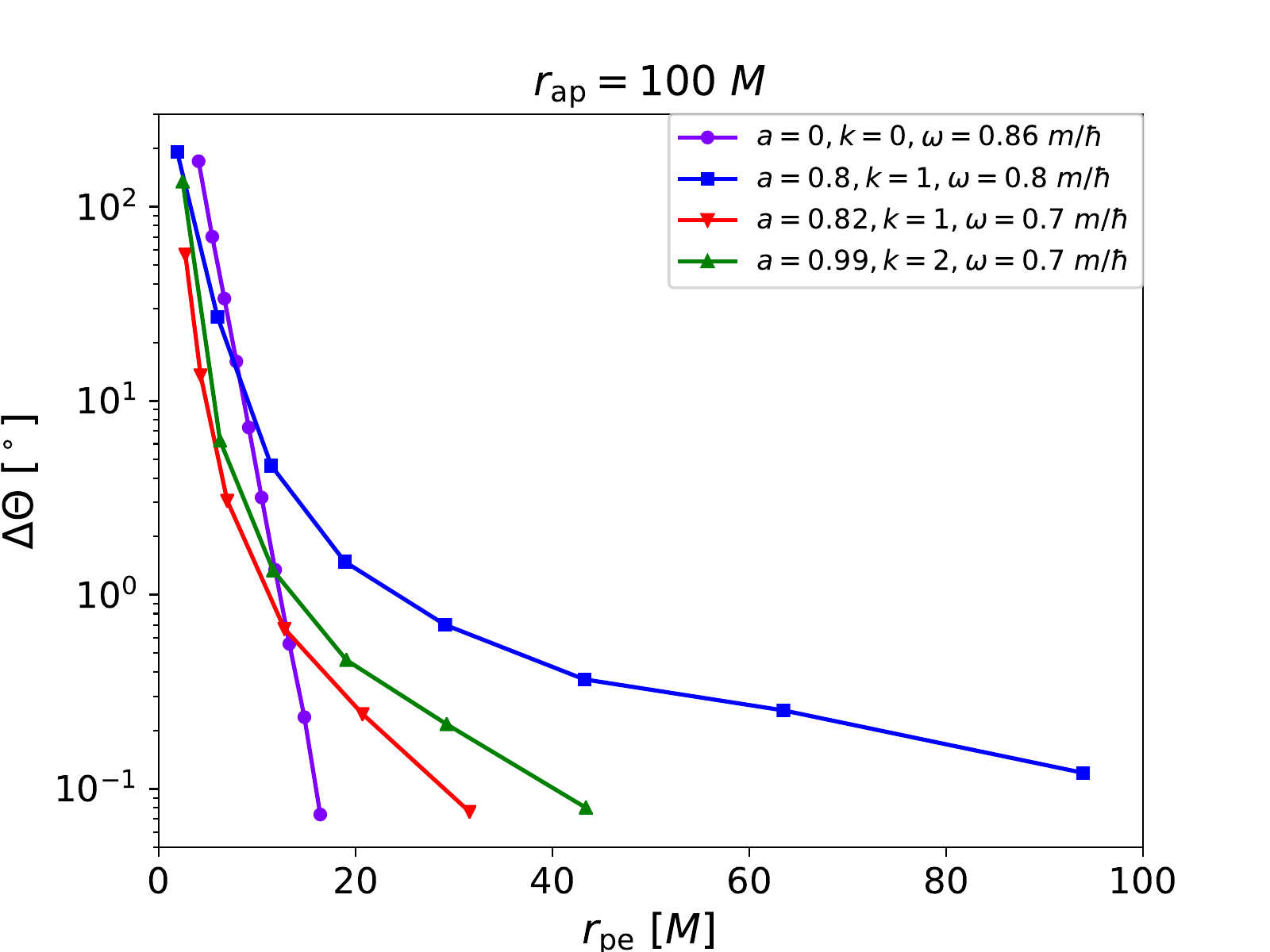}
      \caption{Comparison of relativistic shifts of orbits obtained in a black hole and a boson-star metrics, and considering a star initialized at two different apocenters.} 
       \label{fig:LO2}
\end{figure}

All the results presented here show that even considering a star initialized very far from the compact object, the trajectories computed in both metrics are different and can propagate in opposite sense (see the left plot of Fig.~\ref{fig:LOK1W08_2}). We showed that this behavior is directly linked to the pericenter distance, we therefore decide to now focus on the minimal distance of the star from the compact objects allowing to recover a negligible difference between relativistic shifts. 
For doing so, we measure the offset $\Delta \Theta$ for various angular momentums $l_\mathrm{BH}$ and considering a star initialized at two different apocenters: $r_\mathrm{ap}=50~M$ and $r_\mathrm{ap}=100~M$. By varying $l_\mathrm{BH}$, the pericenter of the star in the black hole metric will vary. Fig.~\ref{fig:LO2} shows $\Delta \Theta$ with respect to the pericenter. We specify that the pericenters plotted on this figure are those obtained in the black hole metric, however, they are nearly similar in the boson-star metric.

\begin{figure}[t]
\centering
       \includegraphics[scale=0.47]{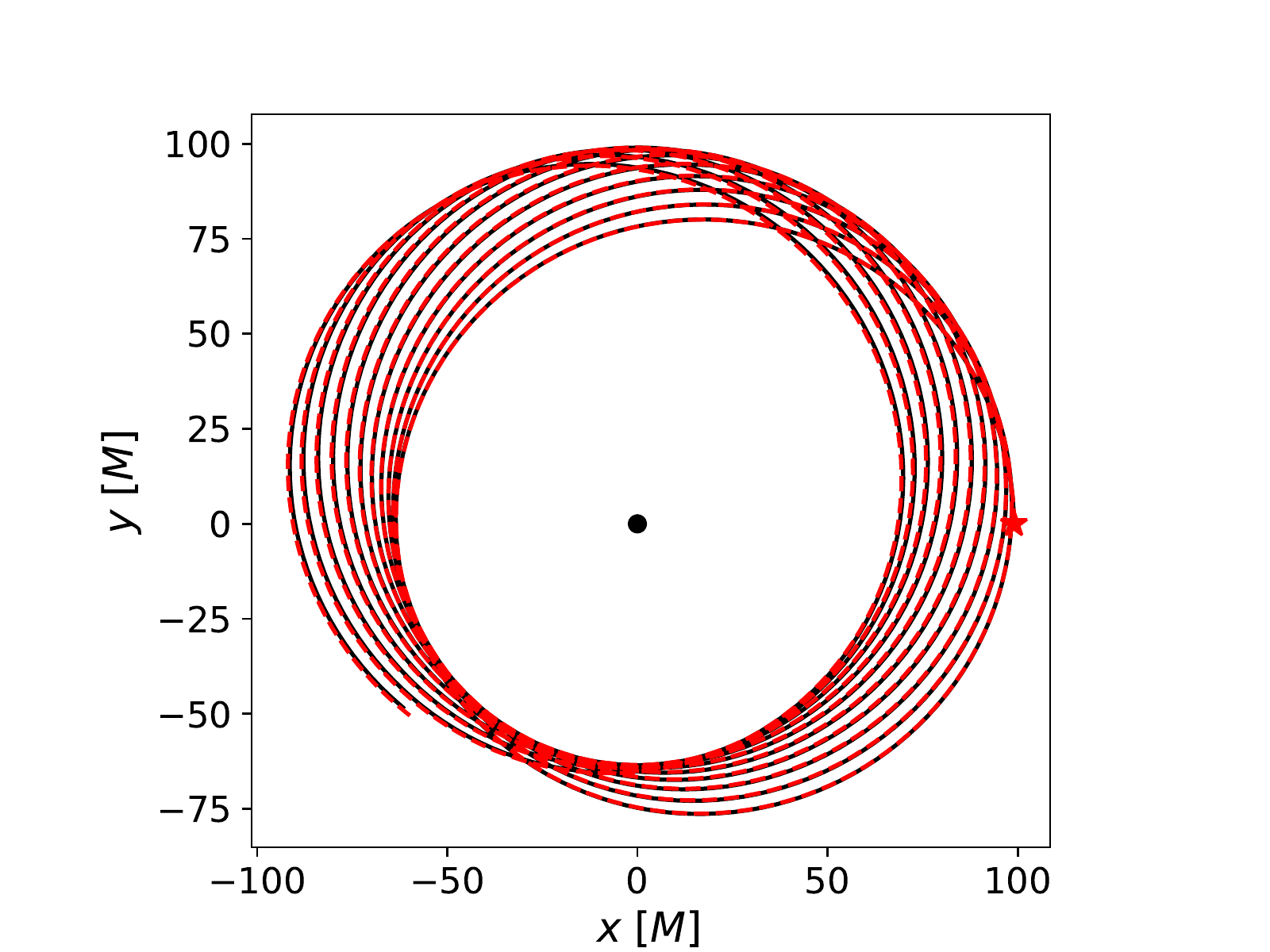}
       \includegraphics[scale=0.48]{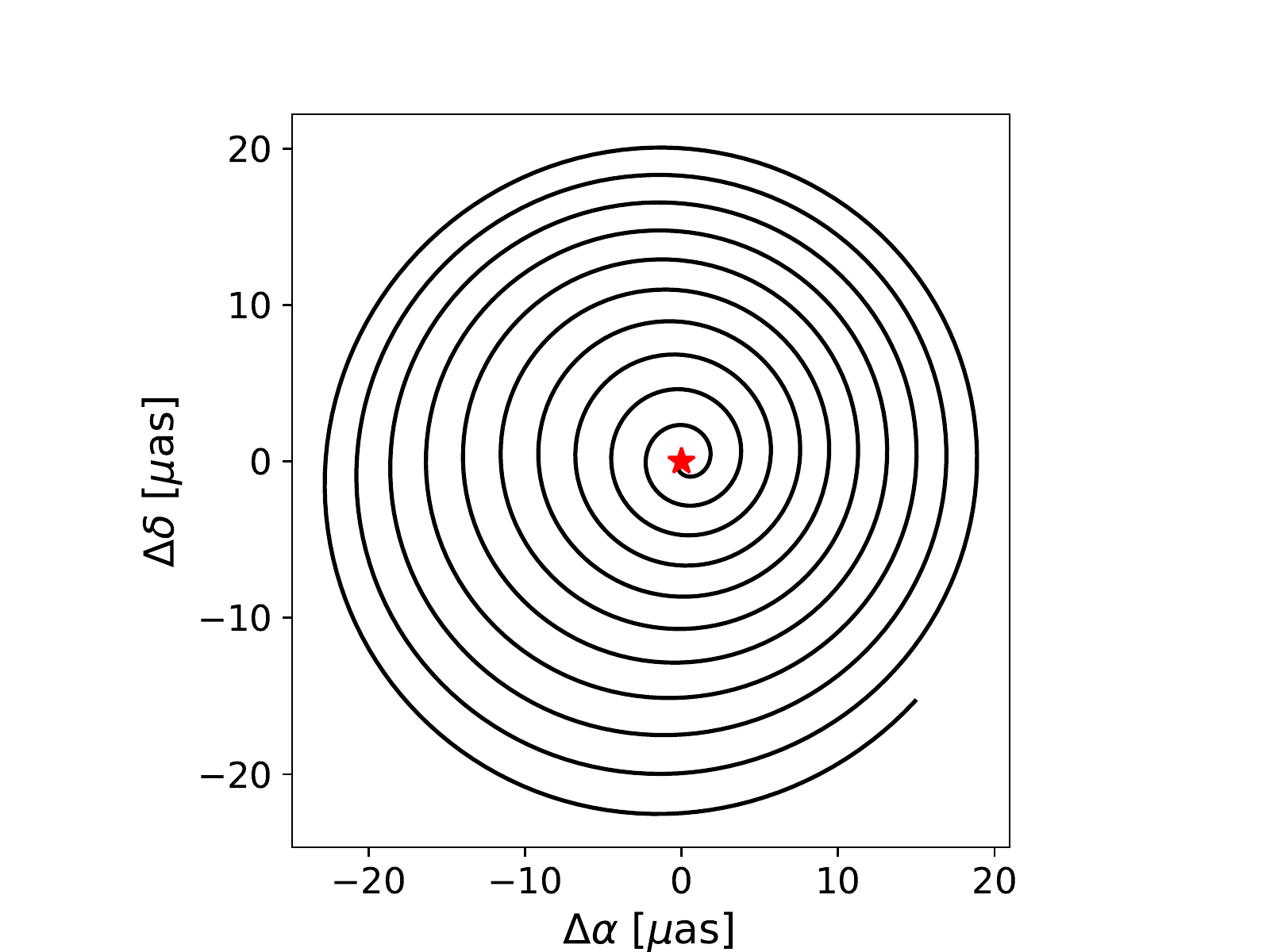}
      \caption{Comparison of orbits obtained in the Kerr and boson-star metrics with $a=0.802$, and a star with a pericenter at 60 $M$ and an apocenter at $100~M$. \textit{Left:} orbits obtained with a black hole (black solid line) and a boson star (dash red line). The black dot denotes the geometrical center of both metrics. \textit{Right:} difference of the apparent positions of the star computed in each metric when observed face on. The apparent positions are obtained by neglecting the relativistic effects on the photon trajectory, only the star trajectory is relativistic. The red star denotes the starting point.}
       \label{fig:DiffOrbit}
\end{figure}

First, we note that the global behavior of each curve obtained for orbits with an apocenter at $r_\mathrm{ap}=50~M$ are close to those obtained with an apocenter at $r_\mathrm{ap}=100~M$. Considering the case at $a=0$, we note a rapid decrease of $\Delta \Theta$ when increasing the pericenter of the star. This is obvious since non-rotating boson-star metrics satisfy the Birkhoff theorem as non-rotating black hole metrics. More precisely, the scalar field of the boson star decreases exponentially, the metric thus tends more rapidly to the non-rotating black hole metric (e.g. for $\omega=0.86~m/\hbar$ the scalar field modulus varies between $0.07$ and $10^{-3}$ from $r=0$ to $r=15~M$, respectively). For rotating compact objects the offset decreases less rapidly, in particular for $a = 0.8$. For both apocenters considered, small $\Delta \Theta$ inferior or equal to $1^{\circ}$ are obtained for $r_\mathrm{pe} \gtrsim 10 - 30 ~M$ (depending on the spin of the compact objects). To get an idea of the observed offset on Earth, we can compute the quantity $r_\mathrm{ap} \Delta \Theta / R_0$ where $R_0$ is the distance of the Earth from the Galactic center equals to 8 kpc. In such a case the plane of the orbit is observed face on, thus, the observed offset is maximal. We find that the offset is superior or equal to $10~\mu$as, corresponding to the astrometric accuracy of the GRAVITY instrument \citep{2003SPIE.4841.1548E}, for $r_\mathrm{pe} \lesssim 10 - 30~M$ and both apocenters. Thus, differences between the Kerr and boson-star metrics can be measured by GRAVITY for stars with pericenters verifying $\lesssim 10 - 30~M$, and the boson stars considered. Note that in the current study we consider apocenters satisfying $r_\mathrm{ap} \leqslant 100~M$, however, higher apocenters also allow to get important differences between orbits. Indeed, as it is visible on the lower plot of Fig.~\ref{fig:LOK1W08}, and Fig.~\ref{fig:LOK1W08_2}, high offsets reach a plateau meaning that even orbits with an apocenter larger than 100 $M$ can get a pericenter sufficiently close to the compact objects to observe strong deviations between both metrics. We also point out that the relativistic shift of the orbits is obtained by considering the two first orbital periods, by increasing the observation time the difference between both orbits will increase. It can be seen on Fig.~\ref{fig:DiffOrbit} where the pericenter considered is 60 $M$ and the maximal astrometric shift reached is $\approx 30~\mu$as after twelve days of monitoring. Thus, larger pericenters will also allow to observe a non-negligible difference between orbits when considering several orbital periods. Note that even considering an observer edge on, high differences can be highlighted. This can be illustrated on the right plot of Fig.~\ref{fig:DiffOrbit} where the astrometric offset in such configuration will be $\Delta \delta = 0$, $\Delta \alpha = [-15, 32]~\mu$as.

We mention that in the case of the current known closest star to the Galactic center, S2, the pericenter reached is around 3000 $M$. Measurements of deviations from the Kerr metric seem obviously impossible with observations of this star obtained by GRAVITY. In particular, if we consider a star with a periastre ten times closer to the compact object than S2, we find a maximal astrometric difference $< 10~\mu$as over seven orbital periods, corresponding to three years of observation. 

\subsection{Unbound orbits: $\varepsilon>1$}

\begin{figure}[t]
\centering
       \includegraphics[scale=0.45]{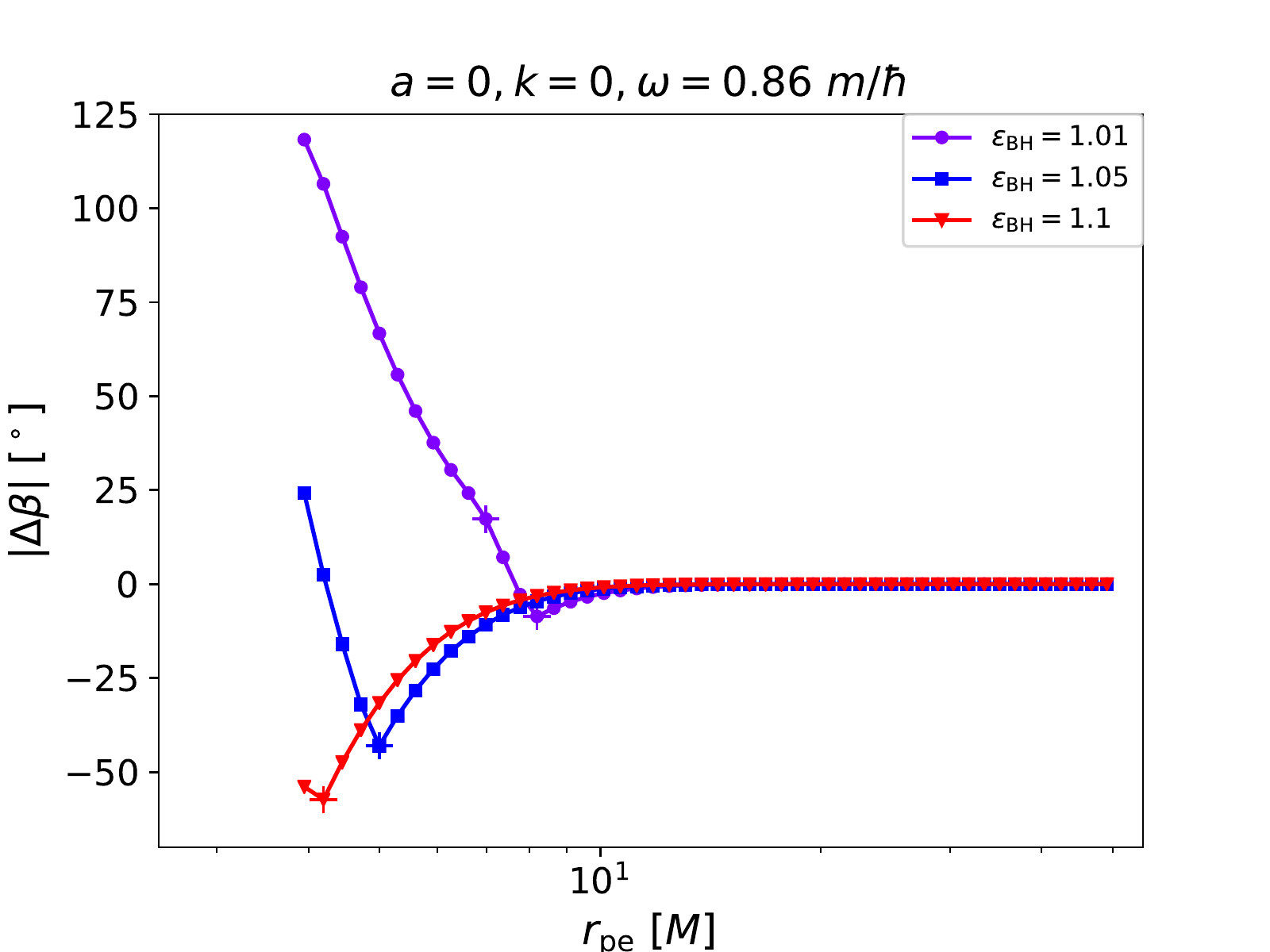}
       \quad
       \includegraphics[scale=0.45]{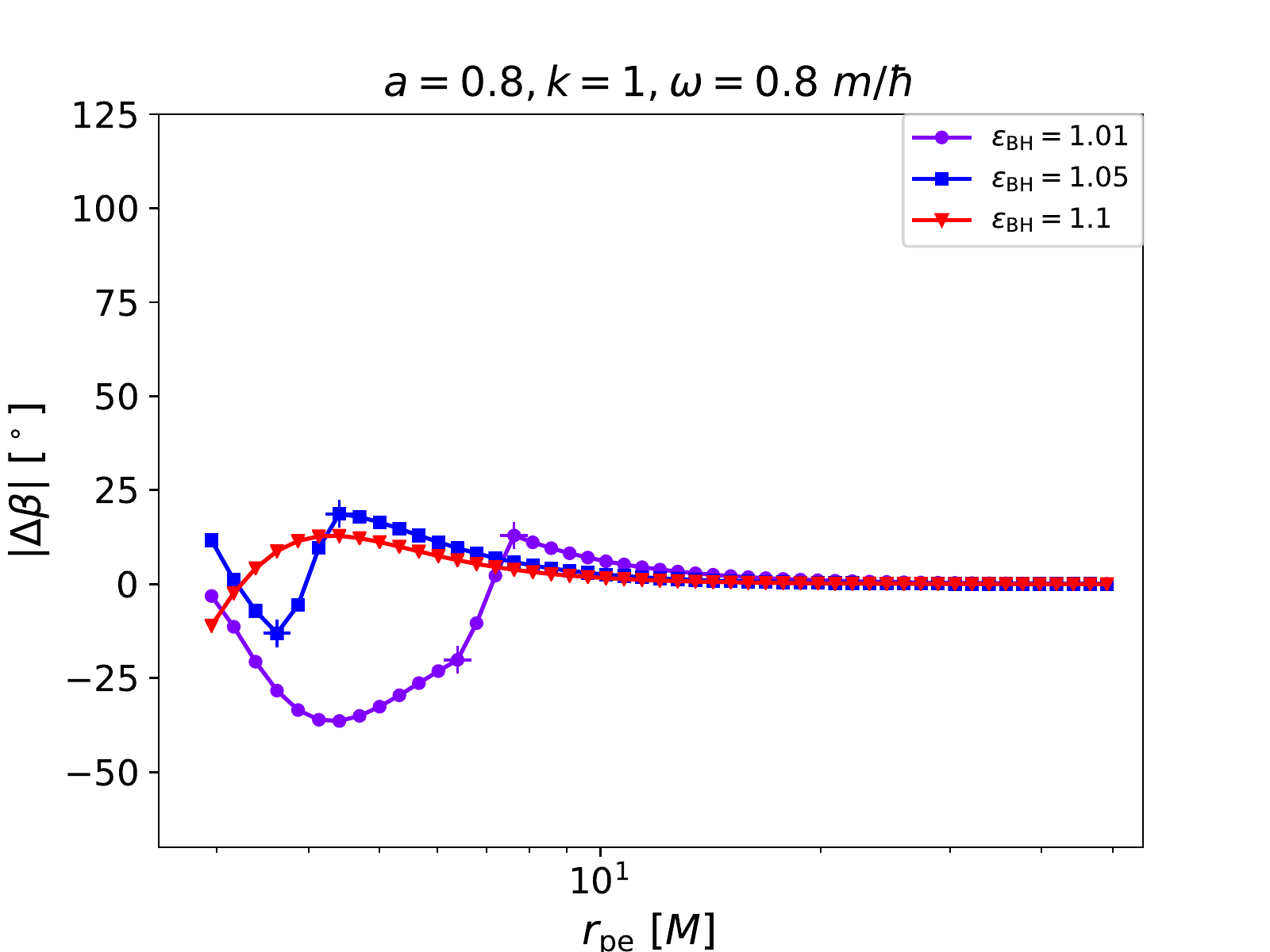}
       \quad
       \includegraphics[scale=0.45]{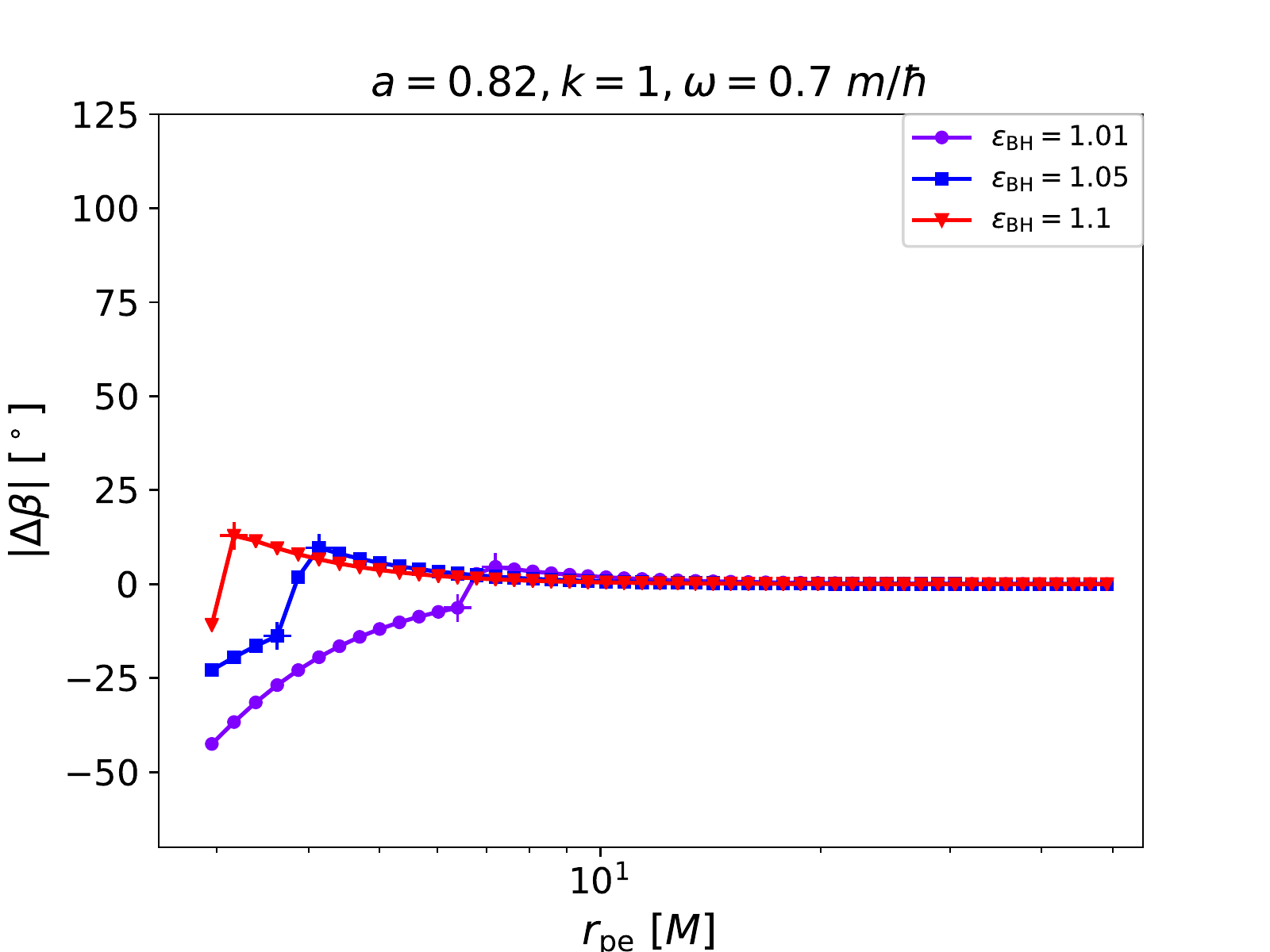}
       \quad
       \includegraphics[scale=0.45]{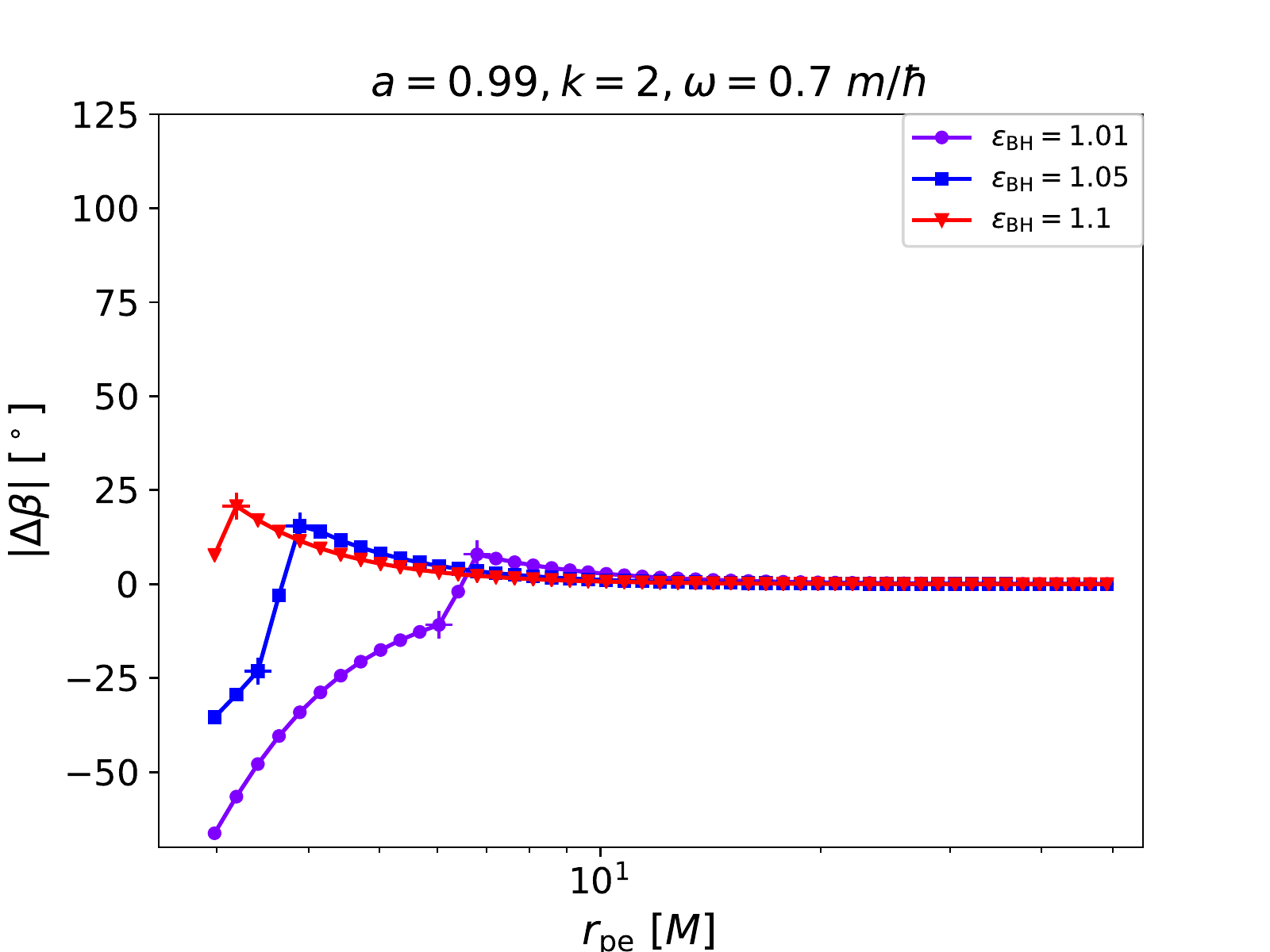}
       \caption{Comparison of aperture angles of the orbits obtained in the black hole and boson-star metrics. Four various spins for the compact objects are considered. The thin crosses on each plot delineate different curve behaviors.}
       \label{fig:NLO}
\end{figure}

\begin{figure}[t]
\centering
       \includegraphics[scale=0.4]{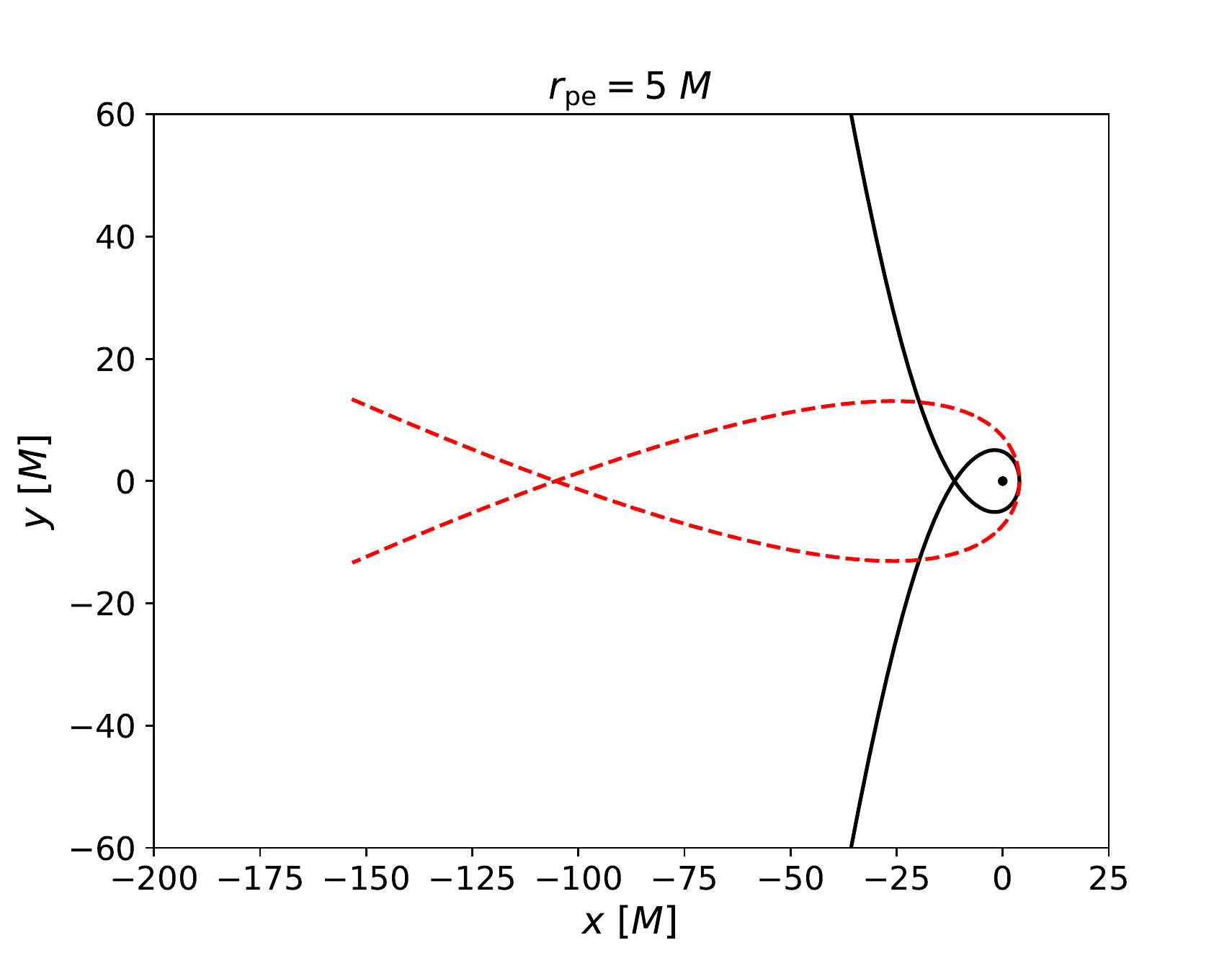}
       \quad
       \includegraphics[scale=0.4]{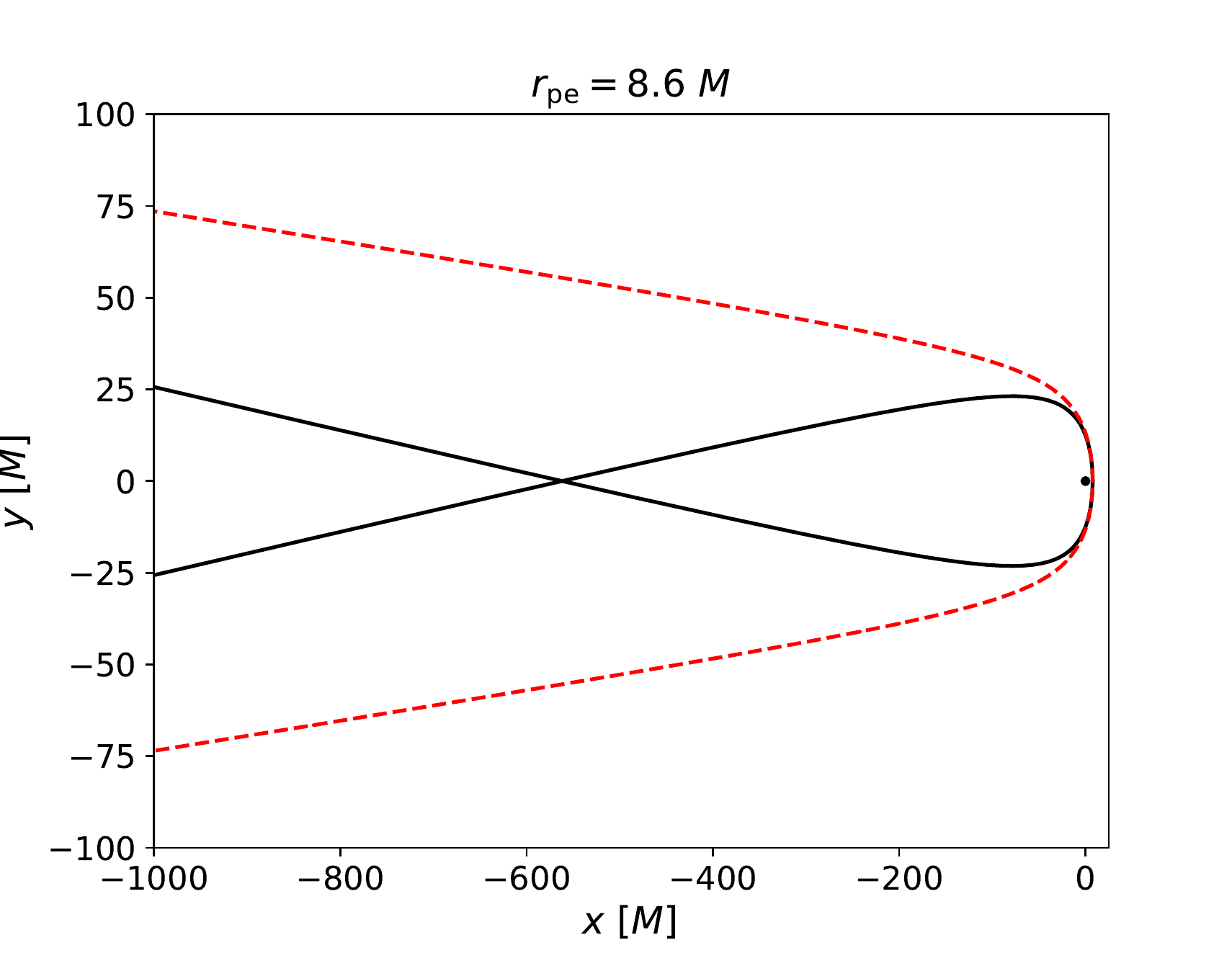}
       \includegraphics[scale=0.4]{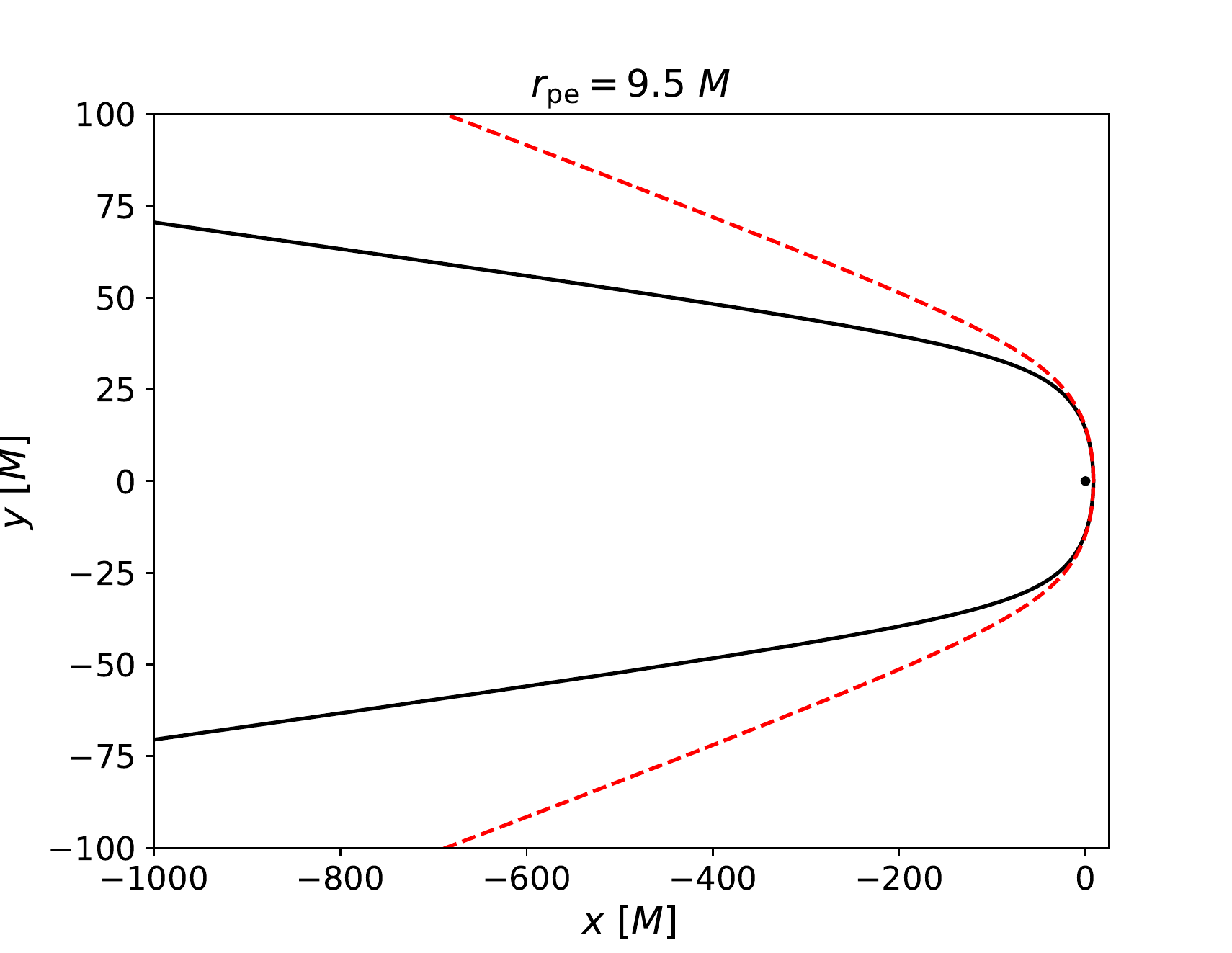}
       \caption{Unbound orbits obtained in the non-rotating black hole metric (solid black lines) and the boson-star metric with $k=0$ and $\omega=0.86$ (dash red lines). The upper left, upper right and lower panels correspond to the first, second and third regimes, respectively. The black dot denotes the geometrical center of the metrics.}
       \label{fig:NLO2}
\end{figure}

In this subsection, we focus on the unbound orbits and compare their aperture angle obtained in the black hole and boson-star metrics (see the right illustration of Fig.~\ref{fig:LNLO}). Each orbit is generated considering a star initialized at pericenter. 

Fig.~\ref{fig:NLO} shows the difference $\Delta \beta$ between aperture angles of orbits obtained in the two metrics considering both various pericenters and three energies $\varepsilon_\mathrm{BH}$ of the star to generate the initial coordinates of the star in the metrics. First, small offsets $\Delta \beta \leqslant 1^\circ$ are reached for $r_\mathrm{pe} \gtrsim 10 - 20~M$, depending on the energy of the star and the spin of the compact objects. The offset can reach a maximum of about $125^\circ$ and $70^\circ$ for non-rotating and rotating objects, respectively. Most of the curves have different regimes in which orbits have particular shapes. These regimes are delimited by thin crosses on Fig.~\ref{fig:NLO}. The number of regimes decrease when the energy of the star increases. Focusing on the case at $e_\mathrm{BH} = 1.01$, we oberve the first regime appearing at small pericenters which corresponds to the unbound orbits having the same shape in both metrics: the star passes two times by the same point (see the left plot of Fig.~\ref{fig:NLO2}). In the second regime, the orbits have different shape in each metric: one where the star still passes two times by the same point, and the other where the star never passes by the same point (see the right plot of Fig.~\ref{fig:NLO2}). In the last regime, both orbits have again the same shape but the orbits are similar to the latter (see the lower plot of Fig.~\ref{fig:NLO2}). The second regime shows that we can find different subtypes of orbits in both metrics even considering the same initial coordinates for the star. However, this regime only appears for a narrow range of pericenters.

As found for bound orbits, strong differences between orbits are observed for stars with small pericenters, typically inferior to $30~M$. By increasing the observation time it is also possible to observe non-negligible differences between orbits when considering higher pericenters. Indeed, as illustrated on the lower plot of Fig.~\ref{fig:NLO2} the difference between orbits increases when the star goes away from the compact objects. For instance, if we consider the boson star $k=1$, $\omega=0.8$, and a star initialized at $r_\mathrm{pe} = 60~M$, we find an astrometric offset, as seen from an observer face on, of about 3 $\mu$as and 40 $\mu$as with an observation time of about $10^4~M$ and $10^5~M$, respectively.

\section{Conclusions and discussions}
\label{ConclusionsDiscussions}

The study performed here puts in light important differences between orbits of stars orbiting a Kerr black hole and a boson star. In particular, sustainable closed orbits in the latter metric exist when they do not in the black hole one. For instance, this is the case for stars with zero angular momentum, small angular momentum or stars initially at rest. There are different types of orbits that we do not encounter in the vicinity of a black hole such as the semi, petal and pointy petal orbits. Stars orbiting a boson star can passe through the compact object, by the geometrical center, or very close, which obviously cannot appear in the black hole metric since the star would fall into it. Moreover, at a given initial coordinates of the star, the orbit can be unbound in the black hole metric and bound in the boson-star one. When the orbits are bound or unbound in both metrics, the difference between trajectories is still high. In particular, strong offsets appear for stars with pericenters $\lesssim 30~M$. For instance, we can observe in both metrics a relativistic shift evolving in opposite sense. Moreover, higher pericenters (e.g. at 60 $M$) also allow to observe important deviations from the Kerr metric by increasing the observation time.

This work shows that accurate astrometric observations obtained by the GRAVITY instrument could allow to distinguish a Kerr black hole from a boson star for stars with a pericenter sufficiently close to the compact object, typically $r_\mathrm{pe} < 100~M$, which naturally exclude the current closest star to the Galactic center, S2. However, further analysis need to be performed on boson stars in order to determine whether it is really possible to discriminate such compact object from a Kerr black hole by using fitting models. More precisely, the question to be investigated is: could a stellar-orbits model obtained with a Kerr black hole fail to describe GRAVITY astrometric observations of stars orbiting a boson star?

\begin{figure}[t]
\centering
       \includegraphics[scale=0.45]{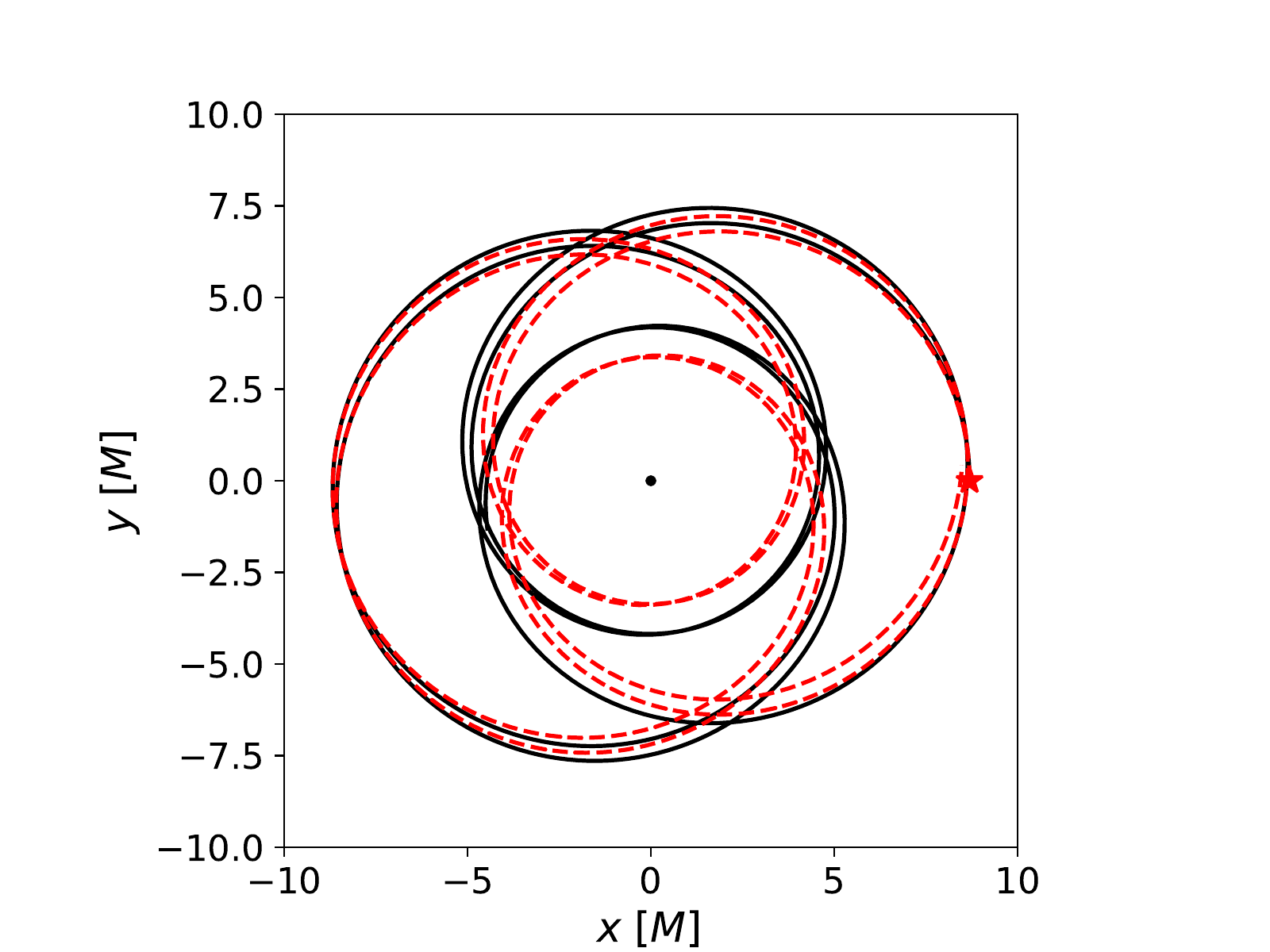}
       \includegraphics[scale=0.45]{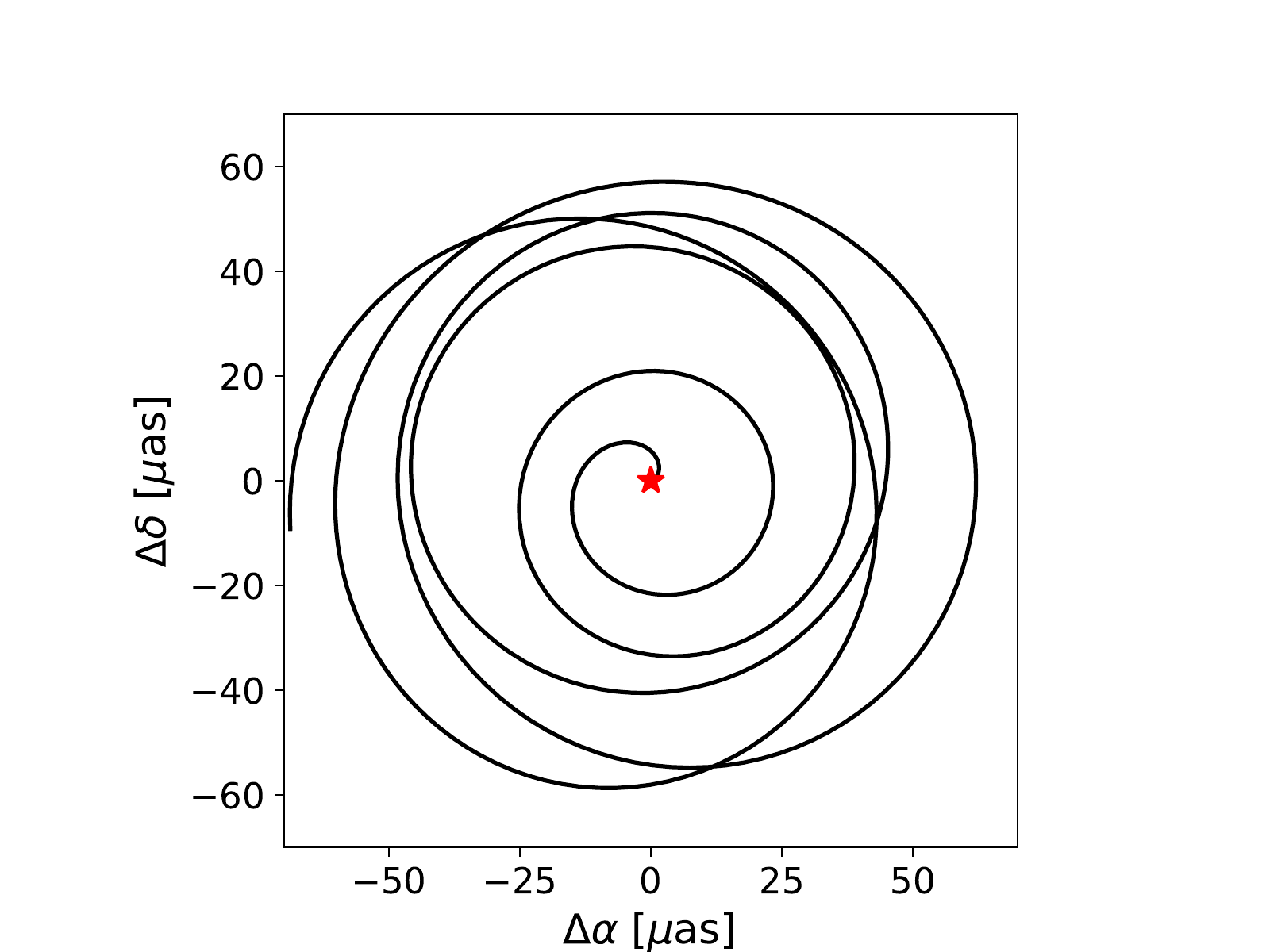}
       \caption{\textit{Left:} orbit of a star in the Kerr metric (black solid line) and the boson-star metric (red dashed line) considering an initial coordinates for the star generated considering $r = 9.7~M$ and $l_\mathrm{BH} = 3~M$. \textit{Right:} astrometric offset between orbits computed in each metrics. The red star on each plot denotes the initial position of the star and the black dot on the left panel denote the geometrical center.}
       \label{fig:Dege}
\end{figure}

Besides, we cannot reject the existence of degeneracies between orbits computed in both metrics or orbits with strong similarities. As illustrated on the left panel of Fig.~\ref{fig:Dege}, even considering a star with a pericenter and an apocenter close to the center, the relativistic shift is the same for both orbits. However, as it is shown on the right panel of Fig.~\ref{fig:Dege} the astrometric difference between orbits is sufficiently high to be detected by GRAVITY since the offset can reach $60~\mu$as. 

Even if timelike geodesics are similar in the Kerr and boson-star metrics for stars with important pericenters, it should be possible to distinguish both compact objects by using null geodesics \citep{2006MNRAS.369..485S,2013arXiv1301.1396B,2017PhRvD..95h4011G}. For instance, \cite{2013arXiv1301.1396B} showed that gravitational lensing is a powerful tool to put in light differences between both spacetimes. In particular, the authors determined that the secondary image of S stars orbiting a boson star can be much more brighter than with a black hole and should be detected by the GRAVITY instrument. Their study and the results obtained in this paper show that observations of stars close to the Galactic center obtained with this instrument will allow to highlight low- and high-order relativistic effects and thus probably bring answers on the nature of the compact source Sgr A*. We also point out that as mentioned in this paper different subtypes of boson stars exist such as those with self-interaction. It is therefore not excluded that boson stars with non-minimal coupling to gravity should be better discriminated from Kerr black holes than mini-boson stars. This is in particular discussed in \cite{2013CQGra..30i5014H} where the authors found that non-minimal coupling significantly affects the shape of null geodesics around boson stars.

In addition to degeneracies that could be encountered between a black hole and a boson star, we need to take into account the fact that alternative theories or alternative exotic objects also described by general relativity could generate orbits similar to those found in the boson-star metric in strong field regime. Furthers studies should thus be performed in this field. 

Another point that can be discussed is the extended mass that is expected to be present at the center of our galaxy. This mass should be composed of stars, stellar remnants or dark matter. The existence of this mass can modify the trajectory of the star. More precisely, \cite{2001A&A...374...95R} showed that in the Schwarzschild metric there is a Newtonien precession of the orbit evolving in the opposite sense of the relativistic precession, which can induced a decrease or a vanishing of the relativistic effect. Distinction between trajectories of stars in the Kerr and boson-star metrics should thus be difficult since the relativistic shift will decrease. However, a study performed by \cite{2010PhRvD..81f2002M} showed that the Lense-Thirring effect affecting the trajectory of the star should be detectable even with an extended mass at the Galactic center if its semi-major axis is inferior to $\approx 0.5$ mpc, which is the case of the stars studied in this paper. Besides, we should mention that it is not excluded that this extended mass could be small or even absent.

\bibliographystyle{plainnat}
\bibliography{biblio}

\end{document}